\documentclass[preprint,12pt]{elsarticle}

\usepackage{amssymb}

\usepackage{amsmath}
\usepackage{indentfirst}
\usepackage{bm}
\usepackage{graphicx}
\usepackage{epsfig}
\usepackage{setspace}
\usepackage{xcolor}
\usepackage{color}
\usepackage{soul}
\usepackage{epstopdf}
\usepackage[section]{placeins}
\usepackage{hyperref}

\journal{Physics Reports}

\begin{document}

\begin{frontmatter}



\title{Vibrational resonance: A review}


\author[label1]{Jianhua Yang}

\address[label1]{Jiangsu Key Laboratory of Mine Mechanical and Electrical Equipment, School of Mechatronic Engineering, China University of Mining and Technology, Xuzhou, 221116, Jiangsu, People's Republic of China}

\author[label2]{S. Rajasekar}

\address[label2]{School of Physics, Bharathidasan University, Tiruchirapalli, 620024, Tamilnadu, India}

\author[label3]{Miguel A. F. Sanju\'an\corref{cor1}}

\ead{miguel.sanjuan@urjc.es}
\cortext[cor1]{Corresponding author}

\address[label3]{Nonlinear Dynamics, Chaos and Complex Systems Group, Departamento de Fisica, Universidad Rey Juan Carlos, Tulipan s/n, Mostoles,             28933, Madrid, Spain}

\begin{abstract}
Over the past two decades, vibrational resonance has garnered significant interest and evolved into a prominent research field. Classical vibrational resonance examines the response of a nonlinear system excited by two signals: a weak, slowly varying characteristic signal, and a fast-varying auxiliary signal. The characteristic signal operates on a much longer time scale than the auxiliary signal. Through the cooperation of the nonlinear system and these two excitations, the faint input can be substantially amplified, showcasing the constructive role of the fast-varying signal. Since its inception, vibrational resonance has been extensively studied across various disciplines, including physics, mathematics, biology, neuroscience, laser science, chemistry, and engineering. Here, we delve into a detailed discussion of vibrational resonance and the most recent advances, beginning with an introduction to characteristic signals commonly used in its study. Furthermore, we compile numerous nonlinear models where vibrational resonance has been observed to enhance readers' understanding and provide a basis for comparison. Subsequently, we present the metrics used to quantify vibrational resonance, as well as offer a theoretical formulation. This encompasses the method of direct separation of motions, linear and nonlinear vibrational resonance, re-scaled vibrational resonance, ultrasensitive vibrational resonance, and the role of noise in vibrational resonance. Later, we showcase two practical applications of vibrational resonance: one in image processing and the other in fault diagnosis. This presentation offers a comprehensive and versatile overview of vibrational resonance, exploring various facets and highlighting promising avenues for future research in both theory and engineering applications.

\end{abstract}





\end{frontmatter}

\section{Introduction}
\label{int}

Multiple excitations with different time scales are frequently utilized across various disciplines.  These excitations may manifest as forces, electric currents, or other signal types.  When a nonlinear system is subjected to simultaneous multiple excitations, the response to the slow varying excitation closely depends on the fast varying excitation. This phenomenon is known as {\it vibrational resonance} discovered by Landa and McClintock in their seminal paper \cite{ref1}. Inspired by another well known nonlinear phenomenon, stochastic resonance \cite{ref2}, \cite{ref3}, Landa and McClintock termed this new phenomenon as vibrational resonance. In their work, the role of noise in stochastic resonance is replaced by a high-frequency harmonic signal in vibrational resonance. Although vibrational resonance shares some similarities with stochastic resonance, particularly regarding dynamical behaviors, they possess distinct essential properties, thus constituting a new subfield of nonlinear dynamics \cite{ref4}. Initially, vibrational resonance primarily focused on the response amplitude of a nonlinear system to a weak low-frequency harmonic excitation.  Landa and McClintock conducted their initial studies of vibrational resonance in a typical bistable Duffing oscillator, examining both overdamped and underdamped versions. They found that the response amplitude of the nonlinear system at low frequencies depends nontrivially on the amplitude of the high-frequency signal. Specifically, the curve depicting the response amplitude at low frequencies versus the amplitude of the high-frequency signal exhibits a resonance-like shape. This shape resembles the well-known frequency-response curve. At the resonance peak, the amplitude of the weak, low-frequency excitation component in the response is significantly amplified by the high-frequency signal. The classical vibrational resonance comprises three basic elements: (i) nonlinear systems, (ii) slowly varying excitation (characteristic signal), and (iii) fast varying excitation (auxiliary signal). It is important to note that the terms "fast" and "slow" are primarily used for comparing the two excitation signals, with the slowly varying excitation typically being weak. Through the interaction of these three basic elements, vibrational resonance emerges. In many cases, valuable information is conveyed through a weak, low-frequency, or slowly varying signal. Therefore, the development of vibrational resonance theory is crucial for detecting and amplifying weak signals.

In their pioneering work, Landa and McClintock introduced vibrational resonance based on numerical simulations.  Subsequently, Gitterman analyzed a second-order bistable system and provided an analytical explanation for the occurrence of vibrational resonance \cite{ref5}. Later, Blekhman and Landa conducted a more specific analytical study on vibrational resonance in overdamped and underdamped Duffing oscillators \cite{ref6}, deriving conditions for single and double resonance patterns. Furthermore, they highlighted errors between analytical and numerical simulations. The analytical technique commonly used to analyze vibrational resonance is known as the method of direct separation of motions \cite{ref7}-\cite{ref9}. This method is typically employed to obtain responses at the excitation frequency of the slowly varying harmonic signal. For more accurate results or results at nonlinear frequency components, such as the response amplitude at subharmonic, superharmonic, or combination frequencies, other analytical methods like the average method, perturbation method, multiscale method, and harmonic balance method can be used \cite{ref7}, \cite{ref10}]. Compared to other methods, the derivation process of the method of direct separation of motions is simpler and more straightforward, making it the preferred choice when responses primarily involve fast and slow motions. Blekhman and Landa extensively explored vibrational resonance using the method of direct separation of motions. They analyzed vibrational resonance with single or multiple resonance patterns in various systems, including quintic oscillators with potential function in different configurations \cite{ref11}-\cite{ref14}, coupled anharmonic oscillators \cite{ref15}, \cite{ref16},  one-way coupled bistable systems \cite{ref17}, rough potential systems \cite{ref18}, Toda potential systems \cite{ref19}, periodic potential systems \cite{ref20}-\cite{ref23}, quantum systems \cite{ref24}-\cite{ref26}, nano-electromechanical systems \cite{ref27}, and chemical reaction models \cite{ref28}, among others.

Effects of time delay have been incorporated into the study of  vibrational resonance. The delay parameter has been observed to induce periodic vibrational resonance in a simple delayed bistable system, with the period of the resonance peak relative to the time delay parameter being equal to the period of the high-frequency excitation signal \cite{ref29}. In coupled systems \cite{ref30}-\cite{ref32}, delay-induced vibrational resonance exhibits two periods when the delay time is sufficiently long, coinciding with the periods of the two excitations \cite{ref30}, \cite{ref32}, \cite{ref33}. Analysis of vibrational resonance considering time delay has been conducted in various systems, including two coupled overdamped anharmonic oscillators \cite{ref15}, the Duffing oscillator \cite{ref34}, \cite{ref35}, an asymmetric bistable system \cite{ref36}, the FitzHugh-Nagumo system \cite{ref37}-\cite{ref39}, the discrete Rulkov neuronal model \cite{ref40}, a genetic toggle switch \cite{ref41}, a gene transcriptional regulatory system \cite{ref42}, and a harmonically trapped potential system  \cite{ref43}, among others. In time-delayed systems, it has been observed that high-frequency excitation alone is not always sufficient to induce resonance in some cases \cite{ref44}, \cite{ref45}. The high-frequency excitation does influence the waveform of the response. When considering time delay in vibrational resonance, it is necessary to account for the Hopf bifurcation induced by the time delay \cite{ref46}. Additionally, due to the coexistence of time delay and multiple excitations, it is essential to consider the combined effect of high-frequency excitation and the delay term on the response simultaneously.

In recent years, fractional nonlinear systems have become an important type of system for addressing various science and engineering problems. Fractional systems incorporate fractional-order derivatives, originating from the long-standing issue of fractional calculus. Specifically, while the traditional derivative order is an integer, in some special cases, it can be non-integer. Fractional nonlinear systems find widespread use in fields such as dynamic modeling, automatic control, electromagnetism, among others, and are even referred to as the the $21^{st}$ century system \cite{ref47}. There are many interesting findings regarding vibrational resonance in fractional systems. A double resonance pattern caused by a fractional-order derivative term has been observed in an overdamped fractional Duffing oscillator \cite{ref48}. Vibrational resonance has also been observed in the Duffing system with fractional-order external and intrinsic damping \cite{ref49}, \cite{ref50}, fractional multistable systems \cite{ref51}, fractional anharmonic coupled systems \cite{ref52}, fractional quintic systems \cite{ref53}-\cite{ref55}, fractional birhythmic biological systems \cite{ref56}, and fractional Toda oscillators \cite{ref57}, among others. Furthermore, the effects of fractional-order damping terms on vibrational resonance are directly related to the static bifurcation of fixed points of the equivalent system \cite{ref58}. Specifically, at critical points of the pitchfork bifurcation \cite{ref48}, \cite{ref59}-\cite{ref63}, the saddle-node bifurcation \cite{ref64}, and the transcritical bifurcation \cite{ref65}, the response amplitude curve exhibits an inflection point, thus influencing the resonance pattern of the curve.

Vibrational resonance in neuronal models and complex networks is currently a topic of significant interest. Ullner et al. \cite{ref66} investigated vibrational resonance in the FitzHugh-Nagumo neuron model. Additionally, vibrational resonance in the FitzHugh-Nagumo model has been studied from different perspectives \cite{ref67}-\cite{ref76}, along with analysis in a neuron-astrocyte coupled model \cite{ref77}. Vibrational resonance is also observed in single or coupled Hindmarsh-Rose neuronal systems \cite{ref78}. These networks may consist of various clusters of neuronal oscillators or other types of oscillators. The influences of different network factors on vibrational resonance have been analyzed, leading to many interesting findings \cite{ref79}-\cite{ref93}.

In a nonlinear system subjected to a harmonic characteristic excitation, vibrational resonance is not limited to linear response but also extends to nonlinear response frameworks. Specifically, vibrational resonance manifests at nonlinear frequencies, including the second harmonic frequency \cite{ref94} and even higher-order frequencies \cite{ref95}-\cite{ref97} of the low-frequency signal. It occurs at subharmonic \cite{ref98}-\cite{ref100}, superharmonic, and combination frequencies \cite{ref98}, as well as at the sum and difference frequencies of the low-frequency and the natural frequency of a system \cite{ref101}. This phenomenon has been investigated in a two-level quantum system for both linear and nonlinear vibrational resonance responses \cite{ref26}.

One type of vibrational resonance is ghost vibrational resonance, which occurs as a high-frequency excitation-induced resonance at a fundamental frequency absent in the input low-frequency excitation \cite{ref102}-\cite{ref105}. Similar to stochastic resonance, ghost vibrational resonance exhibits features akin to ghost stochastic resonance \cite{ref106}. Another variant, entropic vibrational resonance \cite{ref107}, \cite{ref108}, shares similarities with entropic stochastic resonance \cite{ref109}.

Vibrational resonance induced by various characteristic signals has garnered significant attention. Apart from harmonic signals, characteristic signals can take the form of anharmonic periodic signals, aperiodic binary signals, frequency modulated signals, and other complex forms in engineering applications. Notably, binary aperiodic signals can induce aperiodic vibrational resonance. Chizhevsky and Giacomelli \cite{ref110} analyzed aperiodic vibrational resonance induced by aperiodic binary excitation, introducing the cross-correlation coefficient index for analysis.

Furthermore, traditional vibrational resonance typically involves dealing with slowly varying characteristic signals. However, in many scenarios, the characteristic signal manifests as fast-varying. Vibrational resonance has been investigated in the presence of a fast characteristic signal \cite{ref111}-\cite{ref113}, amplitude-modulated signals \cite{ref114}, aperiodic binary signals \cite{ref115}-\cite{ref117}, and frequency-modulated signals \cite{ref118}, \cite{ref119}. Exploring resonance with a complex frequency-modulated signal amidst strong noise background is intriguing \cite{ref120}. To quantify frequency-modulated signal-induced vibrational resonance, spectral amplification factors \cite{ref119} and fusion indices \cite{ref121} have been introduced. The fusion index serves to measure the necessary and sufficient conditions for vibrational resonance and describe amplification and the ``similarity"  of the output waveform compared to the input waveform simultaneously. Additionally, a modified cross-correlation coefficient has been utilized to eliminate errors between the input signal and system output induced by phase delay \cite{ref122}, thereby enhancing performance measures for aperiodic vibrational resonance.

Noise is nearly ubiquitous across various scientific and engineering fields. Understanding how nonlinear systems respond to simultaneous noise, fast, and slow excitations is a crucial concern \cite{ref123}. Particularly, much work has been done to explore the beneficial effects of different types of noise on vibrational resonance. Zaikin et al. \cite{ref124} have investigated vibrational resonance in coupled oscillators excited by parametric noise. They utilized a piecewise linear system for each oscillator, approximating the classical bistable system. Their findings revealed that multiplicative noise applied to each oscillator induces a phase transition, resulting in bistability. Consequently, vibrational resonance occurs, enhancing the weak signal. Essentially, a certain amount of noise fosters vibrational resonance. Subsequent analyses have delved into the interplay between stochastic resonance and vibrational resonance in various nonlinear models \cite{ref125}-\cite{ref147}. Some of these studies have identified instances where vibrational resonance enhances and controls stochastic resonance.

Very recently, there has been a growing interest in exploring vibrational resonance across various engineering applications. One notable area of exploration is vibrational resonance within energy harvesting systems \cite{ref148}-\cite{ref151}. Additionally, research has been done to investigate its potential utility in equipment fault diagnosis \cite{ref152}-\cite{ref157}. Moreover, vibrational resonance in image processing has also been investigated \cite{ref158}, \cite{ref159}. Recognizing the significance of vision research, there is a growing interest in further exploring vibrational resonance in the realm of image perception. We believe that vibrational resonance will continue to play a vital role in various engineering applications, prompting increased attention and research in this domain.

In 2020, to commemorate the 20th anniversary since the inception of the concept of vibrational resonance, as well as McClintock P V E's 80th birthday, Vincent et al. organized a thematic/special issue on nonlinear resonances in driven systems, encompassing vibrational and stochastic resonance \cite{ref160}, \cite{ref161}. This issue was published by {\it Philos. T. R. Soc. A}. The first part of the issue was released online in January 2021, followed by the second part in April 2021. Within this compilation, there are 11 papers dedicated to vibrational resonance \cite{ref26}, \cite{ref77}, \cite{ref93}, \cite{ref159}, \cite{ref162}-\cite{ref168}. These papers delve into vibrational resonance within quantum systems, laser systems, signal processors, logic gates, position-dependent mass systems, and energy harvesters.

Although there are monographs on vibrational resonance \cite{ref58}, \cite{ref169}, a comprehensive review article is still lacking to aid researchers in gaining a better understanding of vibrational resonance and its related topics. This serves as the primary motivation behind this work. The slow-varying excitation may manifest in periodic or aperiodic forms. On one hand, different forms of excitations can induce various resonance patterns. On the other hand, vibrational resonance occurs across a spectrum of nonlinear systems. Therefore, it is essential to provide summaries of these excitations and nonlinear models. In Section~\ref{excitations}, we introduce some typical characteristic signals in periodic or aperiodic forms. Next, we summarize the nonlinear models for vibrational resonance in Section~\ref{models}. Standard performance measures of different kinds of vibrational resonance are presented in Section~\ref{measures}. In Section~\ref{theory}, we delve into the theoretical formulation of the method of direct separation of motions, linear and nonlinear vibrational resonance, and ultrasensitive vibrational resonance. Additionally, the re-scaled vibrational resonance is introduced to address the fast-varying characteristic signal, with a brief analysis of the role of noise in vibrational resonance. In Section~\ref{Appl}, two applications of vibrational resonance are presented. Finally, in Section~\ref{conc}, we offer the main conclusions of the review paper and provide some future outlooks.

\section{The excitations}
\label{excitations}
Typically, vibrational resonance excitation consists of two signals: the characteristic signal and the auxiliary signal, which can take various forms. The characteristic signal in the following common signal types is often considered.

\subsection{Periodic signal}
A periodic signal is the most frequently used in various fields. Among them, the commonly used characteristic or auxiliary signal is a harmonic function. The harmonic signal with amplitude $A$ can be in a sine waveform $s(t)= A\sin(\omega t + \phi)$ or $s(t)= A\sin(2\pi f t + \phi)$, and in a cosine waveform $s(t)= A\cos(\omega t + \phi)$ or $s(t)= A\cos(2\pi f t + \phi)$. For $\omega$ and $f$, their units are $rad/s$ and $Hz$, respectively. The initial phase is $\phi$.  Additionally, there are anharmonic periodic signals, each with different effects on vibrational resonance and related phenomena. Below, we provide examples of typical anharmonic periodic signals, including:

\noindent {\bf(\romannumeral1) Square signal}
\begin{equation}
\label{eq1}
   s(t) = \left\{ \begin{array}{rl}
 A, & (2n - 2) \pi /\omega  < t < (2n - 1) \pi /\omega , \\
  - A, & {\rm{(}}2n - 1)\pi /\omega  < t < 2n\pi /\omega , \\
 \end{array} \right. \;\;n = 1,2, \cdots .
\end{equation}

\noindent {\bf(\romannumeral2) Asymmetric sawtooth signal}
\begin{equation}
 \label{eq2}
   s(t) = \left\{ \begin{array}{rl}
 \frac{2At}{T}  , & (2n - 2)\pi /\omega  < t < (2n - 1)\pi /\omega , \\
 \frac{2At}{T} - 2A , & (2n - 1) \pi /\omega  < t < 2n\pi /\omega , \\
 \end{array} \right. \;\; n = 1,2, \cdots .
\end{equation}

\noindent {\bf(\romannumeral3) Symmetric sawtooth signal}
\begin{equation}
 \label{eq3}
  s(t) = \left\{ \begin{array}{rl}
 \frac{4At}{T}, & (2n - 2)\pi / \omega  < t < (\frac{4n - 3}{2}) \pi /\omega , \\
  - \frac{4At}{T} + 2A, & (\frac{4n - 3}{2}) \pi /\omega  < t
          < (\frac{4n - 1}{2} ) \pi / \omega , \\
 \frac{4At}{T} - 4A,  &  (\frac{4n - 1 }{2}{)}\pi /
         \omega  < t < 2n\pi /\omega , \\
 \end{array} \right. \;\; n = 1,2, \cdots .
\end{equation}

\noindent {\bf(\romannumeral4) Modulus of sine signal}
\begin{equation}
 \label{eq4}
   s(t) = A\left| \sin \left( \frac{\omega }{2} t \right) \right|.
\end{equation}

\begin{figure}[!h]
\begin{center}
\includegraphics[width=0.7\linewidth]{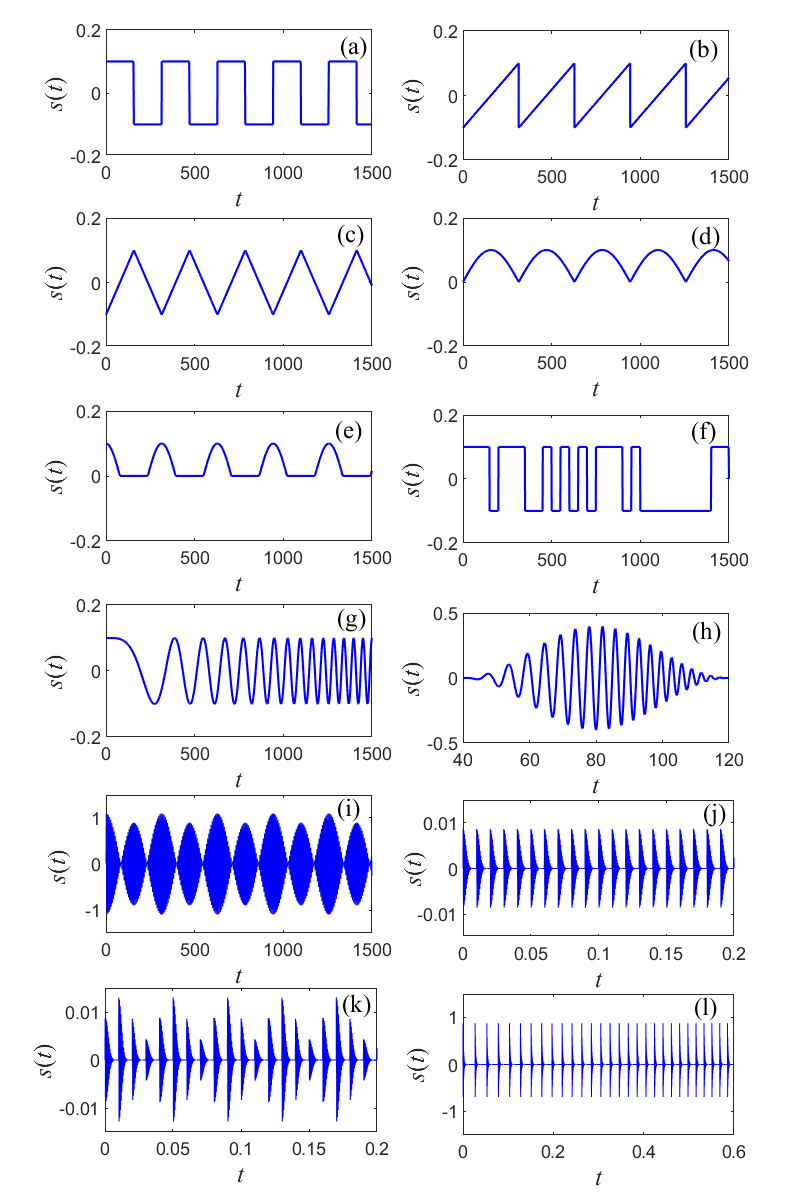}
\end{center}
\caption{Waveforms of different types of characteristic signals. (a) square signal, (b) asymmetric sawtooth signal, (c) symmetric sawtooth signal, (d) modulus of sine signal, (e) rectified cosine signal, (f) aperiodic binary signal, (g) frequency modulated signal, (h) echo chirp signal, (i) amplitude modulated harmonic signal, (j) vibration signal of a bearing with fault in out race (constant speed condition), (k) vibration signal of a bearing with fault in inner race (constant speed condition), (l) vibration signal of a bearing with fault in out race (time-varying speed condition).}
\label{figsignal}
\end{figure}

\noindent {\bf(\romannumeral5) Rectified sine signal}
\begin{equation}
 \label{eq5}
     s(t) = \left\{ \begin{array}{rl}
   A \sin \omega t, & (2n - 2)\pi / \omega  < t < (2n - 1)\pi /\omega , \\
     0 , & (2n - 1) \pi / \omega  < t < 2n\pi /\omega , \\
 \end{array} \right.  \;\; n = 1,2, \cdots .
\end{equation}
\noindent In the above formulas, \( T \) is the period of the excitation that is equal to \( 2\pi/\omega \). The slow excitation in anharmonic periodic form has been used in some works of stochastic resonance in nonlinear systems \cite{ref170}, \cite{ref171}, however not much work on vibrational resonance has been done. The form of the periodic signal is found to influence directly the static bifurcation and the appearance of the vibrational resonance \cite{ref172}. By Fourier series expansion, an anharmonic excitation can be viewed as the superposition of constant and infinite harmonic terms. Moreover, an anharmonic periodic excitation may induce much more complex dynamical behaviors \cite{ref142}, \cite{ref173}-\cite{ref176}. This suggests that it is crucial to conduct more in-depth analysis of vibrational resonance induced by various anharmonic periodic excitations, encompassing both the characteristic and auxiliary signals in an anharmonic periodic form. The waveforms of a harmonic and anharmonic periodic signals mentioned above and certain other signals are given in Fig.~\ref{figsignal}. Among them, in Fig.~\ref{figsignal} (a)-(e), the periodic signals corresponding to \textbf{(\romannumeral1)}-\textbf{(\romannumeral5)} are sequentially presented. The parameters of these signals, such as the period, the duty ratio, etc., have important effects on vibrational resonance and related dynamic phenomena.
%

\subsection{Aperiodic binary signal}
The aperiodic vibrational resonance is reported in \cite{ref110}, \cite{ref115}, \cite{ref116}. The aperiodic binary signal is governed by

\begin{equation}
 \label{eq6}
 s(t) = A \sum \limits_{j =  - \infty }^{ + \infty } {R_j}\Gamma (t - jT),
     \quad \Gamma (\tau) = \left\{ \begin{array}{ll}
    1, & \tau \in [0,\:T] \\
    0, & \tau \notin [0,\:T] \\
    \end{array} \right. .
\end{equation}
and $R_j$ is a random number sequence of $+1$ or $-1$ with an independent stochastic distribution. $A$ is the amplitude of the aperiodic signal. $\Gamma(t)$ is a random pulse sequence with minimal pulse width $T$. The time series of an aperiodic binary signal is shown in Fig.~\ref{figsignal}(f).

The aperiodic vibrational resonance has been used in noisy image enhancement \cite{ref158}, \cite{ref159}. For monochrome image processing, the image is usually transformed to an aperiodic binary signal. Hence, a further study on aperiodic vibrational resonance is important. As an extension, stochastic resonance induced by an M-ary signal has been investigated in certain systems \cite{ref177}-\cite{ref179}. As an example of an application, if we deal with a gray image, we must first transform the image into an M-ary signal. Therefore, the vibrational resonance induced by an M-ary signal is worth exploring.

\subsection{Frequency-modulated signal}

The linear frequency modulated signal is a simple and typical signal in science and engineering, whose form is
\begin{equation}
 \label{eq7}
     s(t) = A\cos (\pi \gamma {t^2} + 2\pi ft + \phi ),
\end{equation}
where $A$ and $\phi$ are the amplitude and the initial phase, respectively. $\gamma$ (the chirp rate) and $f$ (the starting frequency) characterize the modulation characteristics of the frequency. The instantaneous frequency $f_\mathrm{in}(t)$ is the derivative of the phase, i.e., $ f_\mathrm{in} (t) = 2\pi \gamma t + 2 \pi f $. The vibrational resonance caused by the linear frequency modulated signal has been studied in \cite{ref118} and \cite{ref119}.

Another typical frequency modulated signal is the echo chirp signal, which is widely applied in compressed sensing, range measurement, velocity estimation, and so on \cite{ref121}. The expression of each echo chirp component is
\begin{equation}
 \label{eq8}
    e(t) = \left\{ \begin{array}{ll}
 {\mathrm{e}}^{j 2 \pi \left[ (B / (2T))t^2 + (f_\mathrm{c} - B / 2) t \right]},
         & 0 \le t \le T \\
     0, &  \mathrm{otherwise}
 \end{array} \right.   .
\end{equation}
Herein, $B$, $T$ and $f_\mathrm{c} >0$ are the bandwidth, the duration and the central frequency, respectively. The instantaneous frequency of the echo chirp component is
$f_\mathrm{in} = (B/T)t + f_\mathrm{c} - B/2$. The echo chirp signal which is composed by $n$ chirp components is defined as \cite{ref180}
\begin{equation}
 \label{eq9}
   s(t) = \sum \limits_{i = 1}^n {{\alpha ^{i - 1}}
         w(t - id - c)} \Re (e(t - id - c)) ,
\end{equation}
where $\alpha$ is the attenuation rate, $d$ is the delay between adjacent chirp components, and $c$ is the initial delayed time. In addition, the symbol $\Re (\bullet)$ represents the real part of the chirp signal, and $w(t)$ is given by
\begin{equation}
 \label{eq10}
 w(t) = \left\{ \begin{array}{ll}
   0.5[1 - \cos (2\pi t/T)], &  0 \le t \le T  \\
   0, & \mathrm{otherwise}
\end{array} \right\}.
\end{equation}
The plots of the linear frequency modulated signal and the echo chirp signal are shown in Figs.~\ref{figsignal}(g) and \ref{figsignal}(h), respectively. Usually, the frequency modulated signals have a very clear physical or engineering background. For example, the characteristic frequency of a vibration signal of a faulty rolling bearing is in the frequency-modulated form when the equipment operating in a variable speed condition \cite{ref120}. Further, it is interesting to analyze the features of vibrational resonance induced by different forms of the nonlinear frequency modulated signal \cite{ref181}-\cite{ref185} in different engineering fields.

\subsection{Amplitude modulated signal}

The amplitude modulated harmonic signal induced vibrational resonance have been studied in \cite{ref31}, \cite{ref114}, \cite{ref186}-\cite{ref190}. The corresponding signal model is
\begin{equation}
 \label{eq11}
    s(t) = \left[ {A + B\cos \Omega t} \right]\cos \omega t.
\end{equation}
Its plot is given in Fig.~\ref{figsignal}(i). In addition, the amplitude modulated aperiodic binary signal is \cite{ref114}
\begin{equation}
 \label{eq12}
  s(t) = \left[ {A + B \cos \Omega t} \right]
         \sum\limits_{j =  - \infty }^{ + \infty } {{R_j}\Gamma (t - jT)}.
\end{equation}

In fact, the vibration signal of a faulty bearing can be considered as an amplitude modulated signal. For the vibration signal of a bearing with outer race fault, its simulated signal form is \cite{ref191}-\cite{ref193}
\begin{equation}
 \label{eq13}
\left\{ {\begin{array}{*{20}{c}}
   {s\left( t \right) = A\sin \left( {2\pi {f_{\text{n}}}t} \right)\exp \left\{ { - B{{\left[ {t - i\left( t \right)/{f_0}} \right]}^2}} \right\}}  \\
   {i\left( t \right) = {\text{floor}}\left[ {{f_0}t} \right]}  \\

 \end{array} } \right.,
\end{equation}
where $A$ is the amplitude of the pulse signal, $B$ is the attenuation coefficient, $f_\mathrm{n}$ and $f_0$ are the natural frequency and the characteristic fault frequency, respectively. The function $i(t)$ represents the repetitions, and the function $floor [\bullet]$ is the floor function.

With an inner race or a ball element fault, the bearing vibration simulated signal is \cite{ref194}, \cite{ref195}
\begin{equation}
 \label{eq14}
  \left\{ {\begin{array}{*{20}{c}}
   {s\left( t \right) = \sum\limits_i {H\left( t \right)} h\left( {t - iT} \right)}  \\
   {H\left( t \right) = {A_{\text{0}}}\cos \left( {2\pi {f_{\text{r}}}t} \right){\text{ + }}C}  \\
   {h\left( t \right) = A\cos \left( {2\pi {f_{\text{n}}}t} \right)\exp \left( { - B{t^2}} \right)}  \\

 \end{array} } \right.,
\end{equation}
where the characteristic frequency is $1/T$, and $T$ is the period of the pulse. The function $H(t)$ is the modulated signal with amplitude $A_0$,  $C$ is the modulation bias, and $f_\mathrm{r}$ is the rotating frequency.

As shown in Figs.~\ref{figsignal}(j) and \ref{figsignal}(k), for the signals in Eq.~(\ref{eq13}) and Eq.~(\ref{eq14}), the periodic property and the amplitude modulation are apparently manifested. Vibrational resonance is investigated with the signals of Eq.~(\ref{eq13}) and Eq.~(\ref{eq14}) on bearing fault diagnosis \cite{ref152}-\cite{ref157}. Different from other signal amplifier, not only the characteristic frequency is enhanced but other disturbed frequencies are suppressed.

Sometimes, the frequency and the amplitude of the characteristic signal are modulated simultaneously. We still use the bearing fault diagnosis as the engineering background. While the rotating machinery is operating in variable work conditions, the vibration signal of the bearing has both the frequency and the amplitude modulated property. For example, under the variable speed condition, the vibration signal of a bearing with outer race fault is simulated by the following signal \cite{ref196}
\begin{equation}
 \label{eq15}
   \left\{ \begin{gathered}
  s(t) = \sum\limits_{m = 1}^M A \exp \left\{ { - B[t - v(t)]} \right\}\sin \left\{ {2\pi {f_{\text{n}}}[t - v(t)] + \phi } \right\} \hfill \\
  v(t) = \sum\limits_{i = 1}^{m - 1} {{T_i}}  \hfill \\
\end{gathered}  \right.,
\end{equation}
where $m$ is the impulse index, $T_i$ is the repeating period of the $i$th impulse. The function $v(t)$ is the total occurrence time of $(m-1)$ impulses and is a function of time $t$. The plot of this signal is presented in Fig.~\ref{figsignal}(l). The effect of the above signal has been investigated by the stochastic resonance method \cite{ref196}. We still need to analyze the vibrational resonance caused by this signal in the bearing fault diagnosis under different variable working conditions. With the rapid development of mechanical industry, we need to pay enough attention to the research of vibrational resonance in the field of mechanics.

\subsection{Logical signal}

Logical vibrational resonance has been investigated and interesting results have been reported \cite{ref143}, \cite{ref197}-\cite{ref201}. As a result, the logical inputs can be as the characteristic signals, which are random permutations consisting of four logic states (0, 0), (0, 1), (1, 0), (1, 1). Since logical operations are widely used in the field of electronics and computer technology, the theoretical analysis, as well as the analysis oriented towards applications of logical vibrational resonance are of great importance for the future.

\section{Nonlinear models of vibrational resonance}
\label{models}
Depending on whether the system incorporates fractional damping, time delay, or noise, nonlinear models for vibrational resonance are typically classified into five main categories: ordinary differential systems, mapping systems, fractional differential systems, delayed differential systems, and stochastic differential systems. In the literature, authors may adopt different expressions or symbols to label the excitations and the equations. To simplify the system model without the need for a specific notation, we employ $s_\mathrm{L}(t)$, $s_\mathrm{H}(t)$, and $\xi(t)$ to denote the slowly-varying characteristic signal, the fast-varying auxiliary signal, and the noise, respectively.

\subsection{Models with ordinary differential systems}
We present several common nonlinear systems characterized by ordinary differential equation models in the investigation of vibrational resonance. These models are categorized based on the shape of the potential function, the type of damping (linear or nonlinear), the number of subsystems, and the application context of the systems.

\subsubsection{Symmetric bistable oscillators}

The overdamped and underdamped symmetric bistable Duffing oscillator systems are governed by
\begin{equation}
  \label{eq16}
    \dot{x} +  \omega_0^2 x + \beta x^3 =  s_\mathrm{L}(t) + s_\mathrm{H} (t)
\end{equation}
and
\begin{equation}
  \label{eq17}
    \ddot x + \delta \dot x + \omega_0^2 x + \beta x^3
     = s_\mathrm{L}(t) + s_\mathrm{H} (t),
\end{equation}
respectively.

In Eqs.~(\ref{eq16}) and (\ref{eq17}), $\omega_0^2 < 0 $, $\beta >0 $ and $\delta >0$.
The potential function is
\begin{equation}
  \label{eq18}
 V(x) =   \frac{1}{2} \omega_0^2 x^2 + \frac{1}{4} \beta x^4 \,.
\end{equation}
In absence of $s_\mathrm{L}(t)$ and $s_\mathrm{H} (t)$, these two systems have three equilibria $x=0$ and $x_{\pm} = \pm \sqrt {\omega_0^2/\beta}$. Many vibrational resonance works have considered this kind of system \cite{ref1}, \cite{ref5}, \cite{ref6}, \cite{ref94}, \cite{ref96}, \cite{ref98}, \cite{ref101}, \cite{ref102}, \cite{ref111}, \cite{ref119}, \cite{ref121}, \cite{ref127}, \cite{ref132}, \cite{ref141}, \cite{ref152}-\cite{ref155}, \cite{ref202}-\cite{ref205}]. Vibrational resonance in a bistable system is shown in an experimental setup of Eq.~(\ref{eq17}) \cite{ref206}. The symmetric bistable system is a typical model for many physical and engineering systems. In addition, if $\omega_0^2>0$ and $\beta > 0$, the potential function is in a monostable form with only a stable equilibrium $x=0$; if $\omega_0^2>0$ and $\beta<0$, then the system is a softening Duffing system and the potential function has a stable equilibrium $x=0$ and two unstable equilibria $x_{\pm} = \pm \sqrt {\omega_0^2/\beta}$ \cite{ref207}. In current researches on vibrational resonance, the bistable systems of Eqs.~(\ref{eq16}) and (\ref{eq17}) are the most typical and studied models. There is a need for further research on the connection between vibrational resonance and more complex dynamical phenomena, including various bifurcation patterns, safe basin erosion \cite{ref208}, and others, within the framework of this typical bistable system model.

\begin{figure}
\begin{center}
\includegraphics[width=0.7\linewidth]{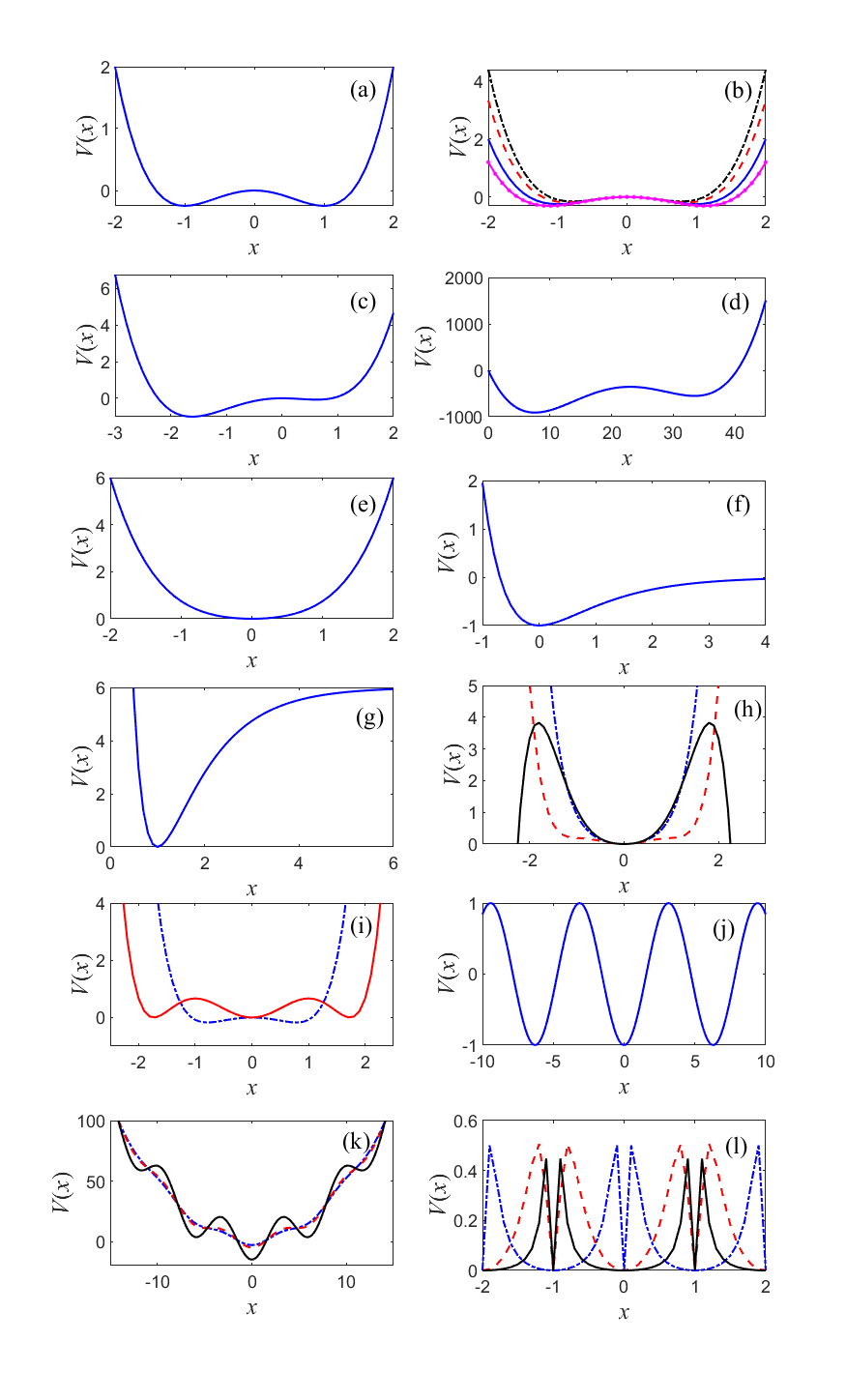}
\end{center}
\caption{Plots of the potential function of some nonlinear oscillators, (a) symmetric bistable potential in Eq~(\ref{eq18}), (b) symmetric bistable potential in Eq.~(\ref{eq20}), (c) asymmetric bistable potential in Eq.~(\ref{eq22}), (d) asymmetric bistable potential in Eq.~(\ref{eq24}), (e) monostable potential in Eq.~(\ref{eq18}), (f) Morse potential function in Eq.~(\ref{eq29}), (g) Tietz-Hua potential in Eq.~(\ref{eq31}), (h) function in Eq.~(\ref{eq33}) with monostable potential, (i) function in Eq.~(\ref{eq33}) with bistable and tristable potentials, (j) periodic potential function, (k) harmonically trapped potential function, (l) asymmetric deformable potential function.}
\label{figpotential}
\end{figure}

Another overdamped symmetric bistable system wherein vibrational resonance is investigated  \cite{ref114}, \cite{ref115}, \cite{ref118}, \cite{ref209}] is
\begin{equation}
 \label{eq19}
   \dot{x} + \omega_0^2 x + \beta x \left| x \right|^{\alpha  - 1}
       = s_\mathrm{L}(t) + s_\mathrm{H}(t) .
\end{equation}
The potential function is
\begin{equation}
 \label{eq20}
     V(x) =   \frac{1}{2} \omega_0^2 x^2 + \frac{\beta}{{\alpha  + 1}}
            \left| x \right|^{\alpha  + 1}, \quad
             \omega_0^2 < 0, \; \beta>0, \; \alpha >0 .
\end{equation}
When $\alpha > 1$, the potential function has two stable equilibria $x_{\pm} =  \pm {\left( { - \frac{{\omega _0^2}}{\beta }} \right)^{\frac{1}{{\alpha  - 1}}}}$ and an unstable equilibrium $x=0$. The parameter $\alpha$ is a real irrational or rational number. In engineering applications,  the value of $\alpha$ is usually related to the material property. The detailed physical background has been explained in previous works \cite{ref210}-\cite{ref212}. For the potential function, the value of $\alpha$ mainly influences its deepness. In the study of vibrational resonance, this kind of system can be viewed as a signal processor. For the case $\alpha=3$, the shape of the potential function presents the typical bistable configuration which is described by Eq.~(\ref{eq18}). In addition, a suitable value of $\alpha$ can cause the system response to be in a stronger resonance state. It is important to investigate this kind of potential further by using analytical, experimental and numerical ways. The plots of the symmetric bistable potential function corresponding to Eq.~(\ref{eq18}) and Eq.~(\ref{eq20}) are given in Figs.~\ref{figpotential}(a) and \ref{figpotential}(b), respectively. Herein, we choose four different values of $\alpha$ to obtain the curves in Fig.~\ref{figpotential}(b).

\subsubsection{Asymmetric bistable oscillators}

There are many interesting asymmetric bistable systems, such as

\begin{equation}
 \label{eq21}
    \ddot{x} + \delta \dot x + \frac{\mathrm{d} V}{\mathrm{d} x}
      = s_\mathrm{L}(t) + s_\mathrm{H}(t) ,
\end{equation}
where the potential function is
\begin{equation}
 \label{eq22}
   V(x) = \frac{1}{2}\omega _0^2 x^2 + \frac{1}{3} \alpha x^3
       + \frac{1}{4}\beta x^4.
\end{equation}
The potential function is in asymmetric double-well form for $\omega_0^2 < 0$, $\beta > 0$ and $\alpha \ne 0$ (see Fig.~\ref{figpotential}(c)). Vibrational resonance also takes place in the asymmetric bistable system of Eq.~(\ref{eq21}) for a wide range of input signal frequencies for which the resonance is impossible in the symmetric bistable system \cite{ref213}.

Another asymmetric bistable potential is
\begin{equation}
 \label{eq23}
   V(x) = \frac{1}{2} \omega _0^2 x^2 + \frac{1}{4}\beta x^4
          + \gamma x, \quad \omega_0^2 <0, \; \beta>0,
\end{equation}
where $\gamma $ is a parameter indicating the asymmetry. Vibrational resonance is investigated also in overdamped systems with the above asymmetric potential function theoretically, numerically and experimentally \cite{ref95}, \cite{ref110}, \cite{ref129}, \cite{ref214}. Interestingly, the asymmetric property is necessary for the occurrence of certain types of resonances. As an illustration, logical vibrational or stochastic resonance typically manifests only in the presence of a bias parameter, and the distinct logical resonances achieved through $AND$, $NAND$, $OR$, and $NOR$ calculations are contingent upon the tilting direction of the potential function. In addition, the asymmetry of the potential also influences the response amplitude curve of vibrational resonance, especially when the initial conditions are located in different potential wells in numerical calculations \cite{ref64}.

Another asymmetric bistable potential exists in the three-level atomic optical bistable system, e.g.,
\begin{subequations}
 \label{eq24}
 \begin{eqnarray}
   \dot{x} & = & -\frac{\mathrm{d} V}{\mathrm{d} x}
      + s_\mathrm{L}(t) + s_\mathrm{H}(t), \\
      V(x) & = &  - yx + {x^2}\left( {\frac{{{c_1}}}{2} + \frac{{{c_2}}}{3}x + \frac{{{c_3}}}{4}{x^2}} \right)
\end{eqnarray}
\end{subequations}
where $y$, $c_1$, $c_2$, $c_3$ are all constants \cite{ref215}. Here, the potential $V(x)$ is an asymmetric bistable form. The curve of the potential function is given in Fig.~\ref{figpotential}(d).

The groundwater-dependent plant ecosystem also has an asymmetric potential with three equilibria (two stable and one unstable) \cite{ref216}
\begin{subequations}
 \label{eq25}
\begin{eqnarray}
     \dot{x} = x ( x_\mathrm{cc} - x),
\end{eqnarray}
where
\begin{eqnarray}
    x_\mathrm{cc} & = &
      \left\{ \begin{array}{ll}
      a \left[ d(t) - d_\mathrm{inf}  \right]
      \left[ d_\mathrm{sup}  - d(t) \right], &
      \mathrm{if} \;\; d_\mathrm{inf}  < d < d_\mathrm{sup}  \\
      0, & \mathrm{otherwise} \\
     \end{array} \right. \\
     d(t) & = &  d_0 + \beta x + s_\mathrm{L}(t) + s_\mathrm{H} (t),
\end{eqnarray}
\end{subequations}
where $x$ is the phreatophyte biomass of the considered species, $x_\mathrm{cc}$ is the carrying capacity of the studied ecosystem, and $d(t)$ is the depth of groundwater level. The two excitations describe the variation of the groundwater table that caused by some factors. The potential function of the system presents an asymmetric bistable form with $x>0$.

\subsubsection{Monostable oscillators}

The nano-electromechanical resonator is a monostable system \cite{ref27}
\begin{equation}
 \label{eq26}
   \ddot{x} + \delta \dot{x} + \omega _0^2x + \beta x^3
      = F [1 + s_\mathrm{L}(t) + s_\mathrm{H}(t)] \cos \omega_\mathrm{f} t,
\end{equation}
where $\omega _0^2>0$, $\beta>0$, and $\omega_\mathrm{f}$ is in the vicinity of the fundamental mode frequency. Its potential function is given by Eq.~(\ref{eq18}) with $\omega_0^2>0$ and $\beta>0$. The plot of the monostable potential function is shown in Fig.~\ref{figpotential}(e).

The monostable system excited by biharmonic signals is \cite{ref217}
\begin{equation}
 \label{eq27}
    \ddot{x} + \delta \dot{x} + \omega_0^2 x +  \alpha x^2 + \beta x^3
        = s_\mathrm{L}(t) + s_\mathrm{H}(t) .
\end{equation}
The potential function is given by Eq.~(\ref{eq22}). When $\alpha=0$, $\omega_0^2$, $\beta>0$, the potential function is monostable and symmetric about $x=0$.  When $0 < \alpha^2<4\omega_0^2 \beta$, the potential function is still in monostable form but asymmetric.  When $\alpha^2>4\omega_0^2 \beta$,  the potential turns to a bistable form. An experimental setup for the system Eq.~(\ref{eq27}) was built and the underlying vibrational resonance was also investigated in this work.

The Morse oscillator can describe the photodissociation and the interatomic potential of molecules. It is a monostable system and the dynamical equation of motion is \cite{ref218}
\begin{equation}
 \label{eq28}
   \ddot{x} + \delta \dot{x} + \beta \mathrm{e}^{-x}
       \left( 1 - \mathrm{e}^{-x} \right) = s_\mathrm{L}(t)
           + s_\mathrm{H}(t) .
\end{equation}
The corresponding potential function is expressed as
\begin{equation}
 \label{eq29}
     V(x) = \frac{1}{2} \beta \mathrm{e}^{-x} \left( 1 - \mathrm{e}^{-x} \right).
\end{equation}
The plot of the potential function of the Morse oscillator is given in Fig.~\ref{figpotential}(f), and $V(x)$ has only one local minimum.

The system with a two-frequency-excited Tietz-Hua quantum well is also a monostable system exhibiting quantum resonance \cite{ref25}. The Hamiltonian of the system is
\begin{equation}
 \label{eq30}
   H = \frac{1}{2m} p_x^2 + V_\mathrm{TH}
      + \lambda x \left( s_\mathrm{L}(t) + s_\mathrm{H}(t) \right) .
\end{equation}
The Tietz-Hua potential function is
\begin{equation}
 \label{eq31}
    V_\mathrm{TH} (x) = V_0 \left[ \frac{
          1 - \mathrm{e}^{-b_h(x - x_\mathrm{e} )}  }
         {1 - c_\mathrm{h} \mathrm{e}^{ - b_\mathrm{h}
          (x - x_\mathrm{e} ) }} \right]^2.
\end{equation}
Its graph of $V_\mathrm{TH}(x)$ is presented in Fig.~\ref{figpotential}(g). The Tietz-Hua potential reduces to the classical Morse potential when its potential constant $c_\mathrm{h}$ is zero.

In the above four kinds of monostable systems, all the excitations are in harmonic form, and yet the monostable system especially has advantages in processing aperiodic pulse signal under noisy background, which has been verified in extracting steel wire rope flaw detection signal based on stochastic resonance \cite{ref219}, \cite{ref220}. At present, there is very little work on vibrational resonance of monostable system under the action of different aperiodic pulse signals. The related theoretical and applied researches are worthwhile to do in the near future.

\subsubsection{Quintic oscillators}

Single or multiple vibrational resonance can occur in a quintic oscillator \cite{ref11}-\cite{ref14}, \cite{ref221}
\begin{equation}
 \label{eq32}
    \ddot{x}  + \delta \dot{x} + \omega_0^2x + \beta x^3
       + \gamma x^5 = s_\mathrm{L}(t) + s_\mathrm{H}(t) .
\end{equation}
The potential function is
\begin{equation}
 \label{eq33}
   V(x) = \frac{1}{2} \omega_0^2 x^2 + \frac{1}{4} \beta x^4
       + \frac{1}{6}\gamma x^6  .
\end{equation}
By choosing different system parameters, the potential may be in a monostable, bistable, or tristable configuration. When $\omega_0^2$, $\beta$, $\gamma > 0$, the potential has a single-well. When $\omega_0^2$, $\gamma>0$, $\beta<0$, and $\beta^2 < 4 \omega_0^2 \gamma$, the potential also has a single well. When $\omega_0^2$, $\beta>0$, and $\gamma<0$, the potential has a double-hump single-well. The potential curves with different single-well shapes are presented in Fig.~\ref{figpotential}(h). When $\omega_0^2<0$, $\beta$, $\gamma>0$, the potential is bistable. When $\omega_0^2$, $\gamma>0$, $\beta<0$, and $\beta^2>16\omega_0^2 \gamma/3$, the potential has a triple-well shape. The plots of the potential with bistable and tristable\emph{} cases are plotted in Fig.~\ref{figpotential}(i). The potential $V(x)$ can be used in different occasions, e.g., modelling dynamics of optical bistability, magnetoelastic beam, folded intermediates of proteins, among some other problems \cite{ref11}.

\subsubsection{Periodic potential oscillators}

There is a series of vibrational resonances appearing in the pendulum system \cite{ref20}, \cite{ref140}, with overdamped version
\begin{equation}
 \label{eq34}
  \dot{x} + \sin x = s_\mathrm{L}(t) + s_\mathrm{H}(t)
\end{equation}
or with underdamped version
\begin{equation}
 \label{eq35}
    \ddot{x} + \delta \dot{x} + \sin x = s_\mathrm{L}(t) + s_\mathrm{H}(t).
\end{equation}
The potential function is $V(x)=-\cos x$ and its form is shown in Fig.~\ref{figpotential}(j). There are also some other modified periodic potential functions, such as $ V(x) =  - (1/k)\left[ \sin kx + (1/4) \sin 2kx \right] $ which represents a rocking ratchet \cite{ref21}. Vibrational resonance has been explored in a system featuring periodic potential, and it has been investigated in the context of cold atoms confined in an optical lattice \cite{ref222}.

Furthermore, vibrational resonance has been studied in a nonlinear dissipative system which has a symmetric periodic potential and a space-dependent nonlinear damping coefficient \cite{ref22}, \cite{ref223}. The system is
\begin{equation}
 \label{eq36}
   \ddot{x} + \gamma_0 \left[ 1 - \lambda \sin (kx + \phi ) \right]
       \dot{x} + V_0 \sin kx = s_\mathrm{L}(t) + s_\mathrm{H}(t),
\end{equation}
which is a more general case of Eq.~(\ref{eq35}).

\subsubsection{Harmonically trapped and deformable potential systems}

The harmonically trapped potential system is \cite{ref224}
\begin{equation}
 \label{eq37}
      \ddot{x} + \delta \dot{x} + \omega_0^2 x + \beta \sin x
         = s_\mathrm{L}(t) + s_\mathrm{H}(t),
          \quad \omega_0^2 > 0,\; \beta > 0 .
\end{equation}
Its potential is $V(x) = (1/2)\omega _0^2x^2 - \beta \cos x$. For different values of $\beta$, the potential may have one well, three wells and even more odd wells, as shown in Fig.~\ref{figpotential}(k).

Another interesting system exhibiting vibrational resonance is \cite{ref225}
\begin{equation}
\label{eq38}
       \ddot{x} + \delta \dot{x}
    + \frac{V_0 \left( 1 - r^2 \right)^2}{\pi}
       \frac{\sin \left( \pi x \right)
          \left[ 2r + \left( 1 + r^2 \right) \cos (\pi x) \right] }
         { \left[ 1 + r^2 + 2r\cos (\pi x) \right]^3}
           =  s_\mathrm{L}(t) + s_\mathrm{H}(t),
\end{equation}
with an asymmetrical deformable (Remoissenet-Peyrard) potential
\begin{equation}
 \label{eq39}
    V(x) = \frac{V_0 \left( 1 - r^2 \right)^2
          \left[ 1 - \cos (2\pi x) \right] }
            {(2\pi )^2 \left[ 1 + r^2 + 2r\cos (\pi x) \right]^2} .
\end{equation}
With different values of $r$, the potential presents different asymmetric deformable potentials, as shown in Fig.~\ref{figpotential}(l). When $r=0$, the system turns to a periodic potential system. The potential described in Eq.~(\ref{eq39}) is commonly employed when examining one-dimensional atomic chains.

\subsubsection{Parametric oscillators}

An oscillator with one or more parameters varying with time is a parametric oscillator, also called a parametrically excited oscillator. The parametric oscillator constitutes a very important class of nonlinear systems. The parametric factor may appear in the inertia, the damping, or the stiffness term.

A bistable system with linear stiffness modulated by a low-frequency signal and excited by an external high-frequency signal is \cite{ref99}
\begin{equation}
 \label{eq40}
    \ddot{x} + \delta \dot{x} + \left( \omega_0^2x + s_\mathrm{L}(t) \right) x + \beta x^3 = s_\mathrm{H}(t).
\end{equation}
In Eq.~(\ref{eq40}), the frequency $s_\mathrm{H}(t)$ is far greater than both the frequency of $s_\mathrm{L}(t)$ and the natural frequency $\omega_0$.

The model of a controllable parametrically excited buckled beam is \cite{ref226}
\begin{equation}
 \label{eq41}
    \ddot x + \delta \dot{x} + \left( \omega_0^2 + s_\mathrm{H1}(t) \right) x + \beta x^3 = s_\mathrm{L}(t) + s_\mathrm{H2}(t) .
\end{equation}
The oscillator is of Mathieu-Duffing type. The natural frequency of the oscillator is adjusted by a high-frequency signal and the oscillator is subjected to a low-frequency signal and another high-frequency signal simultaneously.

Besides, a cantilever beam parametrically excited is described by \cite{ref227}
\begin{equation}
  \label{eq42}
      \left( 1 + \alpha_1 x^2  \right) \ddot{x} + 2 \mu_1 \dot x  + \mu_2 \left| \dot{x} \right| \dot{x} + \alpha_2 x \dot{x}^2 + \left( \omega_0^2 + s_\mathrm{L}(t)+ s_\mathrm{H}(t) \right)  x  + \alpha_3 x^3   =  0 \,.
\end{equation}

A van der Pol-Mathieu-Duffing oscillator with nonlinear damping and two external harmonic excitations with two widely different frequencies is given by \cite{ref228}
\begin{equation}
 \label{eq43}
    \ddot{x} + \gamma \left( x^2 - 1 \right) \dot{x}
        - \omega_0^2 \left( 1 + h \cos \omega_\mathrm{p} t \right) x
        + \beta x^3 = s_\mathrm{L}(t) + s_\mathrm{H}(t).
\end{equation}
The model in Eq.~(\ref{eq43}) can describe the dynamics of a micro-electro-mechanical system.

Not only the linear stiffness but also the damping can be parametrically varied with time, such as the nonlinear vibrational resonance in the van der Pol-Duffing oscillator with a parametric damping \cite{ref100}
\begin{equation}
 \label{eq44}
\ddot x + \gamma (1 + s_\mathrm{L}(t))({x^2} - 1)\dot x - \omega _0^2x + \alpha {x^3} = s_\mathrm{H}(t).
\end{equation}

The role of periodically parametric damping on vibrational resonance was studied also in a modulated signal excited asymmetric mixed Rayleigh-Li\'enard oscillator as follows \cite{ref190}
\begin{equation}
 \label{eq45}
\ddot x + ({a_0} + s_\mathrm{L}(t))\dot x + {b_1}x\dot x + {b_2}{{\dot x}^2} + {b_3}{{\dot x}^3} - {E_0} - x + c{x^3} = ({d_1} + s_\mathrm{H}(t))s_\mathrm{L}(t).
\end{equation}

\subsubsection{Position-dependent mass oscillators}

Another important kind of nonlinear system related to vibrational resonance is the position-dependent mass oscillator. The model with varying mass is usually encountered in many fields such as  aeronautical and astronautical engineering, astronomy, aerology, offshore engineering, civil engineering, condensed matter physics, semiconductor heterogeneous structures, etc.

A Duffing oscillator with position-dependent mass is described by \cite{ref164}
\begin{equation}
 \label{eq46}
 m(x)\ddot x - {m^2}(x)\gamma \lambda x{\dot x^2} + \alpha \dot x + {m^2}(x)\gamma \omega _0^2x + \beta {x^3} = s_\mathrm{L}(t) + s_\mathrm{H}(t).
\end{equation}
Herein, $m(x)$ is a position-dependent function. Roy-Layinde et al. used Eq.~(\ref{eq46}) to study the dynamics of a gas bubble.

Another position-dependent mass oscillator is \cite{ref229}
\begin{equation}
 \label{eq47}
m(x)\ddot x + \frac{1}{2}m'(x){\dot x^2} + \alpha \dot x + m(x)\omega _0^2x + \beta {x^3} = s_\mathrm{L}(t) + s_\mathrm{H}(t).
\end{equation}
that describes vibrational resonance in the model of a NH3 molecule. In Eq.~(\ref{eq46}) and Eq.~(\ref{eq47}), not only the mass but also the damping and stiffness are position-dependent.

The following nonlinear oscillator with position-dependent mass and phase-dependent damping is used to model the current through a Josephson junction \cite{ref23}
\begin{equation}
 \label{eq48}
m(x)\ddot x + \frac{1}{2}m'(x){\dot x^2} + \gamma (x)\dot x + {V_0}\sin (kx) = s_\mathrm{L}(t) + s_\mathrm{H}(t).
\end{equation}

There is another interesting position-dependent mass oscillator which is given by \cite{ref230}
\begin{equation}
 \label{eq49}
(1 + \mu {x^2})\ddot x + \mu x{\dot x^2} + \alpha \dot x + \omega _0^2x = s_\mathrm{L}(t) + s_\mathrm{H}(t).
\end{equation}
that describes a particle that moves on a rotating-parabola system. By some calculations and simplifications, Eq.~(\ref{eq49}) can be transformed into a quintic oscillator which is also a parametrically excited system. Vibrational resonance in a Mathews-Lakshmanan oscillator similar to Eq.~(\ref{eq49}) was also investigated in \cite{ref231}.

\subsubsection{Nonlinearly damped oscillators}

As the name implies, a nonlinearly damped oscillator has a nonlinear damping. Apparently, the systems in Eqs.~(\ref{eq36}), (\ref{eq42})-(\ref{eq49}) are all nonlinearly damped oscillators. In addition to these, there are other nonlinearly damped systems that exhibit vibrational resonance. We summarize them here.

The nonlinearly damped oscillator with a rough potential is given by \cite{ref18}
\begin{equation}
 \label{eq50}
    \ddot{x} - \mu \left( 1 - x^2  + \nu x^4 \right)\dot{x} + \frac{\mathrm{d}V }{\mathrm{d}x} = s_\mathrm{L}(t) + s_\mathrm{H}(t)
\end{equation}
with a potential
\begin{equation}
 \label{eq51}
     V(x) = \frac{1}{2} \omega_0^2 x^2 + \frac{1}{4} \beta x^4 + \frac{1}{6} \gamma x^6 + \epsilon ( \cos \omega_1 x + \sin \omega_2 x ).
\end{equation}
The model comes from a set of quasihydrodynamic equations and describes a two-fluid magnetized plasma oscillator of motion. When $\epsilon=0$, the potential is smooth and has been given in Eq.~(\ref{eq33}). When $\epsilon \ne 0$, the potential is perturbed and increases roughly with the increase of $\epsilon$.

The Toda oscillator in the presence of a dual-frequency forcing is \cite{ref19}
\begin{equation}
 \label{eq52}
    \ddot{x} + k_0 ( 1 + \varepsilon \cos x) \dot x + \mathrm{e}^{x} - 1 = s_\mathrm{L}(t) + s_\mathrm{H}(t) .
\end{equation}
The Toda potential is given by $V(x) = \mathrm{e}^x - x +1$. The Toda oscillator has abundant dynamic behaviors such as bifurcations, chaos, nonlinear resonance, etc. This model comes from the study of different types of lasers. It has been also applied to DNA, and molecular dynamics of muscle contraction \cite{ref232}.

Anharmonic plasma oscillation induced by bi-harmonic forces is expressed as \cite{ref233}
\begin{equation}
 \label{eq53}
    \ddot x + \varepsilon \left(1 + x^2 \right)\dot x + \frac{\mathrm{d}V(x)}{\mathrm{d}x} =  s_\mathrm{L}(t) + s_\mathrm{H}(t) .
\end{equation}
The potential function here is
\begin{equation}
 \label{eq54}
    V(x) = \frac{1}{2}\omega _0^2x^2 + \frac{1}{3} \alpha x^3 +  \frac{1}{4} \beta x^4   .
\end{equation}
This physical model refers to high density plasma interactions caused by high frequency electromagnetic waves. The model in Eq.~(\ref{eq53}) is also used to study the ghost-vibrational resonance when a multi-frequency signal with the form $\sum\limits_{i = 1}^{nf} {{f_i}\cos ({\omega _i} + \Delta {\omega _0})} t$ instead of the slow-varying signal $s_\mathrm{L}(t)$ \cite{ref105}.

A beam with boundary conditions of two ends fixed and prestressed is modeled by \cite{ref234}
\begin{equation}
 \label{eq55}
     \ddot x + \left( \beta_0 + \beta_2 \dot x^2 \right) \dot x + x + \gamma_2 x^2 + \gamma_3 x^3 = s_\mathrm{L}(t) + s_\mathrm{H}(t) .
\end{equation}
The nonlinear damping is the nonlinear dissipation of the Rayleigh type. The background of the biharmonic excitation comes from wind (slow-varying excitation) and earthquake (fast-varying excitation), or flow of people crossing the same bridge (slow-varying excitation) and vehicles crossing the bridge (fast-varying excitation).

A base-excited, tilted cantilever beam is given as \cite{ref235}
%
\begin{subequations}
 \label{eq56}
 \begin{eqnarray}
  \ddot x \! \!+\! \! 2{\mu _1}\dot x\! \! +\! \! {\mu _2}{{\dot x}^3} \! \!+\! \! x\! \! +\! \! {\alpha _1}{x^3}\! \! + \! \!{\alpha _2}({x^2}\ddot x \! \!+ \! \! x{{\dot x}^2}) \! \! & = \! \! & (\eta \cos \beta \! \! - \! \!\lambda x\sin \beta )(\! \! - \! \!{{\ddot z}_p}),\\
    {z_p} & = & s_\mathrm{L1}(t) + s_\mathrm{L2}(t) + s_\mathrm{H}(t),
 \end{eqnarray}
\end{subequations}
%
where $z_p$ is the base excitation. The terms $z_p \sin \beta$ and $z_p \cos \beta$ are two displacements in two mutually perpendicular directions.

Vibrational resonance occurs in the nonlinear Rayleigh-Plesset oscillator \cite{ref236}
\begin{eqnarray}
 \label{eq57}
  & &  \ddot x  + \dot x [ \alpha_0 - \alpha_1 x + \alpha_2 x^2
        - \alpha_3 x^3 + \alpha_4 x^4]
          + \eta \dot x^2 [1 - x + x^2] \nonumber \\
   & & \quad \quad - x \{ \beta  - [s_\mathrm{H}(t) + \varepsilon ]
             s_\mathrm{L}(t) \} + x^2  \{ \gamma
          - [ s_\mathrm{H}(t) + \varepsilon ] s_\mathrm{L}(t)\}  \nonumber \\
     &&  \quad \quad - \gamma x^3 + \lambda x^4
       = [ s_\mathrm{H}(t) + \varepsilon ] s_\mathrm{L}(t) .
\end{eqnarray}
The model is used for the study of the dynamics of a gas bubble under the excitations of complex parametric and external forces.

There is also a dynamical equation of a charged bubble excited by a modulated acoustic field and vibrating in a liquid given by \cite{ref189}
{\small
\begin{subequations}
 \label{eq58}
\begin{eqnarray}
\ddot x & = & {\psi _1}{\psi _2} \\
{\psi _1} \! \! & = & \! \! \left[ {\frac{{{z_1}}}{{{x^{3\Gamma }}}}(1\! \! +\! \! {z_2}\dot x)\! \! - \! \! \frac{{{{\dot x}^2}}}{2}(3\! \! -\! \! \frac{{\dot x}}{c}) + \frac{{{z_3}}}{{{x^4}}}(1\! \! -\! \! \frac{{3\dot x}}{c}) - \frac{{{z_4}\! \! +\! \! {z_5}\dot x}}{x}\! \! -\! \! {\Delta _1}(t)\! \! -\! \! {\Delta _2}(t)} \right] \\
{\psi _2} & = & {\left[ {(1 - \frac{{\dot x}}{c})x + {z_{10}}} \right]^{ - 1}}
 \end{eqnarray}
\end{subequations}
}%
where $\Delta_1 t$ and $\Delta_2 t$ are the functions contained in the parametric modulated excitations
\begin{subequations}
 \label{eq59}
\begin{eqnarray}
 {\Delta _1}(t) & = & (1 + \frac{{\dot x}}{c})\left[ {{z_6} + {z_7}(1 + G\sin (\Omega t))\sin (\omega t)} \right], \\
 {\Delta _2}(t) & = & x\left[ {{z_8}(1 + G\sin (\Omega t))\cos (\omega t) + {z_9}G\sin (\omega t)\cos (\Omega t)} \right].\
\end{eqnarray}
\end{subequations}

Under a dual-frequency excitation, the gyroscope system mounted on a vibrating base is given by \cite{ref237}
\begin{equation}
 \label{eq60}
       \ddot x  + c_1 \dot x + c_2 \dot x^3
            + \alpha^2  \frac{ \left( 1 - \cos x \right)^2}
          { \sin ^3x}
          - \left[ \beta  + s_\mathrm{H}(t) \right] \sin x
        = s_\mathrm{L}(t).
\end{equation}
The system has a nonlinear damping and a parametric excitation, and the potential function may be present in a double-well or a single-well form.

Another typical damping is the resistance force expressed by the signum function. The vibrational resonance is studied in a nonlinear system with a signum nonlinearity \cite{ref238}
\begin{equation}
 \label{eq61}
     \ddot x + \delta \dot x + \omega_0^2 x
          + \beta \mathrm{sgn} (x) = s_\mathrm{L}(t) + s_\mathrm{H}(t) .
\end{equation}
The sign $\mathrm{sgn}(\bullet)$ is the signum function. Here, the nonlinear damping depends on the displacement of the coordinate. In fact, the nonlinear damping that depends on the velocity of the coordinate is much more widely used especially in mechanical devices. The nonlinear damping $\mathrm{sgn} (\dot x)$ is usually viewed as a simple model of the dry friction \cite{ref239}. It is significant to study vibrational resonance in friction systems due to the fact that a fast-varying excitation can quench the vibration and can be viewed as an antiresonance. There are many mathematical models belonging to different friction damping \cite{ref240}. Further, the high-frequency excitation has a very important role on the friction induced mechanical vibration. It can change the properties of the stiffness, bias, and smoothness of the system \cite{ref7}, \cite{ref241}-\cite{ref243}. Vibrational resonance in much more complex friction systems is almost blank. Investigating works in this area will contribute to nonlinear vibration and control significantly.

\subsubsection{Coupled oscillators}

Vibrational resonance and antiresonance in unidirectionally coupled overdamped oscillators have been studied under low-frequency and high-frequency excitations in different oscillators, as described in \cite{ref244}
\begin{subequations}
 \label{eq62}
\begin{eqnarray}
    \dot x  & = & x - x^3 + s_\mathrm{H}(t), \\
    \dot y  & = & y - y^3 + \gamma x + s_\mathrm{L}(t),
\end{eqnarray}
\end{subequations}
and mutually coupled overdamped oscillators
\begin{subequations}
 \label{eq63}
\begin{eqnarray}
    \dot x  & = & x - x^3 + \gamma ( x - y ) + s_\mathrm{L}(t), \\
    \dot y & = & y - y^3 + \gamma ( y - x ) + s_\mathrm{H}(t) .
\end{eqnarray}
\end{subequations}

An anharmonic coupled overdamped oscillator subjected to both low-frequency and high-frequency excitations is \cite{ref245}
\begin{subequations}
 \label{eq64}
\begin{eqnarray}
    \dot x & = &  a_1 x - b_1 x^3 + \gamma x y^2
             + s_\mathrm{L}(t) + s_\mathrm{H}(t), \\
    \dot y & = &  a_2 y - b_2 y^3 + \gamma  x^2 y .
\end{eqnarray}
\end{subequations}
The system is viewed as a dynamical system modeling the competition between two different species. Here, $a_1 \ne a_2$ and the potential function is
\begin{equation}
 \label{eq65}
      V(x, y) =  - \frac{1}{2}a_1x^2 + \frac{1}{4}b_1x^4
       - \frac{1}{2}a_2y^2 + \frac{1}{4}b_2y^4
         - \frac{1}{2} \gamma x^2 y^2 .
\end{equation}
If $a_1$, $a_2$, $b_1$, $b_2$, $\gamma >0$, the potential presents a four-well shape. The system has nine or five fixed points corresponding to $\gamma^2 < b_1 b_2$ and $\gamma^2 > b_1 b_2$, respectively. In the study of this kind of oscillators, one usually chooses $a_1 \ne a_2$, and the potential has an asymmetric potential structure. Vibrational resonance of the coupled system in Eq.~(\ref{eq65}) was also studied when the excitation is in the amplitude modulated form \cite{ref186}.

The chemical reaction model between four molecules is described by \cite{ref28}
\begin{subequations}
 \label{eq66}
\begin{eqnarray}
     \dot x & = & - 2 k_1 x^2 y^2 + k_2 x ( 1 - x - y)
                   + ( F + s_\mathrm{H}(t) ) s_\mathrm{L}(t), \\
     \dot y & =  & k_1 x^2 y^2 - k_3 y (1 - x - y) .
\end{eqnarray}
\end{subequations}
This oscillator has abundant nonlinear dynamics such as hysteresis, vibrational resonance, multistability and chaos.

Vibrational antiresonance is investigated in the following coupled nonlinear oscillators \cite{ref16}
\begin{subequations}
 \label{eq67}
\begin{eqnarray}
     \ddot x_1 + 2 \gamma_1 \dot x_1 - 2 g_1 x_2 + \beta_1x_1^3
        - \omega_1^2 x_1 & = & s_\mathrm{L}(t) + s_\mathrm{H}(t), \\
     \ddot x_2 + 2 \gamma_2 \dot x_2 - 2 g_2 x_1 + \beta_2 x_2^3
         - \omega_2^2 x_2 & = &  s_\mathrm{H}(t) .
\end{eqnarray}
\end{subequations}

Vibrational resonance and its extend application in signal transmission are investigated in a one-way coupled bistable system \cite{ref17}
\begin{subequations}
 \label{eq68}
\begin{eqnarray}
     \dot x_1 & = &  a_1 x_1  - b_1 x_1^2 + s_\mathrm{L}(t)
                     + s_\mathrm{H}(t) , \\
     \dot x_i & = &  a_1x_i - b_1 x_i^2 + \epsilon x_{i - 1},
                      \quad i = 2,3, \cdots, n .
\end{eqnarray}
\end{subequations}

In Eq.~(\ref{eq68}), when the signal $s_\mathrm{L}(t)$ is added on all elements $x_i$, there is a vibrational resonance phenomenon. When the signal $s_\mathrm{L}(t)$ is added only on $x_1$, it is a signal transmission phenomenon.

Ferroelectric liquid crystal is also modeled as a coupled system \cite{ref246}
\begin{equation}
 \label{eq69}
     \dot x_i
        =  x_i - x_i^3 + \sum_{j=1}^{n} J_{ij} (x_j
         - x_i)  + s_\mathrm{L}(t) + s_\mathrm{H}(t)
          + x_i w_i,   \quad i=1,2,\cdots,n ,
\end{equation}
where $J_{ij}$ means the coupling strength and $w_i$ is the local liquid crystal polymer interaction. Vibrational resonance appears in this physical system as well.

\subsubsection{Neural models and complex networks}

Vibrational resonance has garnered increasing attention within various neural models and complex networks. Below, we provide a list of some of these models.

The single FitzHugh-Nagumo model
\begin{subequations}
 \label{eq70}
\begin{eqnarray}
    \varepsilon \dot x & = &  x - \frac{1}{3}x^3 - y, \\
     \dot y & = & x + a + s_\mathrm{L}(t) + s_\mathrm{H}(t) ,
\end{eqnarray}
\end{subequations}
is a typical system exhibiting vibrational resonance \cite{ref66}-\cite{ref73}. In Eq.~(\ref{eq70}b), without the excitations, $a$ is usually considered as the bifurcation parameter. Specifically, when $a > 1$, the FitzHugh-Nagumo model has a stable equilibrium; whereas when $a<1$, the system undergoes a supercritical Hopf bifurcation, producing a stable limit cycle, and then the equilibrium point becomes the focus of instability. The variable $\varepsilon$ means the time scale separation parameter which is considered to be small. Under different response patterns, the subthreshold vibrational resonance or suprathreshold vibrational resonance will be present in the FitzHugh-Nagumo system \cite{ref69}. In addition, vibrational resonance in a modified FitzHugh-Nagumo dynamical system has also been studied and the resonance patterns are discussed in \cite{ref76}. When the excitation in Eq.~(\ref{eq70}b) is replaced by $[S_1(t)+S_2(t)][S_L(t)+S_H(t)]$, where $S_1$ and $S_2$ represent two logical signals, a comprehensive analysis of logical vibrational resonance has been carried out in detail in \cite{ref201}.

Vibrational resonance also occurs in small-word neuronal networks with different types of synapses \cite{ref74}, \cite{ref80}, \cite{ref86}, \cite{ref90}, \cite{ref97}
\begin{subequations}
 \label{eq71}
\begin{eqnarray}
      \varepsilon \dot x_i & = & x_i - \frac{1}{3} x_i^3
           - y_i - I_i^{\mathrm{syn}}, \\
      \dot y_i & = & x_i + a + s_\mathrm{L}(t) + s_\mathrm{H}(t),
\end{eqnarray}
where $I_i^{\mathrm{syn}}$ is the synapse coupling expressing the synaptic current. For the case of an electrical synapse coupling, $I_i^{\mathrm{syn}}$ is
\begin{eqnarray}
     I_i^{\mathrm{syn}}
        = g_{\mathrm{syn}} \sum_j C_\mathrm{e} (i,j) (x_i - x_j ) ,
\end{eqnarray}
and for case of the chemical synapse coupling, $I_i^{\mathrm{syn}}$ is
\begin{eqnarray}
      I_i^{\mathrm{syn}}
     & =  & g_{\mathrm{syn}} \sum_j C_\mathrm{c} (i,j)
                 s_j( x_i - x_{\mathrm{syn}} ),  \\
      \dot s_j & = &  \frac{1}{ \varepsilon } \alpha (x_j)  ( 1 - s_j )
           - \frac{s_j}{ \tau_{\mathrm{syn}} } , \quad
    \alpha(x_j)  =  \frac{\alpha _0}
             {1 + \exp ( - x_j / x_{\mathrm{shp}} ) } ,
 \end{eqnarray}
\end{subequations}
where $\tau_{\mathrm{syn}}=1/\delta$ represents the decay rate of the synapse. Usually, the chemical synaptic coupling is much more efficient on the low-frequency signal transmission than that of the electrical coupling due to its selective coupling.

Vibrational resonance is analyzed in small-world neuronal networks that have spike-timing-dependent plasticity \cite{ref81}. This kind of networks are described as
\begin{subequations}
 \label{eq72}
\begin{eqnarray}
      \varepsilon \dot x_i
     & = & x_i  - \frac{1}{3} x_i^3  - y_i + I_i^{\mathrm{syn}}
            + s_\mathrm{L}(t) + s_\mathrm{H}(t) , \\
    \dot y_i & = & x_i + a - b_i y_i ,
\end{eqnarray}
with
\begin{eqnarray}
   I_i^{\mathrm{syn}}
       =  - \sum_{j = 1,\;(j \ne i}^N
            g_{ij} C_{ij} s_j (t) ( x_i - x_\mathrm{syn} ),
       \quad \dot s_j  =  \alpha ( x_j ) (1 - s_j ) - \beta s_j ,
\end{eqnarray}
and
\begin{eqnarray}
        \alpha ( x_j ) & = &  \alpha_0  /
             \left( 1 + \mathrm{e}^{ - x_j / x_{\mathrm{shp}} } \right), \\
        g_{ij}^{\mathrm{ex}}  & = &  g_{ij}^{\mathrm{ex}} + \Delta g_{ij}^{\mathrm{ex}}, \quad
     \Delta g_{ij}^{\mathrm{ex}}
    =   g_{ij}^{\mathrm{ex}} F(\Delta t), \\
     F(\Delta t) & = & \left\{
          \begin{array}{ll}
               A_+ \exp ( - \left| \Delta t \right| / \tau_ + ),
               & \mathrm{if} \; \Delta t > 0,  \\
         - A_-  \exp(-\left| \Delta t \right| / \tau_- ),  &
          \mathrm{if} \; \Delta t > 0, \\
              0 , &  \mathrm{if} \; \Delta t = 0 .
               \end{array} \right.
\end{eqnarray}
\end{subequations}
It was found that the inhibitory synapses may weaken vibrational resonance in this model.

Vibrational resonance was also studied in a feedforward neuron network in \cite{ref87}
\begin{subequations}
\label{eq73}
\begin{eqnarray}
       \varepsilon \dot x_{i,j}
    & = & x_{i,j} - \frac{ 1}{3} x_{i,j}^3
           - y_{i,j} + I_{i,j}^{\mathrm{syn}}(t), \\
     \dot y_{i,j} & = & x_{i,j} + a_{i,j} - y_{i,j} + S_{i,j}(t), \\
      I_{i,j}^{\mathrm{syn}} & = &
        \sum_{k = 1}^{N_{\mathrm{syn}}}
          g_{\mathrm{syn}} \alpha (t - {t_{i - 1,k}})
           (V_{i,j} - V_{\mathrm{syn}}),
\end{eqnarray}
\end{subequations}
where, $\alpha (t) = (t/\tau) \mathrm{e}^{-t/\tau} $. While in a feedforward neuron network with unreliable synapses \cite{ref82}
\begin{subequations}
\label{eq74}
\begin{eqnarray}
 \varepsilon \dot x_{i,j}
    & = & x_{i,j} - \frac{1}{3} x_{i,j}^3
            - y_{i,j} + I_{i,j}^{ \mathrm{syn} }, \\
     \dot y_{i,j} & = & x_{i,j} + a + I_{i,j}, \quad
        I_{i,j} = s_\mathrm{L} (t) + s_\mathrm{H} (t)
               + \varphi_{i,j}, \\
    I_{i,j}^{ \mathrm{syn} } & = & \frac{1}{N} \sum_{k = 1}^N
           G(i,j;k,j - 1) [ E_{ \mathrm{syn} } - x_{i,j} ], \\
     G(i,j;k,j\! \! - \! \!1) \! \! & \leftarrow & \! \! G(i,j;k,j\! \! -\! \! 1)\! \! +\! \! J(i,j;k,j\! \! -\! \! 1)h(i,j;k,j\! \! -\! \! 1),
\end{eqnarray}
\end{subequations}
and  in a heterogeneous scale free neuron network \cite{ref83}
\begin{subequations}
\label{eq75}
\begin{eqnarray}
     \varepsilon \dot x_i
     & = &  x_i - \frac{1}{3} x_i^3 - y_i + \sum_j
                 g_{ij}  ( x_j - x_i ) ,  \\
     \dot y_i & = & x_i + a_i + I_{\mathrm{ex}}, \quad
    \langle a_i \rangle  =  a_0, \quad \langle (a_i - a_0) (a_j - a_0)
            \rangle  = \delta_{ij} \sigma ^2 .
\end{eqnarray}
\end{subequations}
Different network connections are important for weak signal propagation based on the vibrational resonance mechanism.

Vibrational resonance can also be caused by a special heterogenous aperiodic fast-varying signal in a FitzHugh-Nagumo neuronal parallel array without connections between the neurons \cite{ref89}
\begin{subequations}
\label{eq76}
\begin{eqnarray}
    \varepsilon \dot x_i & = & x_i - \frac{1}{3} x_i^3  - y_i + s(t), \\
    \dot y_i & = & x_i + a + S_i(t),
\end{eqnarray}
\end{subequations}
or with locally coupled neurons
\begin{subequations}
\label{eq77}
\begin{eqnarray}
    \varepsilon \dot x_i  & = &   x_i - \frac{1}{3} x_i^3
           - y_i + g [ x_{i + 1} + x_{i - 1} - 2x_i ]  + s(t), \\
     \dot y_i  & = & x_i + a + S_i(t) ,
\end{eqnarray}
\end{subequations}
where $g$ is the strength of the coupling. Herein, $s(t)$ represents the slow-varying harmonic signal, and $S_i (t)$ denotes the heterogeneous aperiodic fast-varying signal. $S_i (t)$ is generated by randomly modulating the amplitude and frequency of a sinusoidal signal. The general expression of the aperiodic stimuli is defined as $S_i (\tau_{i,j - 1} + t) = A_{i,j} \sin  ( 2\pi t / T_{i,j} ) $ with amplitude $A_{i,j}$ and period $T_{i,j}$. Both $A_{i,j}$ and $T_{i,j}$ obey a certain stochastic distribution.

Besides the standard FitzHugh-Nagumo neuron models, vibrational resonance was also studied in a modified FitzHugh-Nagumo neuron model \cite{ref247}, where the model equation is described by
\begin{subequations}
\label{eq78}
\begin{eqnarray}
 \varepsilon \dot x & = & x - \frac{{{x^3}}}{3} - y - {k_1}\rho (\varphi )x + s_\mathrm{L}(t) + s_\mathrm{H}(t), \\
 \dot y & = & x + a - by, \\
 \dot \varphi  & = & {k_2}x - 0.5\varphi  + {\varphi _{ext}},
\end{eqnarray}
\end{subequations}
$- {k_1}\rho (\varphi )x$ is the modified term, and $\varphi _{ext}$ is a high-low-frequency electromagnetic radiation. They also considered vibrational resonance in the coupled modified FitzHugh-Nagumo models, i.e.,
\begin{subequations}
\label{eq79}
\begin{eqnarray}
 \varepsilon {{\dot x}_1} & = & {x_1} - \frac{{x_1^3}}{3} - {y_1} + s_\mathrm{L}(t) + s_\mathrm{H}(t) - {k_1}\rho ({\varphi _1}){x_1} + g({x_1} - {x_2}), \\
 {{\dot y}_1} & = & {x_1} + a - b{y_1}, \\
 {{\dot \varphi }_1} & = & {k_2}{x_1} - 0.5{\varphi _1} + {\varphi _{ext}}, \\
 \varepsilon {{\dot x}_2} & = & {x_2} - \frac{{x_2^3}}{3} - {y_2} + s_\mathrm{L}(t) + s_\mathrm{H}(t) - {k_1}\rho ({\varphi _2}){x_2} + g({x_2} - {x_1}), \\
 {{\dot y}_2} & = & {x_2} + a - b{y_2}, \\
 {{\dot \varphi }_2} & = & {k_2}{x_2} - 0.5{\varphi _2} + {\varphi _{ext}}.
\end{eqnarray}
\end{subequations}

Vibrational resonance is analyzed in a synthetic gene network \cite{ref91}
\begin{subequations}
\label{eq80}
\begin{eqnarray}
     \dot x & = & \frac{ 1 + x^2 + \alpha \sigma x^4 }
           { [1 + x^2 + \sigma x^4 ]  [ 1 + y^4 ] } - \gamma_x x , \\
     \tau_y \dot y & = & \frac{ 1 + x^2 + \alpha \sigma x^4 }
        { [ 1 + x^2 + \sigma x^4 ] [ 1 + y^4 ] } - \gamma_y y
             + s_\mathrm{L}(t) + s_\mathrm{H}(t) .
\end{eqnarray}
\end{subequations}
It is interesting that the ratio of the amplitude to the frequency of the fast-varying excitation is always a definite constant when the optimal vibrational resonance occurs in this system.

The Hodgkin-Huxley neuron model excited by a biharmonic signal is studied in the equation \cite{ref248}
\begin{subequations}
\label{eq81}
\begin{eqnarray}
 C_m \dot V & = &  - \left[ g_k n^4 (V - V_K ) + g_\mathrm{Na} m^3
               h(V - V_\mathrm{Na} ) + g_1 (V - V_\mathrm{l} )  \right]  \nonumber \\
     & & \quad
              + I_\mathrm{aut} + I_0 + s_\mathrm{L}(t) + s_\mathrm{H}(t) , \\
    \dot m  & = &  \alpha _m (1 - m) - \beta_m m, \\
    \dot n  & = &  \alpha_n ( 1 - n) - \beta_n n, \\
    \dot h  & = &  \alpha_h (1 - h) - \beta_h h .
\end{eqnarray}
\end{subequations}
In Eqs.~(\ref{eq81}), $C_m$ is the cell capacitance, and $V$ is the membrane potential. The parameters $g_k$, $g_\mathrm{Na}$ and $g_l$ represent the maximum conductances of the potassium, the sodium and leak currents, respectively. In addition, $V_K$, $V_\mathrm{Na}$ and $V_\mathrm{l}$ stand for the potassium, the sodium, and the leakage reversal potentials in turn. $I_0$ represents the stimulus current. Based on the vibrational resonance mechanism, the inhibitory autapse significantly enhances the low-frequency subthreshold signal, despite its typically suppressive role in neuronal dynamics.

A neuron model related to the dynamics of Na$^+$ and K$^+$ is described in detail in \cite{ref249}
\begin{equation}
\label{eq82}
     C \dot V
          =  - \left( I_\mathrm{Na} + I_\mathrm{K}
           + I_\mathrm{sLeak} + I_\mathrm{pump}
             + I_\mathrm{sd} \right) + s_\mathrm{L}(t) + s_\mathrm{H}(t) ,
\end{equation}
where $C$ is the soma capacitance and $V$ is the membrane potential. $I_\mathrm{Na}$ and $I_\mathrm{K}$ represent the sum of Na$^+$ currents and the sum of K$^+$ currents, respectively. $I_\mathrm{sLeak} $, $I_\mathrm{pump}$ and $I_\mathrm{sd}$ represent the leakage current, the current induced by the Na$^+$ and K$^+$ pump, and the axial current in turn. The continuous changes of the K$^+$ and Na$^+$ concentrations will cause and enhance vibrational multi-resonances.

Multiple instances of vibrational resonance were explored in various neuron systems, including a single Hindmarsh-Rose neuron, a modified Hindmarsh-Rose neuron, and coupled Hindmarsh-Rose neuron systems with or without an electric field, as reported in \cite{ref78}, \cite{ref250}. The different models are listed below.

The single Hindmarsh-Rose model is
\begin{subequations}
\label{eq83}
\begin{eqnarray}
   \dot x  & = & y - a x^3  + b x^2 - z + I_0 + s_\mathrm{L}(t)+s_\mathrm{H}(t), \\
   \dot y  & = & c - d x^2 - y, \\
   \dot z  & = &  r [ s (x + 1.56) - z],
\end{eqnarray}
\end{subequations}
where $x$ is the membrane potential, $y$ is the recovery variable, $z$ is the adaptation current, and $I_0$ represents a direct current stimulation. Multiple vibrational resonances have been observed in this single Hindmarsh-Rose neuron model.

For coupled Hindmarsh-Rose neurons without electric field, the dynamical model is
\begin{subequations}
\label{eq84}
\begin{eqnarray}
   \dot x_1 & = &  y_1 - ax_1^3 + bx_1^2 - z_1 + g_1 ( x_1  - x_2 )
                      + s_\mathrm{L}(t), \\
   \dot y_1 & = &  c - d x_1^2 - y_1 , \\
   \dot z_1 & = &  r [ s ( x_1 + 1.56) - z_1 ], \\
   \dot x_2 & = &  y_2 - ax_2^3 + bx_2^2 - z_2 + g_2 ( x_2 - x_1)
               + s_\mathrm{H}(t), \\
   \dot y_2 & = &  c - d x_2^2 - y_2, \\
   \dot z_2 & = & r [ s ( x_2 + 1.56) - z_2 ] .
\end{eqnarray}
\end{subequations}
Herein, the coupling between the two neurons are $g_1(x_1-x_2)$ and $g_2(x_2-x_1)$ with coupling strength $g_1$ and $g_2$. The improved Hindmarsh-Rose neuron model with external electric field is
\begin{subequations}
\label{eq85}
\begin{eqnarray}
   \dot x & = &   y - a x^3 + b x^2 - z + I_0 + s_\mathrm{L}(t)+s_\mathrm{H}(t), \\
   \dot y & = &   c - d x^2 - y + RE ,\\
   \dot z & = &  r [s ( x + 1.56) - z] , \\
   \dot E & = &   m y + E_\mathrm{ext} ,
\end{eqnarray}
\end{subequations}
where $R$ is the radius size of the cell, $E$ is the intensity of the uniform electric field, $m$ is a parameter corresponding to the polarization property, and $E_\mathrm{ext}$ is the external electric field.

For coupled Hindmarsh-Rose neurons with electric field, the model is
\begin{subequations}
\label{eq86}
\begin{eqnarray}
   \dot x_1 & = & y_1 - ax_1^3 + bx_1^2 - z_1 + s_\mathrm{L}(t)
                + g_1 ( x_1 - x_2 ) ,\\
   \dot y_1 & = & c - d x_1^2 - y_1 + R E_1  , \\
   \dot z_1 & = & r [ s ( x_1 + 1.56) - z_1 ] , \\
   \dot E_1 & = & m y_1 + E_\mathrm{ext1} , \\
   \dot x_2 & = &  y_2 - ax_2^3 + bx_2^2 - z_2
                 + s_\mathrm{H}(t) + g_2 ( x_2 - x_1 ) , \\
   \dot y_2 & = &  c - d x_2^2 - y_2 + R E_2 ,\\
   \dot z_2 & = &  r[s(x_2 + 1.56) - z_2] ,\\
   \dot E_2 & = & m y_2 + E_\mathrm{ext2} .
\end{eqnarray}
\end{subequations}
Multiple vibrational resonances are prominent regardless of the presence of an electric field. Moreover, the electric field appears to attenuate the impact of multiple vibrational resonances in the single Hindmarsh-Rose neuron. Conversely, it enhances the effect of multiple vibrational resonances in the case of bidirectional coupling between two Hindmarsh-Rose neurons.

\subsubsection{Circuit systems}

Chua's circuit is a renowned nonlinear model known for its complex dynamical behaviors. Vibrational resonance has been observed through simulation or experimentation in numerous studies involving Chua's circuit. Below, we outline the Chua's circuit models that have been extensively discussed in the context of vibrational resonance.

The model of a single Chua's circuit is \cite{ref251}
\begin{subequations}
\label{eq87}
\begin{eqnarray}
     C_1 \dot v_1 & = & ( 1 /R) ( v_2 - v_1 ) - f( v_1 ), \\
    C_2 \dot v_2  & = & ( 1 /R) ( v_1 - v_2 + i_L ), \\
    L \dot i_L    & = &  - v_2 + s_\mathrm{L}(t) + s_\mathrm{H}(t),
\end{eqnarray}
where
\begin{eqnarray}
  f( v_1 )  =   G_b v_1 + 0.5 ( G_a  - G_b )
                 \left[ \left| v_1 + B P_1 \right|
               - \left| v_1 - B P_1  \right|  \right] ,
\end{eqnarray}
\end{subequations}
and its $n$-coupled form
\begin{subequations}
\label{eq88}
\begin{eqnarray}
  C_1 \dot v_1^{(i)}
     &  =  &  ( 1 / R) ( v_2^{(i)} - v_1^{(i)} ) - f (v_1^{(i)}), \\
 C_2 \dot v_2^{(i)}
     & = & (  1 / R) ( v_1^{(i)} - v_2^{(i)} + i_L^{(i)}), \\
   L \dot i_L^{(i)}
    & = &  - v_2 + \delta_i [  s_\mathrm{L}(t) + s_\mathrm{H}(t)]
           + \varepsilon_i \left( v_1^{(i - 1)} - i_L^{(i)} R_C \right) .
\end{eqnarray}
\end{subequations}
There are two stable equilibria for the single Chua's circuit. Vibrational resonance appears when the response switches between the two stable equilibrium state. In the coupled Chua's circuits, the low-frequency signal propagation is discussed for a wide range of coupling parameter by carrying out a vibrational resonance analysis.  The response amplitude at the low frequency of the $i$-th circuit increases with its size $i$, eventually reaching saturation.

The modified Chua's circuit model is \cite{ref103}, \cite{ref252}
\begin{subequations}
\label{eq89}
\begin{eqnarray}
 \dot x & = & \alpha y - \alpha F(x) + s_\mathrm{L}(t)
                 + s_\mathrm{H}(t), \\
 \dot y  & = & x - y + z, \\
 \dot z & = &  - \beta y,
\end{eqnarray}
where
\begin{eqnarray}
     F(x) = \epsilon x + \epsilon A \mathrm{sgn} (x)
             - \epsilon A \sum_{j = 0}^{n - 1}
             \left[ \mathrm{sgn} (x + 2jA) + \mathrm{sgn} (x - 2jA) \right]
\end{eqnarray}
or
\begin{eqnarray}
       F(x) = \epsilon x - \epsilon A \sum_{j = 0}^{n - 1}
                \left[ \mathrm{sgn} (x + (2j + 1)A)
                  + \mathrm{sgn} (x - (2j + 1)A) \right]
\end{eqnarray}
with
\begin{eqnarray}
     \mathrm{sgn} (x) = \left\{ \begin{array}{rl}
     1, & \mathrm{if} \; x > 0 \\
     0, & \mathrm{if} \; x = 0 \\
    - 1, & \mathrm{if} \;x < 0 \end{array} \right.
\end{eqnarray}
\end{subequations}
and $\alpha$, $\beta$, $\epsilon$, $A>0$ and $n \ge 1$. Both expressions of $F(x)$ represent sawtooth functions, yet they have distinct equilibrium points and breakpoints. Typically, the number of resonance peaks corresponds to the number of breakpoints.

The piecewise linear nonautonomous Murali-Lakshmanan-Chua circuit is \cite{ref200}
\begin{subequations}
\label{eq90}
\begin{eqnarray}
 \dot x  & = &  y - h(x), \\
 \dot y  & = &  - \beta (1+v) y - \beta x + s_\mathrm{L}(t) + s_\mathrm{H}(t),
\end{eqnarray}
where
\begin{eqnarray}
   h(x) = \left\{ \begin{array}{ll}
    b x + (a -b), & \mathrm{if} \; x > 1 \\
    a x,  & \mathrm{if} \;  \left| x \right| < 1 \\
    b x - (a - b),  & \mathrm{if} \; x < -1
         \end{array} \right.
\end{eqnarray}
\end{subequations}
Both vibrational resonance and logical vibrational resonance occurs in this kind of Chua's circuit.

In addition to the Chua's circuit mentioned above, there are various models within the Chua's circuit family \cite{ref253}. The dynamics induced by fast-varying and slow-varying excitations in these variants of Chua's circuits may reveal novel vibrational resonance phenomena with practical implications.

\subsubsection{Energy harvesting system}

Vibrational resonance occurs in an energy harvesting system which is governed by \cite{ref148}, \cite{ref149}
\begin{subequations}
\label{eq91}
\begin{eqnarray}
 \ddot x + 2\delta \dot x - k x \left( 1 - x^2 \right)  - \chi y
     & = & s_\mathrm{L}(t) + s_\mathrm{H}(t), \\
 \dot y + \lambda y + \kappa \dot x & = & 0,
\end{eqnarray}
\end{subequations}
where $x$ is the transverse displacement of the beam, while $\delta$ and $k$ describe the damping ratio the stiffness, respectively. Regarding the other parameters, $y$, $\chi$, $\kappa$ denote the voltage across the load resistor, the piezoelectric coupling strength of the mechanical equation, and the piezoelectric coupling strength of the electrical equation, respectively. Additionally, $\lambda \propto 1/RC$ represents the reciprocal of the electrical circuit, where $R$ denotes the load resistance and $C$ represents the capacitance.

Another study on vibrational resonance involves a tri-stable energy harvester connected to a standard rectifier circuit, as discussed in \cite{ref150}
\begin{subequations}
\label{eq92}
\begin{eqnarray}
  \ddot x + \delta \dot x + \frac{{dV}}
{{dx}} + \chi y & = & {s_L}(t) + {s_H}(t), \\
  \dot y + I & = & \dot x,
\end{eqnarray}
\end{subequations}
where $x$, $\delta$, $\chi$ and $y$ have the same meaning as in Eq.~(\ref{eq89}), while $I$ represents the current supplied to the standard rectifier circuit. The expression for $V(x)$ remains identical to the one presented in Eq.~(\ref{eq33})."

Besides the external form, both slow-varying and fast-varying excitations may also manifest in parametric form within an energy harvesting system, as illustrated in the model proposed by \cite{ref151}
\begin{subequations}
\label{eq93}
\begin{eqnarray}
 \ddot x + 2\delta \dot x - k x \left( 1 - x^2 \right)  - \chi y + [s_\mathrm{L}(t) + s_\mathrm{H}(t)] x
     & = & 0, \\
 \dot y + \lambda y + \kappa \dot x & = & 0,
\end{eqnarray}
\end{subequations}
Here, the parametric excitation is the axial load onto the piezoelectric buckled beam harvester.

Investigation of vibrational resonance in an energy harvesting system holds promise for enhancing the efficiency of converting vibrational energy into electrical energy, offering significant engineering applications.

\subsubsection{Laser system}
The optimal Bloch vibrational resonance model is \cite{ref24}, \cite{ref26}, \cite{ref254}
\begin{subequations}
\label{eq94}
\begin{eqnarray}
   \dot x & = &  - \omega_0 y, \\
   \dot y & = & \omega_0 x + \chi z f(t), \\
   \dot z & = &  - \chi y f(t) , \\
     f(t) & = & f_0 + s_\mathrm{L}(t) + s_\mathrm{H}(t),
\end{eqnarray}
\end{subequations}
where $x^2 + y^2 + z^2 = 1$,  $\omega_0$ is the transition frequency of the atom, $\chi $ and $f_0$ are constants. Vibrational resonance in  vertical-cavity surface-emitting lasers are also studied mainly experimentally \cite{ref162}, \cite{ref255}-\cite{ref258}. In laser systems, vibrational resonance has the capability to amplify not just the weak low-frequency signal at the fundamental frequency, but also to enhance the response at subharmonic or superharmonic frequencies by appropriately selecting a high-frequency signal.

\subsection{Models with maps}

Depending on the complexity of the map, the response exhibits not only the conventional vibrational resonance phenomenon but also a broader range of additional dynamical phenomena.

Vibrational resonance was analyzed in the one-dimensional Bellows map \cite{ref259}
\begin{equation}
\label{eq95}
      x_{n+1} = \frac{rx_n}{1+x_n^b} + s_\mathrm{L}(n) + s_\mathrm{H}(n),
\end{equation}
which describes the dynamical evolution of the population density of an organism, and it was analyzed also in the two-dimensional Rulkov map
\begin{subequations}
\label{eq96}
\begin{eqnarray}
      x_{n+1} & = & \frac{\alpha}{1+x_n^2} +y_n + s_\mathrm{L}(n) + s_\mathrm{H}(n), \\
      y_{n+1} & = & y_n - \beta x_n - \sigma,
\end{eqnarray}
\end{subequations}
that mimics the behavior of complex continuous time neuronal models.

The sine square map, which characterizes the hybrid optical bistable interferometer, and the sine circle map under biharmonic signals are respectively defined as follows \cite{ref260}

\begin{equation}
  \label{eq97}
   x_{n + 1} = A \sin ^2 \left( x_n - b \right)
      + s_\mathrm{L}(n) + s_\mathrm{H}(n)
\end{equation}
and
\begin{equation}
  \label{eq98}
    y_{n + 1} = y_n + \mu \sin (y_n)  + s_\mathrm{L}(n) + s_\mathrm{H}(n) .
\end{equation}
These two kinds of maps exhibit vibrational resonance induced by high-frequency excitation.

Vibrational resonance was analyzed in detail in coupled neuronal maps in \cite{ref79}
\begin{subequations}
  \label{eq99}
\begin{eqnarray}
  x_{n + 1}  & = &  a_i / \left( 1 + x_n^2 \right)
                      + y_n + s_\mathrm{L}(n) + s_\mathrm{H}(n), \\
   y_{n + 1}  & = &  y_n - \beta x_n - \gamma,
\end{eqnarray}
\end{subequations}
and in small-word networks
\begin{subequations}
  \label{eq100}
\begin{eqnarray}
    x_{i,n + 1}\! \! & = & \! \! a_i / \left( 1\! \! + \! \!x_{i,n}^2 \right)\! \! + \! \!y_{i,n}\! \! +\! \! \varepsilon \sum_j C(i,j) \left( x_{j,n}\! \! - \! \! x_{i,n} \right)
               \! \! + \! \!s_\mathrm{L}(n)\! \! +\! \! s_\mathrm{H}(n) ,   \\
    y_{i,n + 1} & = & y_{i,n} - \sigma _i x_{i,n} - \beta_i x_n
\end{eqnarray}
\end{subequations}
and in modular networks
\begin{subequations}
  \label{eq101}
\begin{eqnarray}
    x_{I,i, n + 1}
     &  = &  \alpha / \left( 1 + x_{I,i,n}^2 \right)  + y_{I,i,n}
             + I_{I,i,n}^{\mathrm{syn}}
               + s_\mathrm{L}(n) + s_\mathrm{H} (n) , \\
    y_{I,i,n + 1}
     & = &  y_{I,i,n} - \beta x_{I,i,n} - \gamma ,
\end{eqnarray}
where
{\small
\begin{eqnarray}
 I_{I,i,n}^{\mathrm{syn}}
        =  \varepsilon_{\mathrm{intra}} \sum_j
               A_I(i,j) \left( x_{I,j,n}\! \! -\! \! x_{I,i,n} \right)
              \! \!  + \! \! \varepsilon_{\mathrm{inter}} \sum_j
                  B_{I,J}(i,j) \left( x_{J,j,n}\! \! -\! \! x_{I,i,n} \right) .
\end{eqnarray}
}
\end{subequations}

The delayed discrete Rulkov neuron model is \cite{ref40}
\begin{subequations}
  \label{eq102}
\begin{eqnarray}
     x_{n + 1} & = & \frac{\alpha }{ 1 + x_n^2 } + y_n
              + s_\mathrm{L}(n) + s_\mathrm{H} (n) , \\
     y_{n + 1} & = &  y_n - \beta x_{n - \tau } - \gamma.
\end{eqnarray}
\end{subequations}
Vibrational resonance in the above Rulkov neuron model has also been explored.

\subsection{Models with fractional differential systems}

Three commonly used definitions for the fractional-order derivative are the Riemann-Liouville, Caputo, and Gr\"unwald-Letnikov definitions \cite{ref261} (see Appendix for details). Vibrational resonance has been investigated in certain systems involving fractional-order derivatives. The presence of fractional-order damping may lead to new vibrational resonance patterns. We outline these systems below.

There are different fractional Duffing oscillators, such as it in the overdamped version \cite{ref48}, \cite{ref112}
\begin{equation}
 \label{eq103}
      \frac{ \mathrm{d}^\alpha  x }{ \mathrm{d} t^\alpha }
          + \omega _0^2 x + \beta x^3  = s_\mathrm{L}(t) + s_\mathrm{H}(t),
\end{equation}
or in the underdamped version
\begin{equation}
 \label{eq104}
    \frac{ \mathrm{d}^2 x }{ \mathrm{d} t^2}
       + \delta \frac{ \mathrm{d}^\alpha x }{ \mathrm{d} t^\alpha }
          + \omega _0^2 x + \beta x^3  = s_\mathrm{L}(t) + s_\mathrm{H}(t),
\end{equation}
or with both external and intrinsic fractional-order damping terms \cite{ref49}
\begin{equation}
 \label{eq105}
    \frac{ \mathrm{d}^\beta x }{ \mathrm{d} t^\beta}
         + \delta \frac{ \mathrm{d}^\alpha x }{ \mathrm{d} t^\alpha }
          + \omega _0^2 x + b x^3  = s_\mathrm{L}(t) + s_\mathrm{H}(t).
\end{equation}
The fractional Duffing oscillator with asymmetric bistable potential is \cite{ref64}
\begin{equation}
 \label{eq106}
    \frac{ \mathrm{d}^\alpha  x}{ \mathrm{d} t^\alpha }
         - \omega_0^2 x + a x^2 + b x^3 = s_\mathrm{L}(t) + s_\mathrm{H}(t) .
\end{equation}

Vibrational resonance is analysed also in a quadratic oscillator \cite{ref65}
\begin{equation}
 \label{eq107}
     \frac{\mathrm{d}^\alpha x}{ \mathrm{d} t^\alpha }
          - \mu x  + x^2 = s_\mathrm{L}(t) + s_\mathrm{H}(t) .
\end{equation}

The fractional Mathieu-Duffing oscillator is \cite{ref61}
\begin{equation}
 \label{eq108}
      \frac{ \mathrm{d}^2 x }{ \mathrm{d} t^2 }
        + \delta \frac{ \mathrm{d}^\alpha x}{ \mathrm{d} t^\alpha}
          + \left( \omega_0^2 + s_\mathrm{H}(t) \right) x + \beta x^3
              = s_\mathrm{L}(t) .
\end{equation}

There are two kinds of quintic oscillators with fractional-order damping, as shown in \cite{ref53}, \cite{ref62}
\begin{equation}
 \label{eq109}
     \frac{ \mathrm{d}^2 x }{ \mathrm{d} t^2}
       + \delta \frac{ \mathrm{d}^\alpha x  }{ \mathrm{d} t^\alpha }
        + \omega_0^2 x + \beta x^3  + \gamma x^5
         =  s_\mathrm{L}(t)  + s_\mathrm{H}(t),
\end{equation}
or in the other form \cite{ref54}
\begin{equation}
 \label{eq110}
       \frac{ \mathrm{d}^\beta x }{ \mathrm{d} t^\beta }
       + \delta \frac{ \mathrm{d}^\alpha x }{ \mathrm{d} t^\alpha }
        + \omega_0^2 x + \beta x^3 + \gamma x^5
        =  s_\mathrm{L}(t)  + s_\mathrm{H}(t).
\end{equation}

Under the excitations of a fast-varying parametric signal and a slow-varying external signal, the model of an oscillator with two fractional-order damping terms is given by \cite{ref50}
\begin{equation}
 \label{eq111}
    \frac{ \mathrm{d}^q x }{ \mathrm{d} t^q }
       + \delta \frac{ \mathrm{d} ^p x  }{ \mathrm{d} t^p }
        + \left( \omega_0^2 + s_\mathrm{H}(t) \right) x
          + \beta x  \left| x \right|^{r - 1} = s_\mathrm{L}(t) .
\end{equation}

The fractional-order system with the periodic potential is \cite{ref51}
\begin{equation}
 \label{eq112}
    \frac{ \mathrm{d}^\alpha x }{ \mathrm{d} t^\alpha }
        + \cos ax + \frac{1 }{2} \Delta \cos 2ax
           =  s_\mathrm{L}(t) + s_\mathrm{H}(t).
\end{equation}

Coupled fractional anharmonic oscillators are given as \cite{ref52}
\begin{subequations}
 \label{eq113}
\begin{eqnarray}
      \frac{ \mathrm{d} ^\alpha  x }{ \mathrm{d} t^\alpha }
       & = & a_1 x - b_1 x^3  + \gamma x y^2  + s_\mathrm{L}(t)
            + s_\mathrm{H}(t), \\
      \frac{ \mathrm{d} ^\alpha  y }{ \mathrm{d} t^\alpha }
       & = & a_2 y - b_2 y^3  + \gamma x^2 y .
 \end{eqnarray}
\end{subequations}

\subsection{Models with delayed differential systems}

There are some different delayed models corresponding to vibrational resonance. A very simple example system is the overdamped bistable oscillator
\begin{equation}
 \label{eq114}
   \dot x + \omega_0^2 x(t-\tau) + \beta x^3 + r
                   = s_\mathrm{L}(t) + s_\mathrm{H}(t) .
\end{equation}
The cases $r=0$ and $r \ne 0$ are considered in \cite{ref29}, \cite{ref33}, \cite{ref45} and \cite{ref36}, respectively.

Vibrational resonance has been explored in the overdamped version of the delayed Duffing oscillator
\begin{equation}
 \label{eq115}
    \dot x + \omega_0^2 x  +  \beta x^3 + \gamma x(t-\tau)
            = s_\mathrm{L}(t) + s_\mathrm{H}(t),
\end{equation}
and in underdamped version
\begin{equation}
 \label{eq116}
     \ddot x + \delta \dot x + \omega_0^2 x + \beta x^3
           + \gamma x(t-\tau) = s_\mathrm{L}(t) + s_\mathrm{H}(t),
\end{equation}
respectively \cite{ref34}.
In these two systems, the potential depends on the system parameters and may present in single-well, double-well or double-hump configuration.

The delayed harmonically trapped potential system is \cite{ref43}
\begin{equation}
 \label{eq117}
      \ddot x + \delta \dot x + \omega_0^2x + \beta \sin x
             + r x(t - \tau ) = s_\mathrm{L}(t) + s_\mathrm{H}(t) .
\end{equation}

The delayed coupled anharmonic oscillators are described by \cite{ref15}, \cite{ref30}, \cite{ref31}
\begin{subequations}
 \label{eq118}
\begin{eqnarray}
    \dot x & = &  a_1 x(t - \tau ) - b_1 x^3 + \delta x y^2
                  + s_\mathrm{L}(t) + s_\mathrm{H}(t) , \\
    \dot y & = &  a_2 y - b_2 y^3 + \delta x^2 y
 \end{eqnarray}
\end{subequations}
and the globally coupled oscillators are given as \cite{ref32}
\begin{subequations}
 \label{eq119}
\begin{eqnarray}
    \dot x_1
        & = & x_1 - x_1^3 + \frac{\varepsilon}{N}
               \sum_{j = 1}^N [x_j(t - \tau ) - x_1]
                 + s_\mathrm{L}(t) + s_\mathrm{H}(t) , \\
    \dot x_i
       \! \! & = & \! \! x_i \! \!-\! \! x_i^3 \! \!+\! \! \frac{\varepsilon }{N}
                  \sum_{j = 1}^N [x_j(t\! \! -\! \! \tau)\! \! -\! \! x_i]
                     \! \! + \! \! s_\mathrm{H}(t) , i = 2,3, \cdots ,N.
 \end{eqnarray}
\end{subequations}
Vibrational resonance was also investigated in the same coupled oscillators described by Eq.~(\ref{eq118}), albeit with excitations replaced by an amplitude-modulated signal, as studied in \cite{ref31}.

Considering multiple time delays \cite{ref35}, vibrational resonance was reported in a single Duffing oscillator
\begin{equation}
 \label{eq120}
     \ddot x + \delta \dot x  + \omega_0^2 x  + \beta x^3
              + \frac{\gamma }{M} \sum_{m = 1}^M x(t - m\tau )
               = s_\mathrm{L}(t) + s_\mathrm{H}(t)
\end{equation}
and in coupled Duffing oscillators
\begin{subequations}
 \label{eq121}
\begin{eqnarray}
    \ddot x_1 + \delta \dot x_1  + \omega_0^2 x_1 + \beta x_1^3
         & = & s_\mathrm{L}(t) + s_\mathrm{H}(t) , \\
    \ddot x_i \! \! + \! \! \delta \dot x_i \! \! + \! \! \omega_0^2 x_i \! \! + \! \! \beta x_i^3
          \! \! & = &\! \! \frac{\gamma }{M}\sum_{l = 1}^M x_{i \! \! - \! \! 1}(t \! \! - \! \! m\tau ),
                    i = 2,3, \cdots ,N .
\end{eqnarray}
\end{subequations}

Vibrational resonance appears in the delayed single FitzHugh-Nagumo model \cite{ref37}, \cite{ref38}
\begin{subequations}
 \label{eq122}
\begin{eqnarray}
     \varepsilon \dot x  & = & x  - \frac{1}{3} x^3  - y , \\
     \dot y & = &  x + a + K \left[ y(t - \tau ) - y \right]
             + s_\mathrm{L}(t) + s_\mathrm{H}(t),
\end{eqnarray}
\end{subequations}
and in the delayed coupled FitzHugh-Nagumo model \cite{ref39}
\begin{subequations}
 \label{eq123}
\begin{eqnarray}
    \varepsilon \dot x_1
      & = & x_1 - \frac{1}{3} x_1^3  - y_1, \\
    \dot y_1
      & = & x_1 + a + \frac{K}{N} \sum_{j = 1}^N
         \left[ y_j (t - \tau ) - y_1 \right]
           + s_\mathrm{L}(t) + s_\mathrm{H}(t), \\
    \varepsilon \dot x_i
      & = & x_i - \frac{1}{3} x_i^3 - y_i , \\
    \dot y_i
     \! \! & = & \! \! x_i + a + \frac{K}{N} \sum_{j = 1}^N
         \left[ y_j (t - \tau ) - y_i \right]  + s_\mathrm{H}(t), i = 2,3, \cdots , N .
\end{eqnarray}
\end{subequations}
In \cite{ref38}, $\tau$ is time varying and $\tau  = \tau _0 + \varepsilon_0\sin \omega _0t$. In \cite{ref37}, \cite{ref39}, $\tau$ is a constant.

The delayed genetic toggle switch is \cite{ref41}
\begin{subequations}
 \label{eq124}
\begin{eqnarray}
    \dot u & = &  \frac{\alpha }{1 + v^\beta } - u(t - \tau )
                  + s_\mathrm{L}(t) + s_\mathrm{H}(t), \\
    \dot v & = & \frac{\alpha }{1 + u^\beta } - v(t - \tau ) .
\end{eqnarray}
\end{subequations}
Herein, $u$ and $v$ indicate the concentrations of two considered transcription factors. The excitations come from some factors such as the oscillating temperature, the experimental setting variation of the chemical inductor, etc.

The gene transcriptional regulatory system with a linear delayed term is modeled by \cite{ref42}
\begin{equation}
 \label{eq125}
     \dot x = \frac{k_\mathrm{f} x^2 }{x^2 + K_\mathrm{d} }
                 - k_\mathrm{d} x(t - \tau ) + R_\mathrm{bas}
            + s_\mathrm{L}(t) + s_\mathrm{H}(t).
\end{equation}
When the delayed terms are in nonlinear form, the dynamical equation is
\begin{equation}
 \label{eq126}
      \dot x  =  \frac{ k_\mathrm{f} x^2(t - \tau ) }
           {x^2(t - \tau ) + K_\mathrm{d} } - k_\mathrm{d}x
            + R_\mathrm{bas} + s_\mathrm{L}(t) + s_\mathrm{H}(t) .
\end{equation}
In the absence of time delay and external perturbations, the system's potential exhibits an asymmetric shape similar to the curve shown in Fig.~\ref{figpotential}(d).

Some systems also consider both fractional-order damping and time delay simultaneously. The fractional Duffing oscillators incorporating a time delay term are described in \cite{ref59}
\begin{equation}
 \label{eq127}
        \frac{ \mathrm{d}^\alpha x }{ \mathrm{d} t^\alpha} + \omega_0^2x
           + \beta x^3 + \gamma x(t-\tau) = s_\mathrm{L}(t) + s_\mathrm{H}(t),
\end{equation}
and
\begin{equation}
 \label{eq128}
      \frac{\mathrm{d}^2 x }{ \mathrm{d} t^2}
          + \delta \frac{ \mathrm{d}^\alpha x }{ \mathrm{d} t^\alpha }
        - \omega_0^2x + \beta x^3  + \gamma x(t - \tau )
         = s_\mathrm{L}(t) + s_\mathrm{H}(t) ,
\end{equation}
and \cite{ref262}
\begin{equation}
 \label{eq129}
      \frac{ \mathrm{d}^\alpha x }{ \mathrm{d} t^\alpha} + \omega_0^2x + a x^2
           + \beta x^3 + \gamma x(t-\tau) = s_\mathrm{L}(t) + s_\mathrm{H}(t).
\end{equation}

The Duffing system with a generalized time delay term is \cite{ref263}
\begin{equation}
 \label{eq130}
      \frac{ \mathrm{d}^2 x }{ \mathrm{d} t^2 }
          + \delta \frac{ \mathrm{d}^\alpha x }{ \mathrm{d} t^\alpha }
          + \omega_0 x + \beta x^3 + \gamma \mathrm{D}^\alpha x(t - \tau )
         = s_\mathrm{L}(t) + s_\mathrm{H}(t) ,
\end{equation}
where $\mathrm{D}^\alpha x(t - \tau )$ is the delayed feedback in a generalized fractional-order differential form.

The fractional-order quintic oscillator with a linear time delay is investigated in the equation \cite{ref264}
\begin{equation}
 \label{eq131}
        \frac{ \mathrm{d}^2 x }{ \mathrm{d} t^2 }
         + \delta \frac{ \mathrm{d}^\alpha x }{ \mathrm{d} t^\alpha }
         + \omega_0 x + \beta x^3 + \gamma x^5 + \xi x(t - \tau )
           = s_\mathrm{L}(t) + s_\mathrm{H}(t) .
\end{equation}

The fractional Mathieu-Duffing oscillator modeled with distributed time delay is \cite{ref265}
\begin{equation}
 \label{eq132}
     \frac{ \mathrm{d}^2 x }{ \mathrm{d} t^2 }
        + \delta \frac{ \mathrm{d}^\alpha x }{ \mathrm{d} t^\alpha }
         + [ a + s_\mathrm{H}(t) ] x + b x^3
          + \gamma \int_{ - \infty }^t  h(t - \tau ) x (\tau )
            \mathrm{d} \tau  = s_\mathrm{L}(t),
\end{equation}
while with fixed time delay is
\begin{equation}
 \label{eq133}
     \frac{ \mathrm{d}^2 x }{ \mathrm{d} t^2 }
          + \delta \frac{ \mathrm{d}^\alpha x }{ \mathrm{d} t^\alpha }
      + [a + s_\mathrm{H}(t) ] x + bx^3 + \gamma x(t - \tau )
             = s_\mathrm{L}(t) .
\end{equation}

The fractional Duffing oscillator with distributed time delay and excited by multi-frequency excitations is described as \cite{ref266}
\begin{eqnarray}
 \label{eq134}
 & &   m \ddot x(t) + c\dot x(t) + \delta \frac{{{d^\alpha}x}}{{d{t^\alpha}}} + kx(t) + s_\mathrm{L1}(t)x(t) + \beta {x^3}(t) \nonumber \\
  & &  + r\int_{ - \infty }^t {\eta (t - \tau )} x(\tau )d\tau  = s_\mathrm{L2}(t) + s_\mathrm{H}(t).
\end{eqnarray}

A generalized fractional Duffing-van der Pol oscillator with distributed time delay and excited by slow-varying external  signal and fast-varying parametric signal was investigated in the following system \cite{ref63}
\begin{eqnarray}
 \label{eq135}
& & m\ddot x(t) + [{c_1} + {c_2}{x^2}(t)]\dot x(t) + {k_1}x(t) + {\delta _1}D_{0,t}^qx(t) + {\alpha _1}{x^3}(t) \nonumber \\
& & + {\lambda _1}\int_{ - \infty }^t {r(t - \tau )} x(\tau )d\tau  = s_\mathrm{L}(t) + x(t)s_\mathrm{H}(t),
\end{eqnarray}
where $D_{0,t}^qx(t)$ is the fractional-order derivative with respect to the variable $x(t)$ in the Caputo definition.

\subsection{Models with stochastic differential systems}
Due to the significant influence of noise in various scenarios, we consider in this subsection the equation of vibrational resonance in the presence of noise.

\subsubsection{Monostable systems}
Vibrational resonance in a monostable system subjected to multiplicative noise in \cite{ref134}
\begin{equation}
 \label{eq136}
  \dot x + \omega_0^2 x \xi (t) + \beta x^3 + r = \eta (t)
         + s_\mathrm{L1} (t) + s_\mathrm{L2} (t) + s_\mathrm{H}(t) .
\end{equation}
There are a fast-varying excitation, two slow-varying excitations, and two random excitations in the above equation, where $\xi (t)$ and $\eta (t)$ are uncorrelated Gaussian white noises and $r$ is the bias parameter.

\subsubsection{Bistable systems}
The bistable system excited by both a slow- and fast-varying harmonic signals and a Gaussian white noise is given by \cite{ref125}, \cite{ref130}
\begin{equation}
 \label{eq137}
    \dot x + \omega_0^2 x + \beta x^3 = s_\mathrm{L} (t) + s_\mathrm{H}(t) + \xi (t).
\end{equation}

The coupled piecewise linear system is a symmetric bistable system \cite{ref124}
\begin{subequations}
 \label{eq138}
 \begin{eqnarray}
     \dot x_i  = f(x_i) + g(x_i) \xi_i(t)
       + \frac{D}{2d} \sum_{j \in nn(i)} (x_j - x_i)
            + s_\mathrm{L} (t) + s_\mathrm{H}(t)
 \end{eqnarray}
with
\begin{eqnarray}
      \langle \xi _i (t) \xi_j (t') \rangle
          & = &  \sigma_m^2 \delta_{ij} \delta (t - t') , \\
 f(x) & = &  \left\{ \begin{array}{ll}
          - G_b x - ( G_a  - G_b ) B_\mathrm{p} & \mathrm{if} \;
                 x \le  - {B_p}, \\
           - G_a x & \mathrm{if}\;
                \left| x \right| < B_\mathrm{p}, \\
          - G_b x + ( G_a - G_b ) B_\mathrm{p} & \mathrm{if}\;
                x \ge B_\mathrm{p}   .\end{array}
                  \right.
\end{eqnarray}
\end{subequations}
The piecewise linear system can be viewed as an approximation of a typical bistable system. A similar topic was also investigated in the conventional globally coupled bistable system
\begin{equation}
 \label{eq139}
{\dot x_i} = {x_i} - x_i^3 + \frac{\varepsilon }{N}\sum\limits_{j = 1}^N {({x_j} - {x_i})}  + \sqrt {2D} {\xi _i}(t) + s_\mathrm{L} (t) + s_\mathrm{H}(t).
\end{equation}

The effects of the noise and the bi-harmonic signal on the logical response of the system are investigated in the model \cite{ref143}
\begin{equation}
 \label{eq140}
     \dot x + \omega_0^2  x(t) + \beta x^3(t) = I_1 + I_2 + r + \xi (t)
            +  s_\mathrm{L} (t) + s_\mathrm{H}(t) ,
\end{equation}
where $I_1$ and $I_2$ are two logical inputs, and $\xi(t)$ is a Gaussian white noise.

Vibrational ratchet which is very similar to vibrational resonance was investigated in the system \cite{ref144}-\cite{ref146}
\begin{equation}
 \label{eq141}
    \dot x  =  - V'(x) + s_\mathrm{L} (t) + s_\mathrm{H} (t) + \xi (t) , \quad
          V(x) = d(1 - \cos x) .
\end{equation}
The model describes the dynamics of a Brownian particle diffusing on a one-dimensional periodic substrate. The case with
\begin{equation}
 \label{eq142}
    V(x) = \sum_{n = 1}^\infty  a_n\cos nx  + \sum_{n = 1}^\infty
                b_n \sin nx
\end{equation}
has also been explored. The transport of the particle presents a vibrational ratchet effect with the cooperation of the system and the excitations. Certainly, vibrational resonance can also occur in this kind of systems.

\subsubsection{The Schmitt trigger circuit}

The Schmitt trigger circuit is \cite{ref267}
\begin{equation}
 \label{eq143}
    U_\mathrm{out} = \mathrm{sgn} \left[ \Delta U
          - s_\mathrm{L} (t) - s_\mathrm{H}(t) - \xi (t) \right] ,
\end{equation}
where $\Delta U = \frac{{{R_1}}}{{{R_1} + {R_2}}}{U_{out}}$ is the threshold value. $U_{out}$ is either the positive or negative voltage $\pm {U_0}$ and adjusted by external signals.  Here, $\xi (t)$ is a colored noise.

\subsubsection{Neural models and complex networks}

The role of noise on vibrational resonance was discussed in the FitzHugh-Nagumo model by adding the noise term $\xi(t)$ to the right-side of Eq.~(\ref{eq70}b) \cite{ref66} and also in the FitzHugh-Nagumo model with different coupling \cite{ref75}
\begin{subequations}
 \label{eq144}
\begin{eqnarray}
 \varepsilon \dot x_i & = & x_i - \frac{1}{3} x_i^3 - y_i - I_i^\mathrm{syn}, \\
    \dot y_i & = & x_i  + a + s_\mathrm{L}(t) + s_\mathrm{H}(t) + \xi (t) .
\end{eqnarray}
\end{subequations}
Here, $I_i^\mathrm{syn}$ represents different kinds of coupling.

The feed-forward multilayer neural network corresponding to the Hodgkin-Huxley neuron model is governed by \cite{ref84}
\begin{subequations}
 \label{eq145}
\begin{eqnarray}
     C_m \dot V_{i,j}
      & = &  - \left[ g_k n_{i,j}^4 \left( V_{i,j} - V_k \right)
               + g_\mathrm{Na} m_{i,j}^3 h_{i,j}
               \left( V_{i,j} - V_\mathrm{Na} \right)
                + g_l \left( V_{i,j}  - V_l \right) \right] \nonumber \\
      &  & \quad   + I_0 - I^\mathrm{syn}_{i,j}
              + a_{i,j} \left[ s_\mathrm{L}(t) + s_\mathrm{H}(t) \right], \\
      \dot x_{i,j}
      & = & \alpha_x (1 - x_{i,j} ) - \beta_x x_{i,j}
            + \xi_{i,j} x_{i,j} , \quad x_{i,j} = m_{i,j} , n_{i,j}, h_{i,j},
\end{eqnarray}
where
\begin{eqnarray}
     \alpha_{m_{i,j}} & = & 0.1 \left( V_{i,j} + 40 \right) /
                    \left( 1 - \mathrm{e}^{ - (V_{i,j} + 40)/10 } \right) , \\
     \beta_{ m_{i,j}} & = & 4 \mathrm{e}^{-( V_{i,j} +  65) /18} , \\
     \alpha_{ n_{i,j} } & = &  0.01 \left( V_{i,j} + 55 \right) /\left( 1 - \mathrm{e}^{ - ( V_{i,j} + 55)/10 } \right),\\
     \beta_{ n_{i,j} } & = & 0.125 \mathrm{e}^{ -( V_{i,j} + 65 ) / 80} , \\
     \alpha_{h_{i,j}} & = & 0.07 \mathrm{e}^{-( V_{i,j} + 65)/20}, \\
     \beta_{h_{i,j} } & = & 1 / \left( 1 + \mathrm{e}^{-(V_{i,j} + 35)/10}\right),
\end{eqnarray}
\end{subequations}
and $\xi _{i,j}(t)$ is a Gaussian white noise.

\section{Performance metrics of vibrational resonance }
\label{measures}

In this section, we primarily provide performance metrics for different types of vibrational resonance. These frequently employed metrics encompass response amplitude and gain factor for harmonic component vibrational resonance, cross-correlation coefficient and bit error rate for binary aperiodic signal-induced vibrational resonance, cross-correlation coefficient and spectrum amplification factor for frequency-modulated signal-induced vibrational resonance, as well as success probability for logical vibrational resonance, among others

\subsection{Response amplitude}

The response amplitude is used in measuring the response at a harmonic frequency. We label the response amplitude as $Q$. We extract the harmonic $\omega$ from the time series $x(t)$ by the Fourier series. The sine component $Q_\mathrm{s}$ and cosine component $Q_\mathrm{c}$ of $x(t)$ at the frequency $\omega$ are respectively calculated by
\begin{equation}
 \label{eq146}
  Q_\mathrm{s} = \frac{2}{mT} \int\limits_0^{mT} x(t) \sin (\omega t)
           \text{d} t, \quad
  Q_\mathrm{c} = \frac{2}{mT} \int\limits_0^{mT} x(t)
             \cos (\omega t) \text{d} t ,
\end{equation}
where $m$ is a large enough positive number. Without a special explanation, $x(t)$ is a steady time series where the initial transient response has been removed. For a discrete series, the Fourier coefficients are
\begin{equation}
 \label{eq147}
  Q_\mathrm{s} = \frac{2 \Delta t}{mT} \sum\limits_{i = 1}^{mT / \Delta t}
           x(t_i) \sin (\omega t_i) , \quad
  Q_\mathrm{c} = \frac{2 \Delta t}{mT} \sum\limits_{i = 1}^{mT / \Delta t}
           x(t_i) \cos (\omega t_i) .
\end{equation}
In Eq.~(\ref{eq147}), $\Delta t$ is the time step for the numerical calculations, and $T$ is the period of the low-frequency signal which is equal to $2 \pi/\omega$. As a result, the response amplitude $Q$ at the frequency $\omega$ is calculated by
\begin{equation}
 \label{eq148}
        Q = \sqrt { Q_\mathrm{s}^2 + Q_\mathrm{c}^2 } / A \,
\end{equation}
where $Q$ indicates the amplification of the weak low-frequency signal after is passes through the nonlinear system. Specifically, the low-frequency signal signal is improved $Q$ times by the vibrational resonance.

\begin{figure}[!h]
\begin{center}
\includegraphics[width=0.7\linewidth]{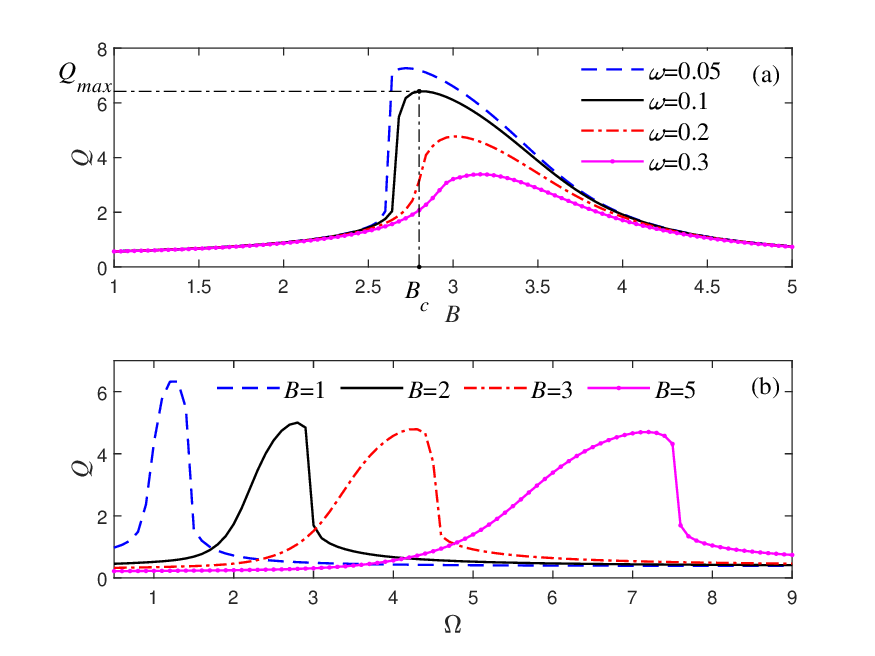}
\end{center}
\caption{Vibrational resonance induced by changing the amplitude $B$ or the frequency $\Omega$ of the fast-varying auxiliary signal in the overdamped bistable system of Eq.~(\ref{eq16}). (a) The response amplitude $Q$ versus the amplitude $B$ presents vibrational resonance for four fixed values of $\omega$. (b) The response amplitude $Q$ versus the frequency $\Omega$ presents vibrational resonance for four fixed values of $B$. The parameters are $\omega_0^2=-1$, $\beta=1$, $A=0.1$, $\Omega=4$ in (a), and $\omega=0.1$ in (b).}
\label{qoverd}
\end{figure}

Consider the system (\ref{eq16}) with
\begin{equation}
\label{eq149}
   s_\mathrm{L}(t) = A \cos (\omega t), \quad s_\mathrm{H}(t) = B \cos (\Omega t),
      \quad \omega \ll \Omega,
\end{equation}
$\omega_0^2 < 0 $ and $\beta > 0 $. By numerical simulations, a typical vibrational resonance is shown in Fig.~\ref{qoverd}(a). With varying the amplitude $B$, the resonance on the $Q-B$ curve is clearly shown. Moreover, when we change the frequency $\Omega$, we also find a vibrational resonance phenomenon, as given in Fig.~\ref{qoverd}(b).
The critical value $B_\mathrm{c}$ at which $Q$ becomes maximum and the corresponding maximum value $Q_\mathrm{max}$ are indicated in Fig.~\ref{qoverd}(a) for $\omega = 0.1$. The two signals in Eq.~(\ref{eq149}) are always used in the following analysis unless otherwise stated.

\begin{figure}[t]
\begin{center}
\includegraphics[width=0.7\linewidth]{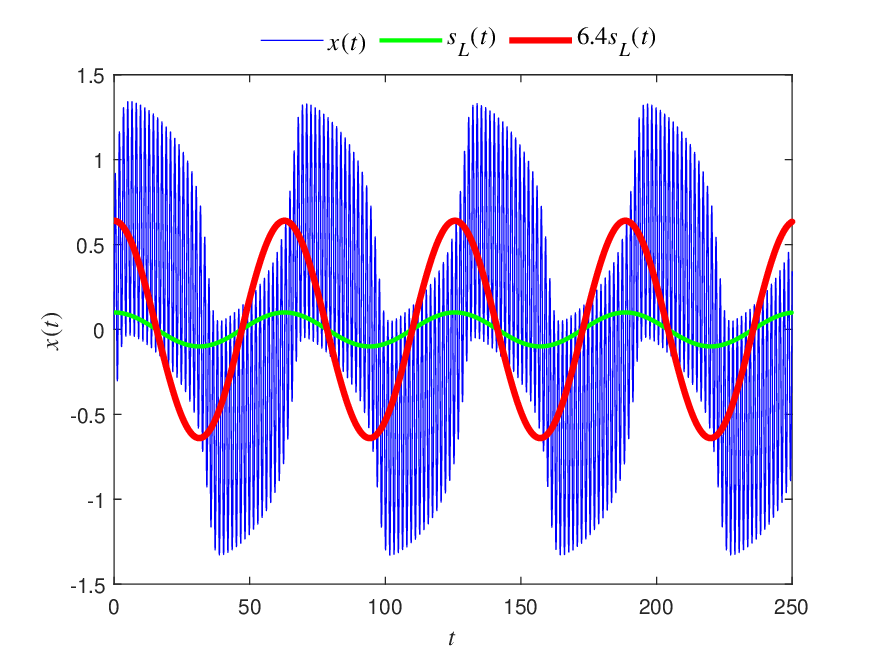}
\end{center}
\caption{Comparison of the time series of the optimal vibrational resonance response and the input low-frequency signal. The parameters are $\omega_0^2=-1$, $\beta=1$, $A=0.1$, $\omega=0.1$, $B=2.8$, $\Omega=4$. }
\label{overtime}
\end{figure}

The output of the time series corresponding to the critical amplitude $B_\mathrm{c} = 2.8 $ for which  $Q = Q_\mathrm{max} =6.4$ is shown in Fig.~\ref{overtime} to present an intuitive understanding. While in Fig.~\ref{qoverd} is given along with $s_\mathrm{L}(t)$ and $6.4(=Q_\mathrm{max})s_\mathrm{L}(t)$. In this figure, we find that the response amplitude at the low-frequency $\omega$ is greatly enhanced. It agrees with the results in Fig.~\ref{qoverd}. It presents the occurrence of vibrational resonance once again in the time domain. As ilustrated by the results, the enhancement of the low-frequency characteristic signal is observed in both Fig.~\ref{qoverd} and Fig.~\ref{overtime}.

\begin{figure}[!h]
\begin{center}
\includegraphics[width=0.7\linewidth]{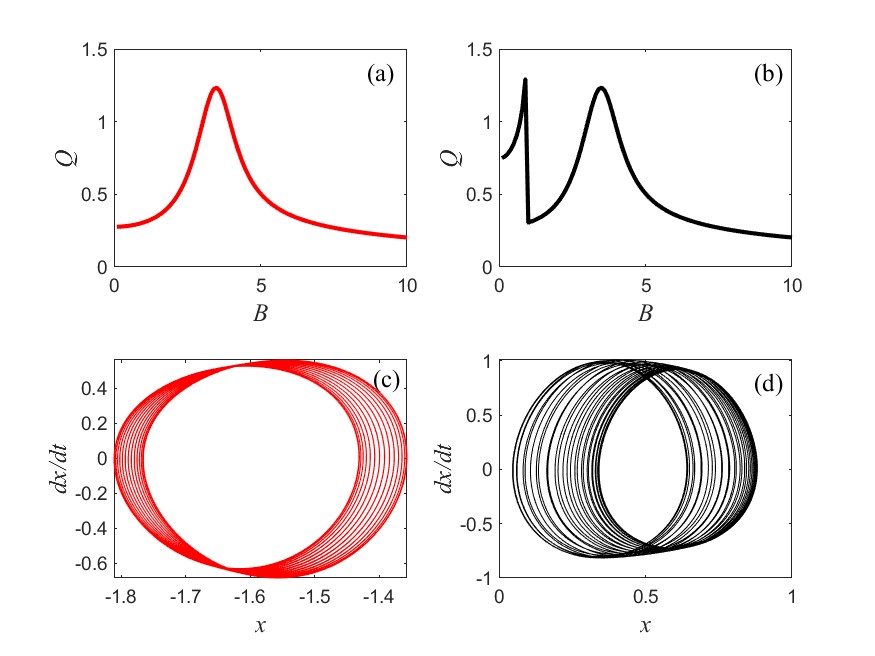}
\end{center}
\caption{Different initial conditions induce the change in the modal of vibrational resonance in the overdamped asymmetric bistable system with potential of Eq.~(\ref{eq22}). (a) The response amplitude $Q$ versus the amplitude $B$ of the fast-varying auxiliary signal presents single resonance. The initial condition is $x(0) = -1$. (b) The response amplitude $Q$ versus the amplitude $B$ of the fast-varying auxiliary signal presents double resonance. The initial condition is $x(0) = 1$. (c) The phase portrait oscillates in the left potential well under the conditions $x(0) = -1$ and $B = 0.9$. (d) The phase portrait oscillates in the right potential well under the conditions $x(0) = 1$ and $B = 0.9$. Other parameters are $\omega_0^2=-1$, $\alpha=1$, $\beta=1$, $A=0.1$, $\omega=0.1$, $\Omega=3$.}
\label{diffic}
\end{figure}

The response may vary depending on the initial conditions when a system has more than one stable state. Taking an asymmetric bistable system in an overdamped version as an example, the potential function is given in Eq.~(\ref{eq22}). The two harmonic excitations are $A\cos(\omega t)$ and $B\cos(\Omega t)$, respectively, where $\Omega\gg\omega$. The influence of the initial conditions on the response is illustrated in Fig.~\ref{diffic}. Clearly, if the initial conditions differ, the response may also differ. Regarding the asymmetric bistable system, the difference is easily understandable. The tilt of the potential function causes the particle to move only in a high or a low potential well when the excitation is weak and there is no cross-well motion.

\begin{figure}[!h]
\begin{center}
\includegraphics[width=0.8\linewidth]{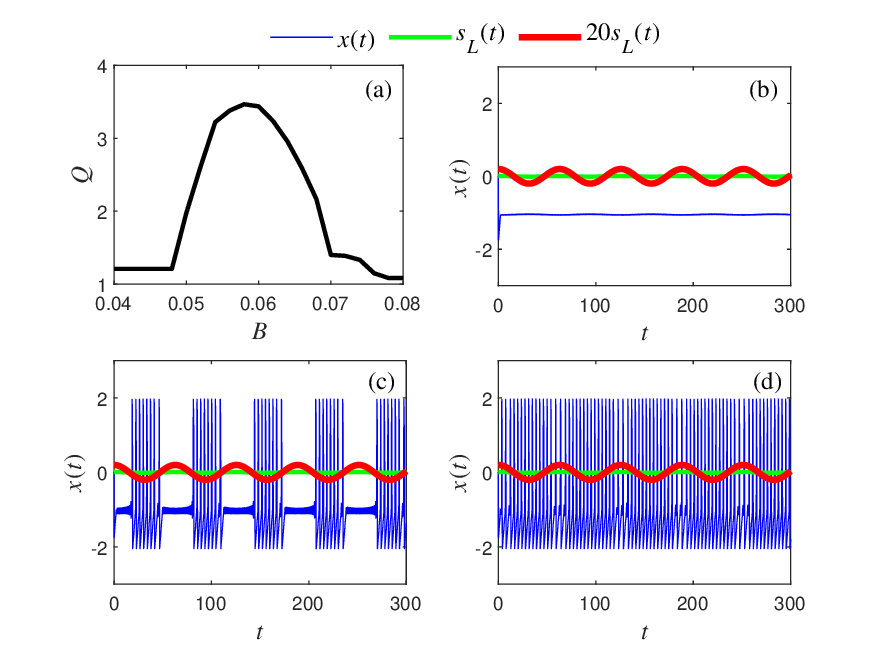}
\end{center}
\caption{The response of the FitzHugh-Nagumo neuron system of Eq.~(\ref{eq70}) to the slow-varying characteristic signal is influenced by the amplitude $B$ of the fast-varying auxiliary signal. (a) The response amplitude $Q$ versus the signal amplitude $B$ presents vibrational resonance. (b)-(d) The time series of the response corresponding to different values of the signal amplitude $B$. The parameters are $\varepsilon=0.01$, $a=1.05$, $A=0.01$, $\omega=0.1$, $\Omega=5$, and $B=0.008$, $0.058$, $0.075$ in (b), (c), (d), respectively.}
\label{FHNQ}
\end{figure}

Next, consider the FitzHugh-Nagumo neuron model in Eq.~(\ref{eq70}) \cite{ref73}, which is an excitable system subjected to biharmonic signals. The curve of $Q$ versus $B$ and some time series are depicted in Fig.~\ref{FHNQ}. To display the periodic relationship between the signal $s_L(t)$ and the output $x(t)$ better, we give $20s_L(t)$ as a comparison in the figure. At the critical value $B=0.058$, as shown in Fig.~\ref{FHNQ}(a), $Q_{max}=3.47$, the input signal $s_L(t)$ and the output $x(t)$ achieve the optimal synchronization. The weak slow-varying signal is amplified remarkably, and the spikes appear just in a half cycle of the slow-varying signal. For a smaller value of $B$, such as $B=0.008$ in the figure, almost no spike emerges. For a larger value of $B$, e.g., $B=0.075$, the spikes occur frequently. Although the output of the FitzHugh-Nagumo neuron model has a different form with the ordinary bistable system, the vibrational resonance still occurs when the system is excited by appropriate slow- and fast-varying signals. In fact, the vibrational resonance in the FitzHugh-Nagumo neuron model is complicated. It will produce the subthreshold and suprathreshold vibrational resonance. The bifurcation parameter $a$ also influences the vibrational resonance pattern. In Fig.~\ref{FHNQ}, the subthreshold vibrational resonance is shown. For other cases, a detailed discussion can be found in the work of \cite{ref69}.

\subsection{Gain factor}

The phenomenon of vibrational resonance results from the interaction between the slow characteristic signal, the nonlinear system, and the fast auxiliary signal. However, a strong resonance can still occur if the system is solely driven by a single characteristic signal. This raises the question: What is the role of the fast auxiliary signal? Is it indispensable for amplifying the weak, slowly varying characteristic signal in a nonlinear system?

\begin{figure}[t]
\begin{center}
\includegraphics[width=0.6\linewidth]{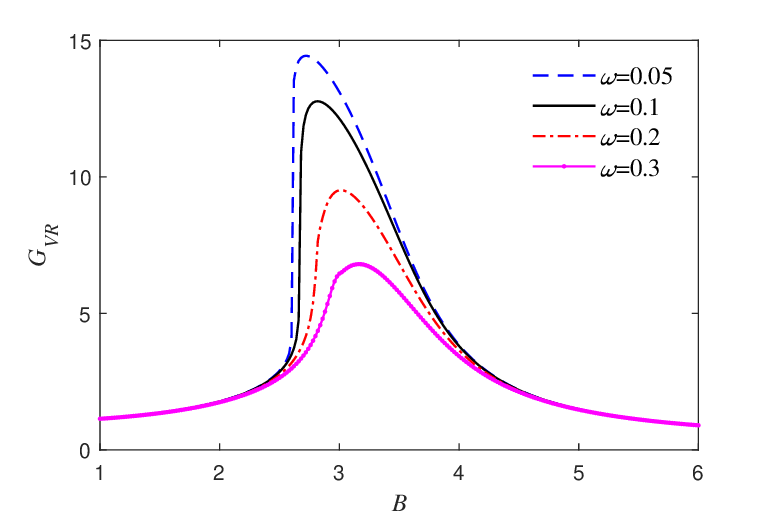}
\end{center}
\caption{The gain factor $G_\mathrm{VR}$ versus the amplitude $B$ of the fast-varying auxiliary signal also presents vibrational resonance. All simulation parameters are the same as that in Fig.~\ref{qoverd}(a).}
\label{gainfactor}
\end{figure}

To reflect the role of the fast auxiliary signal on vibrational resonance directly, Chizhevsky and Giacomelli defined the gain factor of the response amplitude in their work \cite{ref127}, \cite{ref214}. The gain factor $G_\mathrm{VR}$ is calculated by
\begin{equation}
\label{eq150}
   G_\mathrm{VR} = \frac{ Q_\mathrm{L} (\omega) }{ Q_0(\omega) } .
\end{equation}
Herein, $Q_\mathrm{L}(\omega)$ and $Q_0(\omega)$ represent the response amplitude at the frequency of the slow-varying characteristic signal in the presence and absence of the fast-varying auxiliary signal, respectively. $Q_\mathrm{L}(\omega)$ and $Q_0(\omega)$ are calculated using Eq.~(\ref{eq148}). We continue to use the overdamped system in Eq.~(\ref{eq16}) as an example. The curve of $G_\mathrm{VR}$ is depicted in Fig.~\ref{gainfactor}. The effect of the fast-varying auxiliary signal is clearly illustrated in the figure. Due to the presence of the auxiliary signal, similar to the results shown in Fig.~\ref{qoverd}, the weak slow-varying characteristic signal is notably amplified.

In fact, to investigate the role of the fast-varying auxiliary signal on vibrational resonance, one can consider introducing the concept of the gain factor as described in Eq.~(\ref{eq150}). Moreover, the response amplitude $Q(\omega)$ can be substituted with various other metrics such as signal-to-noise ratio (SNR), spectral amplification factor, cross-correlation coefficient, and so on, as deemed suitable.

\subsection{Cross-correlation coefficient}

When investigating vibrational resonance induced by an aperiodic binary signal, a commonly used metric is the cross-correlation coefficient between the system output and the input of a slowly-varying signal, as discussed in \cite{ref110}, \cite{ref115}, \cite{ref116}. The cross-correlation coefficient is defined by
\begin{equation}
\label{eq151}
        C_{sx} = \frac{ \sum_{j = 1}^n  \left[ s(j) - \bar{s} \right]
                   \left[ x(j) - \bar{x} \right] }
                { \sqrt{ \sum_{j = 1}^n \left[ s(j) - \bar{s} \right]^2
                 \sum_{j = 1}^n \left[ x(j) - \bar{x} \right]^2 }} ,
\end{equation}
where $\bar s$ and $\bar x$ are the average values of the input aperiodic signal and the output time series, respectively. The numerical value of $C_{sx}$ lies in $[-1, 1]$. When vibrational resonance occurs, the cross-correlation coefficient usually reaches a maximum. Nevertheless, even though the cross-correlation coefficient is a large enough value, vibrational resonance does not necessarily occur. Specifically, the cross-correlation coefficient is the necessary but not sufficient condition for vibrational resonance. Thus, there is a need to establish novel metrics for assessing aperiodic vibrational resonance. Furthermore, recent studies have revealed that in cases where the system is subjected to a single aperiodic signal, the most pronounced aperiodic resonance can also manifest at the inflection point, where the correlation coefficient reaches its minimum, as reported in \cite{ref268}.

\begin{figure}[t]
\begin{center}
\includegraphics[width=0.7\linewidth]{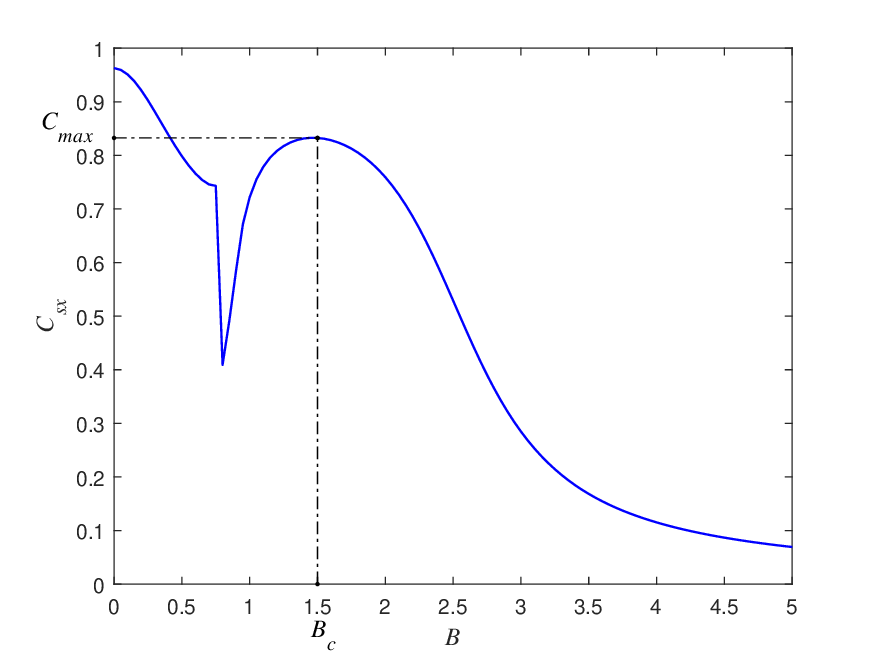}
\end{center}
\caption{The cross-correlation coefficient $C_{sx}$ versus the amplitude $B$ of the fast-varying auxiliary signal presents aperiodic vibrational resonance in the system of Eq.~(\ref{eq152}). The parameters are $\omega_0^2=-1$, $\beta=1$, $A=0.3$, $T=20$ and $\Omega=\pi$.}
\label{crossc}
\end{figure}

\begin{figure}[!h]
\begin{center}
\includegraphics[width=0.7\linewidth]{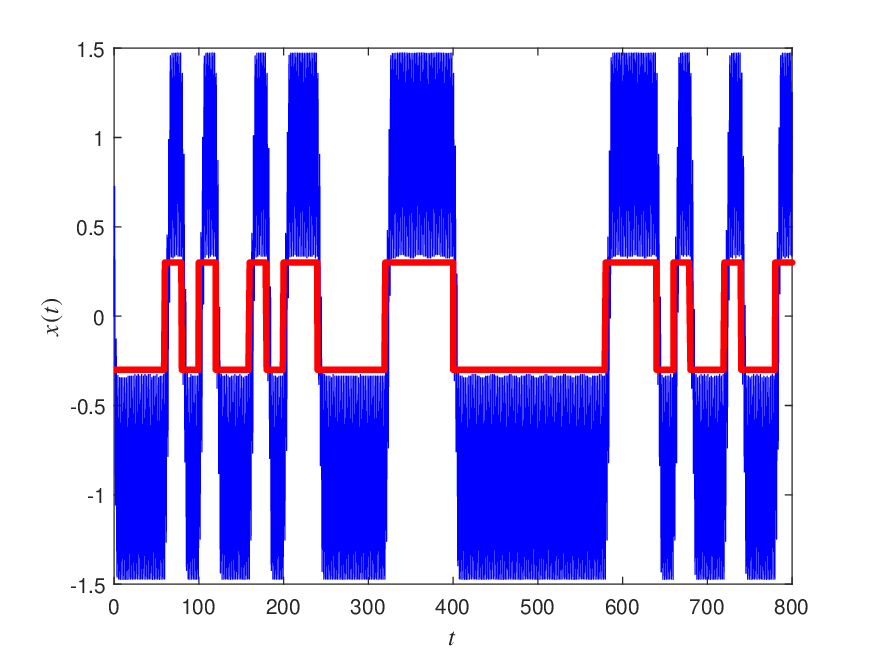}
\end{center}
\caption{Time series of the optimal aperiodic vibrational resonance caused by the binary signal in the system of Eq.~(\ref{eq152}). The thick line (red color) is the aperiodic binary characteristic signal. The thin line (blue color) is the output corresponding to the optimal aperiodic vibrational response. The parameters are $\omega_0^2=-1$, $\beta=1$, $A=0.3$, $B=1.5$, $T=20$ and $\Omega=\pi$.}
\label{crosstime}
\end{figure}

Here, we consider the system described by
\begin{equation}
\label{eq152}
      \dot x + \omega_0^2 x  + \beta x^3 = s(t) + B \mathrm{sgn} ( \cos \Omega t)
\end{equation}
where $s(t)$ is an aperiodic binary signal  defined in Eq.~(\ref{eq6}). The fast-varying auxiliary signal is a periodic square signal with amplitude $B$. In Fig.~\ref{crossc}, the cross-correlation coefficient corresponding to the amplitude of the fast-varying auxiliary signal is presented. At the maximal value $C_\mathrm{max}$, i.e., $B=B_\mathrm{c}$, the response achieves the strongest aperiodic vibrational resonance. The response corresponding to the optimal aperiodic vibrational resonance is given in Fig.~\ref{crosstime}. The amplification of the weak aperiodic signal is evident.

\subsection{Bit error rate}

\begin{figure}[!h]
\begin{center}
\includegraphics[width=0.8\linewidth]{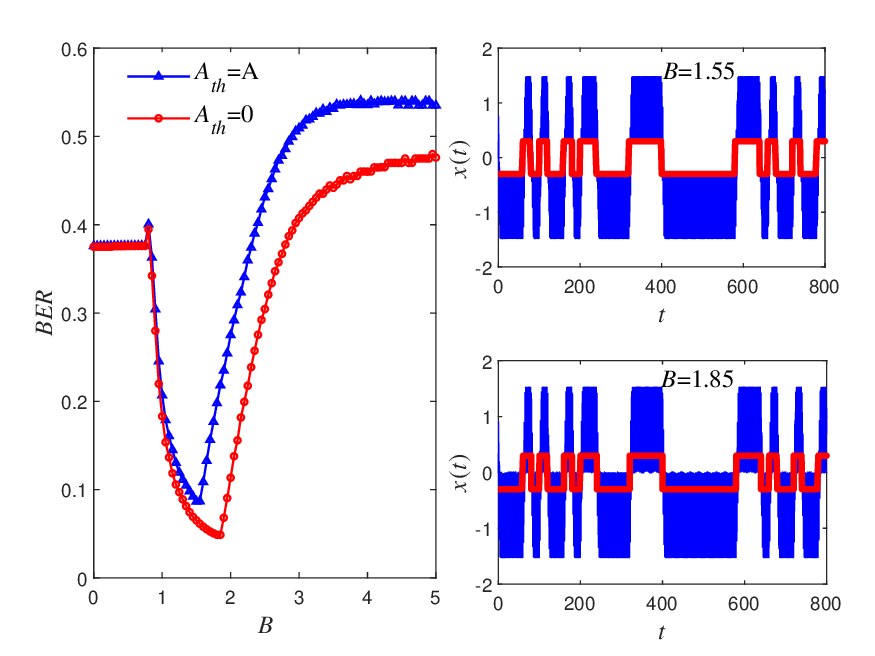}
\end{center}
\caption{The bit error rate versus the amplitude $B$ of the fast-varying auxiliary signal presents aperiodic vibrational resonance in the system of Eq.~(\ref{eq152}) (left) and the optimal vibrational resonance response corresponding to the threshold value $A_{th}=A$ (right, top) and the threshold value $A_{th}=0$(right, bottom), respectively. All simulation parameters are the same as those in Fig.~\ref{crossc}}.
\label{BERFIG}
\end{figure}

The bit error rate is another commonly used metric to quantify the correctness of the output-input. The resonance occurs when the bit error rate reaches the minimum. The bit error rate has been successfully used in the stochastic resonance research \cite{ref269}-\cite{ref271}. Chizhevsky and Giacomelli introduced the bit error rate to quantify the occurrence of vibrational resonance induced by an aperiodic binary signal \cite{ref110}. The bit error rate is the ratio of wrong received bits to the total number of transmitted bits. Specifically, it is the number of the received error bits divided by the total number of all transferred bits during the considered time interval. Usually, the bit error rate is a percentage number. As a result, it is important to define the correct or wrong received bits of the output. Here, there is a little difference to the conventional calculation of the bit error rate. As is well known, the response amplitude does not equal the input signal level in general. Hence, we consider that the output is correct if the absolute value of the discrete response value is greater than the given threshold value $A_{th}$, and vice versa. Apparently, the threshold value $A_{th}$ is important. We still use the system in Eq.~(\ref{eq152}) as an example and give two values to plot the curves of the bit error rate versus the signal amplitude $B$ in Fig.~\ref{BERFIG}. Apparently, with the increase of the amplitude $B$, the minimal value of the bit error rate appears. It indicates the occurrence of aperiodic vibrational resonance. The bit error rate has the same effect as the cross-correlation coefficient. For the case $A_{th}=0$, it indicates that the received bit is correct if the output and the input signal have the same sign. For this case, the bit error rate achieves the minimum at $B=1.85$. It is different from the optimal value $B_c$ in Fig.~\ref{crossc}. We also know that $B=1.85$ can not make the aperiodic vibrational resonance reach the strongest resonance output from the time series in Fig.~\ref{BERFIG}. In addition, if we choose $A_{th}=A$, implies that the output amplitude is enhanced in comparison to the input level. Regarding to $A_{th}=A$, the critical value of $B$ corresponding to the minimal of the bit error rate is $B=1.55$. Compared with the result $B=1.5$ in Fig.~\ref{crossc}, there is only a minor error. This shows that using an error rate to measure resonance is also effective as long as the appropriate threshold is selected. The time series in Fig.~\ref{BERFIG} corresponding to $B=1.55$ illustrates the feasibility of using the bit error rate as the measure of aperiodic vibrational resonance again.

\subsection{Spectrum amplification factor}

\begin{figure}[t]
\begin{center}
\includegraphics[width=0.7\linewidth]{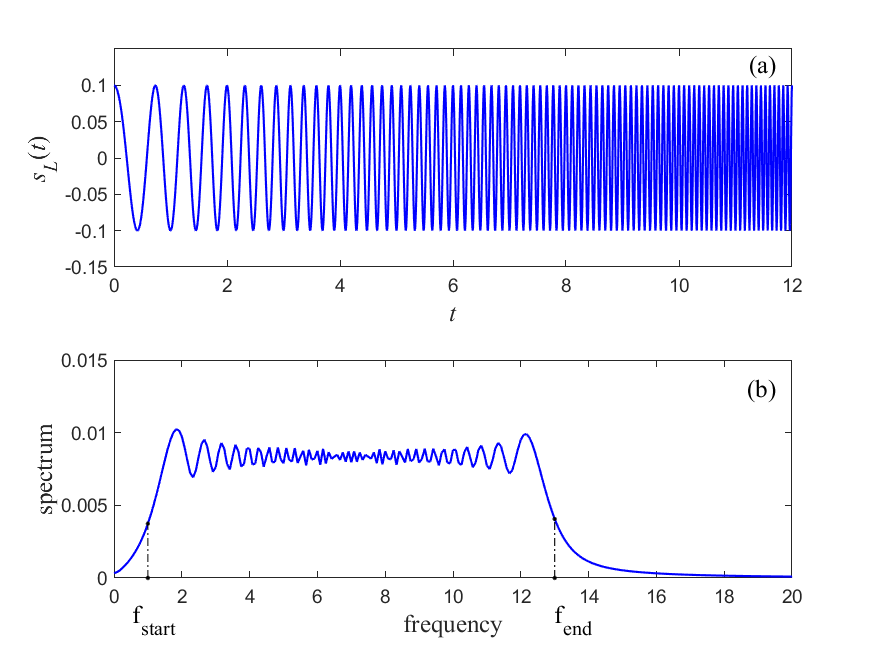}
\end{center}
\caption{The linear frequency modulated signal considered in the system of Eq.~(\ref{eq153}). (a) Time series of the input linear frequency modulated signal, (b) Spectrum of the input linear frequency modulated signal. The parameters are $A=0.1$, $\gamma=1$, $f=1$.}
\label{linearfms}
\end{figure}

Consider the system \cite{ref119}
\begin{equation}
\label{eq153}
      \dot x - \omega_0^2 x + \beta x^3
       = A \cos \left( \pi \gamma t^2 + 2 \pi ft \right)
         + B \cos \left( k \pi \gamma t^2 + 2k \pi ft  \right),
\end{equation}
which is used to investigate vibrational resonance induced by the linear frequency modulated signal. In this equation, $A\cos \left( \pi \gamma t^2 + 2 \pi ft  \right)$ is the characteristic signal $s_\mathrm{L}(t)$, $B \cos \left( k \pi \gamma t^2 + 2k \pi ft \right)$ is the auxiliary signal $s_\mathrm{H}(t)$ with $k$ is a positive number and $k \gg 1$. The instantaneous frequency of the auxiliary signal is much greater than that of the characteristic signal. We provide the plot of the characteristic signal, {\it i.e.}, the linear frequency modulated signal, in Fig.~\ref{linearfms}(a) and its corresponding spectrum in Fig.~\ref{linearfms}(b). Unlike the periodic signal, the spectrum of the linear frequency modulated signal is continuous. The spectrum of a periodic signal is discrete in some specific frequencies. In fact, when the frequency of the signal is varying with time, it is better to analyze the spectrum by using time-frequency analytical tools, such as the short-time Fourier transform \cite{ref272}, the fractional Fourier transform \cite{ref273}, and the empirical mode decomposition \cite{ref274}-\cite{ref276}, {\it etc}.

For studying vibrational resonance induced by the linear frequency modulated signal, Jia {\it et al.} \cite{ref119} proposed the spectrum amplification factor. At first, they used the cross-correlation coefficient to measure aperiodic vibrational resonance. Then, by a new spectrum amplification factor, they found that the new index is better for describing the vibrational resonance performance. The spectrum amplification factor $\eta$ is defined by
\begin{equation}
\label{eq154}
     \eta  =
               \frac{
            {\displaystyle{
              \frac{1}{ f_\mathrm{end}- f_\mathrm{start} } }}
             {\displaystyle{
             \int_{ f_\mathrm{start} }^{ f_\mathrm{end} }
             s_x \left(  f_\mathrm{in} \right) \mathrm{d}
             f_\mathrm{in}      }}   }
             {  {\displaystyle{  \frac{1}{ f_\mathrm{end}  - f_\mathrm{start} } }}
              {\displaystyle{
               \int_{ f_\mathrm{start} }^{ f_\mathrm{end} }
             s_u \left( f_\mathrm{in} \right)
             \mathrm{d} f_\mathrm{in} }  }}
             = \frac{ {\displaystyle{  \sum_{i = 1}^n s_x(i) } }}
                   { {\displaystyle{  \sum_{i = 1}^n s_u(i) }} },
\end{equation}
where $s_u(\bullet)$ is the spectrum of the input linear frequency modulated signal, and $s_x(\bullet)$ is the spectrum of the system output. The spectrum amplification factor is an appropriate measure to quantify vibrational resonance in the case of driving forces considered in Eq.~(\ref{eq153}).

\begin{figure}[t]
\begin{center}
\includegraphics[width=0.7\linewidth]{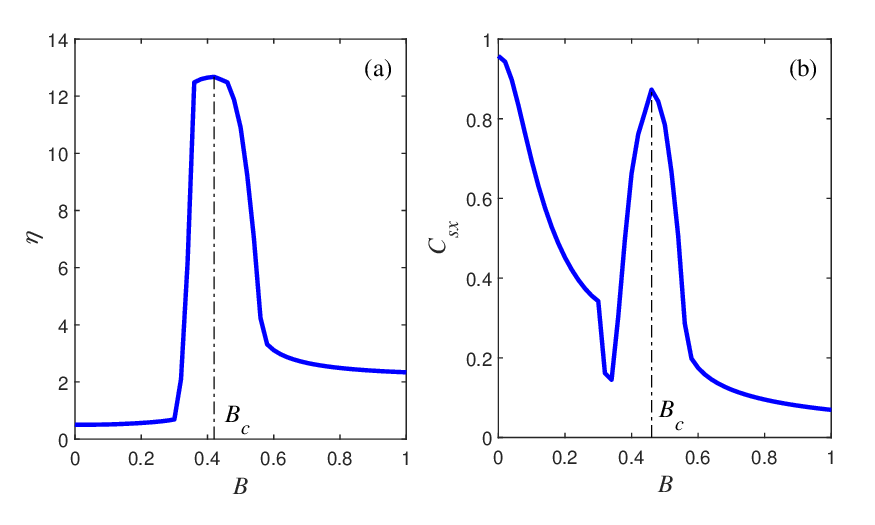}
\end{center}
\caption{Different characterizations of aperiodic vibrational resonance induced by the linear frequency modulated signal in the system of Eq.~(\ref{eq153}). (a) The spectrum amplification factor $\eta$ versus the amplitude $B$ of the fast-varying auxiliary signal presents aperiodic vibrational resonance. (b) The cross-correlation coefficient $C_{sx}$ versus the amplitude $B$ of the fast-varying auxiliary signal presents aperiodic vibrational resonance. The parameters are $\omega_0^2=300$, $\beta=300$,  $A=0.1$, $\gamma=1$, $f=1$, $k=10$. }
\label{spectrumfactor}
\end{figure}

For the system Eq.~(\ref{eq153}), the spectrum amplification factor and the cross-correlation coefficient are given in Fig.~\ref{spectrumfactor}(a) and Fig.~\ref{spectrumfactor}(b), respectively. Apparently, there is only one maximal value corresponding to the peak value of the curve in Fig.~\ref{spectrumfactor}(a). Although these two kinds of curves can prove the occurrence of vibrational resonance, in Fig.~\ref{spectrumfactor}(b), the peak value of the curve does not correspond to the maximal value of the considered index $C_{sx}$. Due to the time modulation of the frequency, this kind of vibrational resonance is essentially an aperiodic vibrational resonance. It may occur at the peak of the $C_{sx}-B$ curve. For the spectrum amplification factor, the peak value of the $\eta-B$ curve corresponds to the maximal response of the system, {\it i.e.}, the aperiodic vibrational resonance. Besides, the spectrum amplification parameter can measure the degree of the amplification of the input characteristic signal. From the value of the optimal amplitude of the fast-varying auxiliary signal, in Fig.~\ref{spectrumfactor}(a), at $B=0.42$, $\eta$ achieves the peak value $12.68$. In Fig.~\ref{spectrumfactor}(b), at $B=0.46$, the peak value is $C_{sx}=0.8729$. That is to say, the optimal amplitudes of $B$ obtained from the two measures are approximately equal. To This method will be detailed in Section~\ref{rescvr}.

\begin{figure}[!h]
\begin{center}
\includegraphics[width=0.8\linewidth]{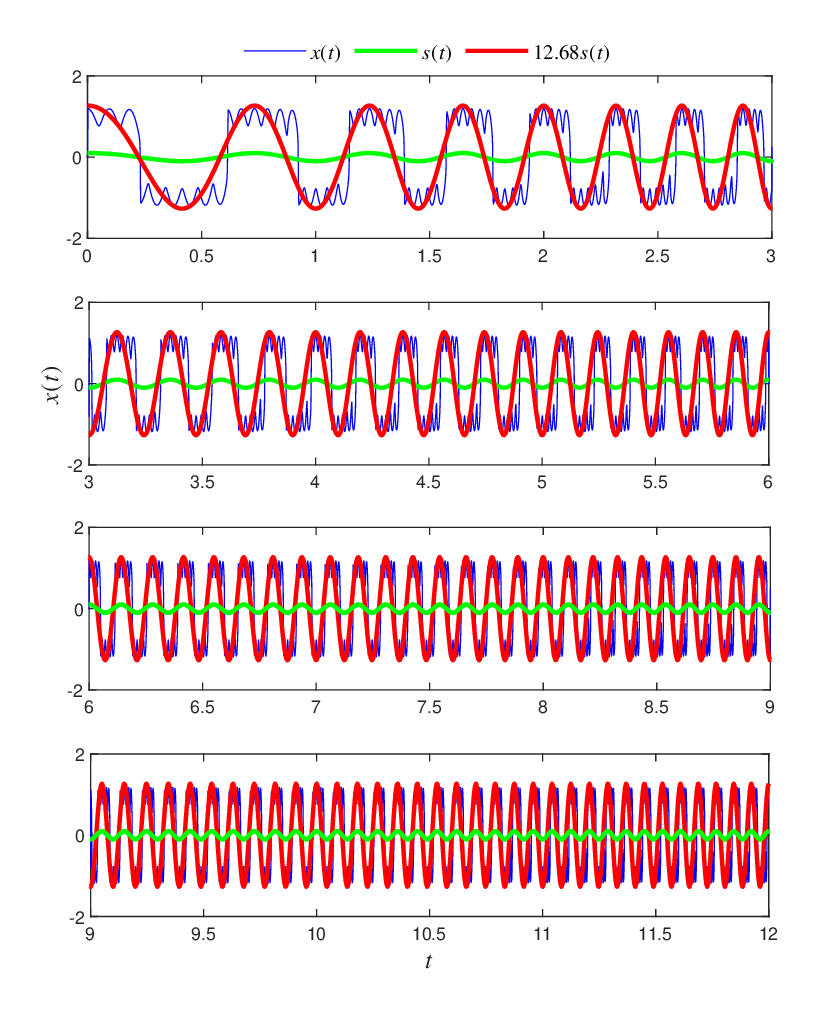}
\end{center}
\caption{Time series of the response when the optimal aperiodic vibrational resonance occurs in system of Eq.~(\ref{eq153}). The parameters are $A=0.1$, $\gamma=1$, $f=1$, $B=0.42$, $k=10$.}
\label{tsaperiodic}
\end{figure}

The response of the system corresponding to strong aperiodic vibrational resonance is illustrated in Fig.~\ref{tsaperiodic}. To enhance clarity, we display the time series within the interval $[0, 12]$ across four subplots. Upon comparison of the time series of $x(t)$, $s(t)$, and $12.68s(t)$, it becomes evident that the weak characteristic signal experiences significant amplification. Furthermore, the curve $12.68s(t)$ exhibits a magnitude nearly identical to that of the curve $x(t)$. This observation aligns with the peak value of the $\eta-B$ curve depicted in Fig.~\ref{spectrumfactor}(a).

\subsection{Success probability}

We define $P$ as the success probability, which serves as the index to measure the logical vibrational resonance
\begin{equation}
\label{eq155 }
    P = \frac{ \text{The correct turns} }{ \text{The total turns} } .
\end{equation}
The specific process for obtaining the success probability $P$ is given as follows.

\renewcommand{\labelenumi}{\theenumi.}
\renewcommand{\theenumi}{\arabic{enumi}}
\begin{enumerate}
\item
Randomly arrange an array of four possible logical states (0,0), (0,1), (1,1), (1,0) and place it into the nonlinear system.
\item
Drive the system by each logical state with the same time. Choose the total turns in a large number in the calculation. The turns mean all trajectories for the calculation.
\item
Analyze the output based on the logical calculation truth table to obtain the success probability.
\end{enumerate}
For the logical $\bf OR$, $\bf AND$, $\bf NOR$, $\bf NAND$, $\bf XOR$, $\bf NXOR$, the logical calculation truth values are given in Table~\ref{t1}.

\begin{table}
\begin{center}
\caption{The logical calculation truth table.}
 \vskip 3pt
 \label{t1}
\begin{tabular}{ c c c c c c c c c}
 \hline    \noalign{\smallskip}
 Input set($I_1$,$I_2$) &  Value of ($I_1+I_2$) & \bf OR & \bf AND & \bf NOR & \bf NAND & \bf XOR & \bf NXOR\\   \noalign{\smallskip}
 \hline    \noalign{\smallskip}
 (0,0) &  $-1$ & 0 & 0 & 1 & 1 & 0 & 1\\  \noalign{\smallskip}
 (0,1)/(1,0) &  \;\;0 & 1 & 0 & 0 & 1 & 1 & 0\\  \noalign{\smallskip}
 (1,1) & \;\;1 & 1 & 1 & 0 & 0 & 0 & 1\\  \noalign{\smallskip}
 \hline
\end{tabular}
\end{center}
\end{table}

\begin{figure}[t]
\begin{center}
\includegraphics[width=0.7\linewidth]{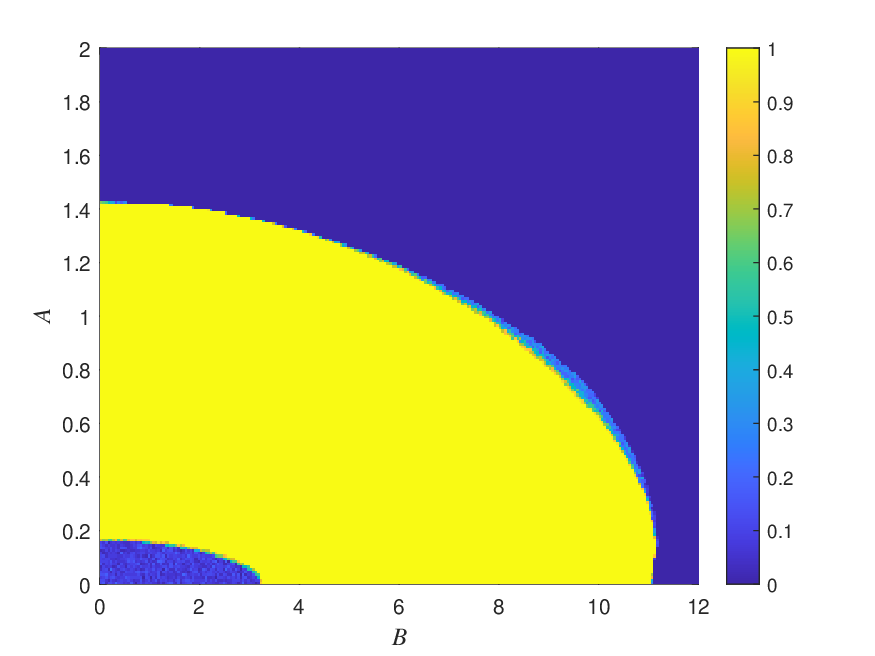}
\end{center}
\caption{There is an obvious logical vibrational resonance region corresponding to the $\bf OR$ logic operation in the two dimensional plane of the success probability with both $A$ and $B$. The parameters are $\omega_0^2 =-2$, $\beta = 4$, $r = 0.5$, $\omega = 1$, $\Omega=20$.}
\label{logres}
\end{figure}

In the absence of noise, the logical computation in the system described by Eq.~(\ref{eq140}) involves two harmonic signals: $A \cos(\omega t)$ and $B \cos(\Omega t)$. Typically, the system output realizes $\bf OR$, $\bf NOR$, $\bf AND$, and $\bf NAND$ logical calculations for $r>0$ and $r<0$, respectively. However, the bistable system cannot achieve $\bf XOR$ and $\bf NXOR$ calculations. To accomplish these operations using the response of a nonlinear system, it is suitable to employ a triple-well system \cite{ref277}. Notably, the logical response exhibits a resonance region when varying the two harmonic signals, as depicted in Fig.~\ref{logres}. This phenomenon is {\it logical vibrational resonance}. In the simulation results shown in Fig.~\ref{logres}, $I_1, I_2= - 0.5$ corresponds to logic $0$, and $I_1, I_2=0.5$ corresponds to logic $1$. In this figure, for a weak slow-varying signal and a strong fast-varying signal, i.e., $B \gg A$, the logical vibrational resonance occurs. It is noteworthy that the same phenomenon occurs for a weak high-frequency signal and a strong low-frequency signal, i.e., $A\gg B$. This behavior is distinct from conventional vibrational resonance, where the amplitude of the slow-varying signal is typically very weak. In summary, appropriately chosen harmonic signals have a positive impact on the logical response of a nonlinear system.

\begin{figure}
\begin{center}
\includegraphics[width=0.7\linewidth]{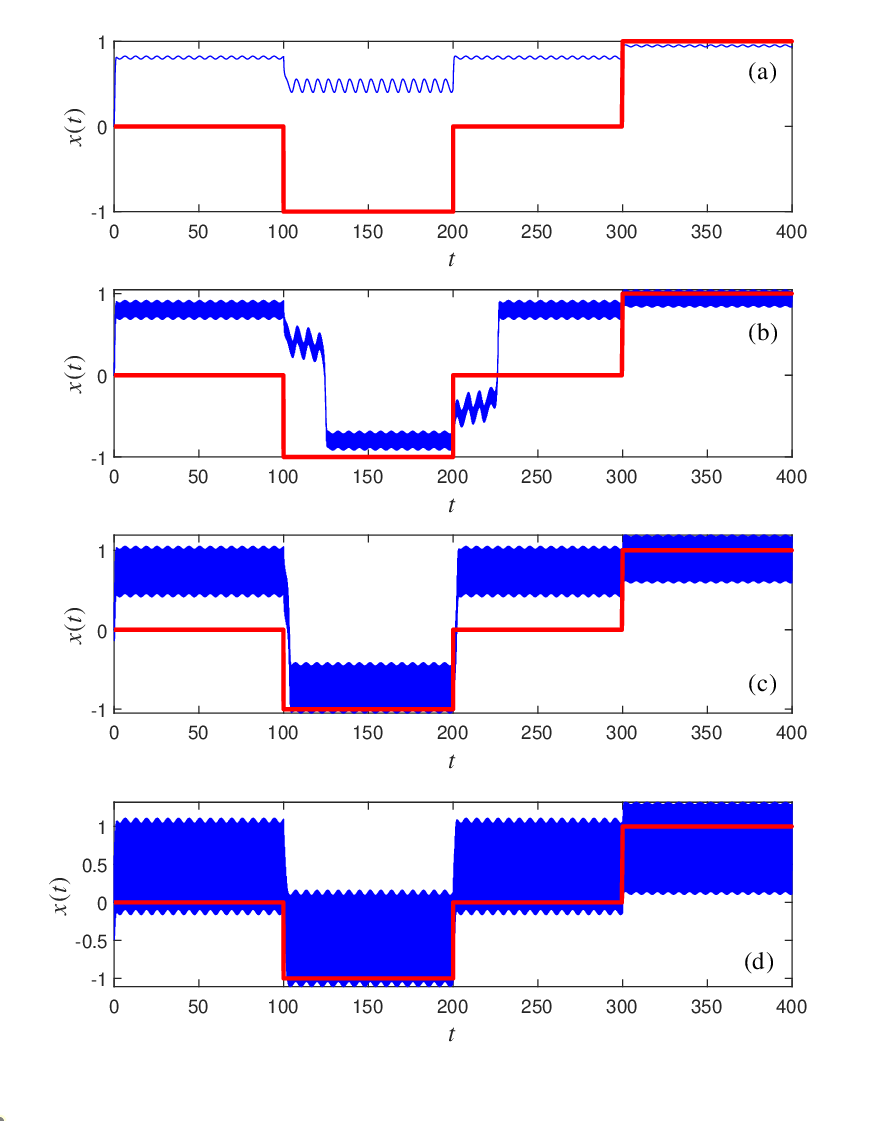}
\end{center}
\caption{Time series of the logic input $I_1+I_2$ (thick line in red color) and logical output (thin line in blue color) of the system of Eq.~(\ref{eq140}) for different values of the amplitude $B$ of the fast-varying auxiliary signal. The parameters are $\omega_0^2 =2$, $\beta = 4$, $r = 0.5$, $A=0.1$, $\omega = 1$, $\Omega=20$, (a) $ B = 0$, (b) $ B = 2 $, (c) $ B = 6$, (d) $ B = 12$.}
\label{logser}
\end{figure}

To illustrate the logical vibrational resonance further, we show a time series plot in Fig.~\ref{logser}. In Figs.~\ref{logser}(a), (b), (d), the time series present a poor logical output. In Figs.~\ref{logser}(c), corresponding to the strong resonance (bright color region) in Fig.~\ref{logres}, the time series presents an excellent logical output.

\section{Theoretical formulation of vibrational resonance}
\label{theory}

\subsection{The linear response theory}
\label{linear}

The method of direct separation of motions either alone or in combination with other methods, serves as a valuable technique for analyzing the effects of a high-frequency excitation within a theoretical framework \cite{ref278}-\cite{ref291}. In this subsection, the method of direct separation of motions serves as the primary theoretical basis. Additionally, unless otherwise specified, we will consistently employ $A \cos(\omega t)$ and $B \cos(\Omega t)$ as the two excitations acting on the system.

\subsubsection{Vibrational resonance in the bistable system}

Taking the overdamped Duffing oscillator system Eq.~(\ref{eq16}) as an example, we introduce the method of direct separation of motions in detail. Letting
\begin{equation}
\label{eq156}
    x(t) = X(t) + \Psi(t),
\end{equation}
where, $X(t)$ is a slow variable with period $2\pi/\omega$, and $\Psi (\tau = \Omega t)$ is a fast variable with period $2\pi$. Assuming the average of $\Psi$ over the period is $\Psi_\mathrm{av} = (1/2 \pi) \int_0^{2\pi} \Psi \mathrm{d} \tau =0$, the equations for $X(t)$ and $\Psi(t)$ are
{\small
\begin{subequations}
 \label{eq157}
\begin{eqnarray}
\dot X \! \! + \! \! \left( \omega_0^2 + 3 \beta \Psi_\mathrm{av}^2 \right) X
       + \beta \left( X^3 + \Psi_\mathrm{av}^3 \right) + 3 \beta X^2 \Psi_\mathrm{av} \! \! & = & \! \! A \cos (\omega t), \\
\dot\Psi \! \! + \! \! \omega_0^2 \Psi \! \! + \! \! 3 \beta X^2 \left( \Psi \! \! - \! \! \Psi_\mathrm{av} \right)
     \! \! + \! \! 3 \beta X \left( \Psi^2 \! \! - \! \! \Psi_\mathrm{av}^2 \right) \! \! + \! \! \beta \left( \Psi^3 \! \! - \! \! \Psi_\mathrm{av}^3 \right) \! \! & = & \! \! B \cos (\Omega t),
\end{eqnarray}
\end{subequations}
}
where $\Psi_\mathrm{av}^n = (1/2\pi) \int_0^{2\pi} \Psi^n \mathrm{d} \tau$.

As $\Psi$  is fast varying, we neglect all nonlinear terms in Eq.~(\ref{eq157}b).  Then, the solution of the linear equation in the limit of $t \to \infty$ is given by
\begin{eqnarray}
\label{eq158}
     \Psi = C \cos (\tau+\phi), \quad
     C = \frac{B}{ \sqrt{\Omega^2+(\omega_0^2)^2} }, \,
       \quad \phi = \tan^{-1} \left( - \frac{\Omega}{\omega_0^2} \right) .
\end{eqnarray}

For $\Omega \gg \omega_0^2$, we can neglect the $(\omega_0^2)^2$ in the above equation and approximate the amplitude of $\Psi$ as $g = B/\Omega$. Since $\Omega \gg \omega$, in the interval $[t, t+2\pi/\Omega]$, $X(t)$ can be viewed as a constant. Using the $\Psi $ given in Eq.~(\ref{eq158}), we find $\Psi_\mathrm{av} = 0$, $\Psi_\mathrm{av}^2 = C^2/2$ and $\Psi_\mathrm{av}^3 =0$.  Then, the equation for the slow variable (\ref{eq157}a) becomes
\begin{equation}
\label{eq159}
  \dot X + \mu X + \beta X^3 = A \cos (\omega t),
\end{equation}
where $\mu = \frac{3}{2} \beta C^2 + \omega_0^2$.

For $\omega_0^2 <0$ and $\beta >0$ (bistable case), the equilibrium points of Eq.~(\ref{eq159}) when $A =0$ are $X^* = 0$ and $X^*_\pm =\pm \sqrt{-\mu/\beta}\,$.  In this scenario, a slow oscillation occurs around the stable equilibrium points. The parameters $B$ and $\Omega$ are typically considered as control parameters for vibrational resonance. By adjusting these parameters, the equilibrium points and their stability can be naturally altered.

To determine the response amplitude at the frequency $\omega$, we define $Y=X-X^*$ as the deviation of $X$ from $X^*$, where $X^*$ represents the stable equilibrium of the equivalent system (\ref{eq159}). Then, we can derive the following expression
\begin{equation}
 \label{eq160}
 \dot Y + \omega_\mathrm{r} Y + 3 \beta X^* Y^2 + \beta Y^3 = A \cos \omega t ,
\end{equation}
where $\omega_\mathrm{r} = \mu + 3\beta X^{*2}$.
For a weak characteristic signal $A \cos \omega t$ with $A \ll 1$, $\vert Y \vert \ll 1$, making it reasonable to neglect all nonlinear terms in the equation. Then, the steady solution of the corresponding linear equation becomes $Y = A_\mathrm{L} \cos (\omega t + \phi)$, where
\begin{subequations}
 \label{eq161}
\begin{eqnarray}
  A_\mathrm{L} & = & \frac{A}{\sqrt{\omega^2 + \omega_\mathrm{r}^2} }, \\
  \phi & = & \tan^{ - 1} \left( - \frac{\omega }{\omega_\mathrm{r} }\right).
\end{eqnarray}
\end{subequations}

Apparently, the amplitude $A_\mathrm{L}$ is a nonlinear function of $B$ and $\Omega$. Then, the response amplitude $Q$ is obtained as%
\begin{equation}
\label{eq162}
     Q = \frac{1}{ \sqrt{ \omega ^2 + \omega_\mathrm{r}^2} } \,.
\end{equation}

\begin{figure}[!h]
\begin{center}
\includegraphics[width=0.8\linewidth]{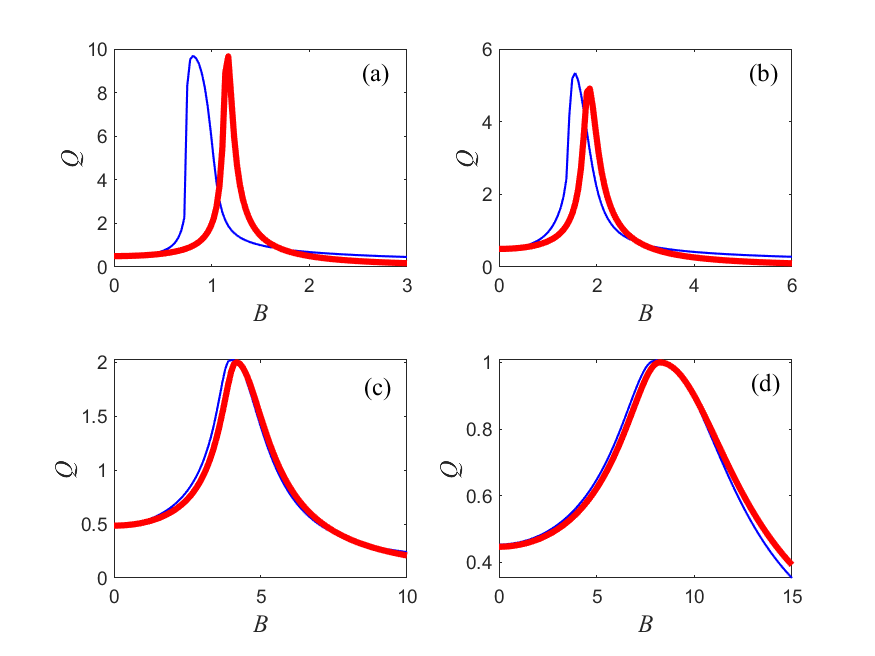}
\end{center}
\caption{Comparison between the analytical results (thick lines) and the numerical simulations (thin lines) of the response amplitude at the low-frequency $\omega$ in the system of Eq.~(\ref{eq16}). (a) $\omega=0.1$, (b) $\omega=0.2$, (c) $\omega=0.5$, (d) $\omega=1$. The parameters are $\omega_0^2=-1$, $\beta=1$, $A=0.1$, $\Omega = 10 \omega$.}
\label{numvsana}
\end{figure}

As mentioned in Section~\ref{int}, the method of direct separation of motions is an approximate method that may introduce errors in certain cases. Blekhman and Landa addressed this issue in their work \cite{ref6}, focusing on vibrational resonance in the conventional bistable system in both overdamped, Eq.~(\ref{eq16}, and underdamped ,Eq.~(\ref{eq17}, versions, respectively. In the overdamped bistable system, they compared analytical results with numerical results, as shown in Fig.~\ref{numvsana}. The curves illustrate that numerical results deviate more significantly from analytical results, especially when $\omega$ is very small. However, as $\omega$ increases, the analytical and numerical results gradually converge. Despite discrepancies in certain cases, the resonance phenomenon is clearly evident in each curve. Moreover, the errors typically remain within an acceptable range for $A \ll 1$. This analytical method, while simple, proves valuable in identifying and understanding the occurrence of vibrational resonance. Consequently, the method of direct separation of motions finds widespread use in vibrational resonance studies. Additionally, another similar linear response theory method is discussed in \cite{ref292}.

The resonance behavior can be obtained analytically by the expression of $Q$. We usually focus on the maximal value $Q_{max}$ and the corresponding value $B_{VR}$. From Eq.~(\ref{eq162}), we note that $Q$ will be maximum when $\omega_\mathrm{r}^2 $ achieves a minimum.  According to the formula of $\omega_\mathrm{r}$ in Eq.~(\ref{eq160}), it is easy to get that $\omega_\mathrm{r}^2 = 0$ when $B = B_{VR} = B_c$, and
\begin{equation}
\label{eq163}
    {B_{VR}} = \sqrt {\frac{{ - 2\omega _0^2\left[ {{\Omega ^2} + (\omega _0^2)} \right]}}{{3\beta }}}.
\end{equation}

In addition, $B_\mathrm{c}$ denotes the pitchfork bifurcation point of Eq.~(\ref{eq159}). Specifically, when $B < B_\mathrm{c}$, the equivalent system has one unstable equilibrium point $X^*=0$ and two stable equilibrium points $X^*_\pm =\pm \sqrt{-\mu/\beta}$. Conversely, when $B \geq {B_c}$, only one stable equilibrium point $X^*=0$ exists in the equivalent system (\ref{eq159}). Furthermore, $\omega_r$ behaves as a decreasing monotone function when $B < B_\mathrm{c}$ and as a monotone increasing function when $B \geq {B_\mathrm{c}}$. Consequently, the response amplitude $Q$ monotonically increases when $B < B_\mathrm{c}$ and monotonically decreases when $B \geq {B_\mathrm{c}}$. Likewise, there is only one peak in the $Q-B$ curve at the pitchfork bifurcation $B=B_\mathrm{VR}=B_\mathrm{c}$, with the corresponding peak value being $Q = \frac{1}{\omega }$. These peak values are corroborated by Fig.~\ref{numvsana} for various values of $\omega$. It is noteworthy that double resonance peaks may emerge if the overdamped bistable system is in a fractional-order version, as in Eq.~(\ref{eq103}). Specifically, double resonance peaks may appear when the fractional order satisfies $\alpha > 1$ \cite{ref48}. However, these double resonance peaks do not appear in the $Q-B$ curve of the ordinary differential system Eq.~(\ref{eq16}).

With regards to the underdamped Duffing oscillator system Eq.~(\ref{eq17}) excited by two cosine signals as another example, the analytical result of $Q$ is
\begin{equation}
\label{eq164}
     Q = \frac{1}{ \sqrt{ (\delta \omega) ^2 + (\omega_\mathrm{r} - \omega ^2) ^2} } ,
\end{equation}
where $\omega_\mathrm{r}$ and $X^*$ have the same expressions with that in the above overdamped system, but the amplitude of the fast variable is $C = \frac{B}{ \sqrt{ (\delta \Omega) ^2 + (\omega_\mathrm{0} ^2 - \Omega ^2) ^2} }$. The pitchfork bifurcation of the equivalent system is
\begin{equation}
\label{eq165}
   {B_c} = \sqrt {\frac{{ - 2\omega _0^2\left[ {{{(\delta \Omega )}^2} + {{(\omega _0^2 - {\Omega ^2})}^2}} \right]}}{{3\beta }}}.
\end{equation}
When $B<B_c$, the equivalent system has two stable equilibrium points $X^*_\pm =\pm \sqrt{-\mu/\beta}$. When $B \geq B_c$, the equivalent system has one stable equilibrium point $X^*=0$. According to the formula of $\omega_r$ in Eq.~(\ref{eq160}), we know that the equivalent system has a different mathematical expression when $B$ is greater or less than $B_c$.

\begin{figure}[t]
\begin{center}
\includegraphics[width=0.7\linewidth]{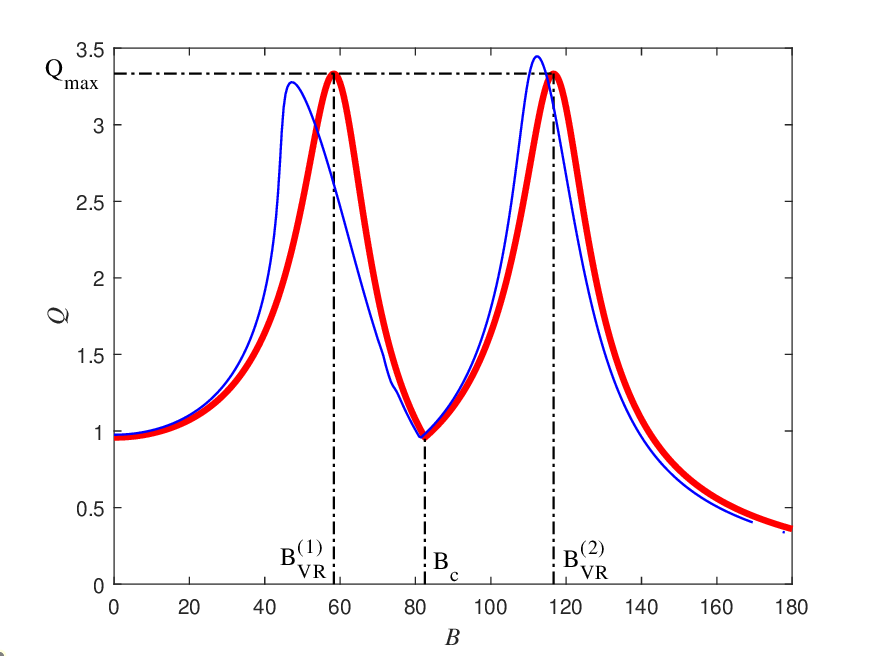}
\end{center}
\caption{The response amplitude $Q$ versus the amplitude $B$ of the fast-varying auxiliary signal presents double vibrational resonance in the underdamped bistable system of Eq.~(\ref{eq17}). The thick solid line is plotted by using analytical results and the thin solid line is plotted by using numerical results. The parameters are $\omega_0^2 = -1$, $\beta=1$, $A=0.1$, $\omega=1$, $\Omega=10$.}
\label{modal}
\end{figure}

Solving $\omega_r=\omega^2$, when $0< B< B_c$, the response amplitude achieves the peak at the location
\begin{equation}
\label{eq166}
   B_{VR}^{(1)} = \sqrt { - \frac{{2\omega _0^2{\text{ + }}{\omega ^2}}}{{3\beta }}\left[ {{{(\delta \Omega )}^2} + {{(\omega _0^2 - {\Omega ^2})}^2}} \right]}.
\end{equation}
The corresponding peak value is $Q_{max}^{(1)}= \frac{1}{{\delta \omega }}$. When $ B > B_c$, the resonance peak occurs at
\begin{equation}
\label{eq167}
   B_{VR}^{(2)} = \sqrt {\frac{{2({\omega ^2} - \omega _0^2)}}{{3\beta }}\left[ {{{(\delta \Omega )}^2} + {{(\omega _0^2 - {\Omega ^2})}^2}} \right]} .
\end{equation}
The corresponding peak value is $Q_{max}^{(2)}=Q_{max}^{(1)}= \frac{1}{{\delta \omega }}$. Hence, at $B_{VR}^{(1)}$ and $B_{VR}^{(2)}$, the resonance corresponding to two different equivalent systems occurs. Further, at the bifurcation point $B=B_c$, there is a valley value ${Q_{val}} = \frac{1} {{\sqrt {{{(\delta \omega )}^2} + {\omega ^4}} }}$. The specific resonance modal is presented in Fig.~\ref{modal}.

\subsubsection{Vibrational resonance in the pendulum system}

\begin{figure}[t]
\begin{center}
\includegraphics[width=0.8\linewidth]{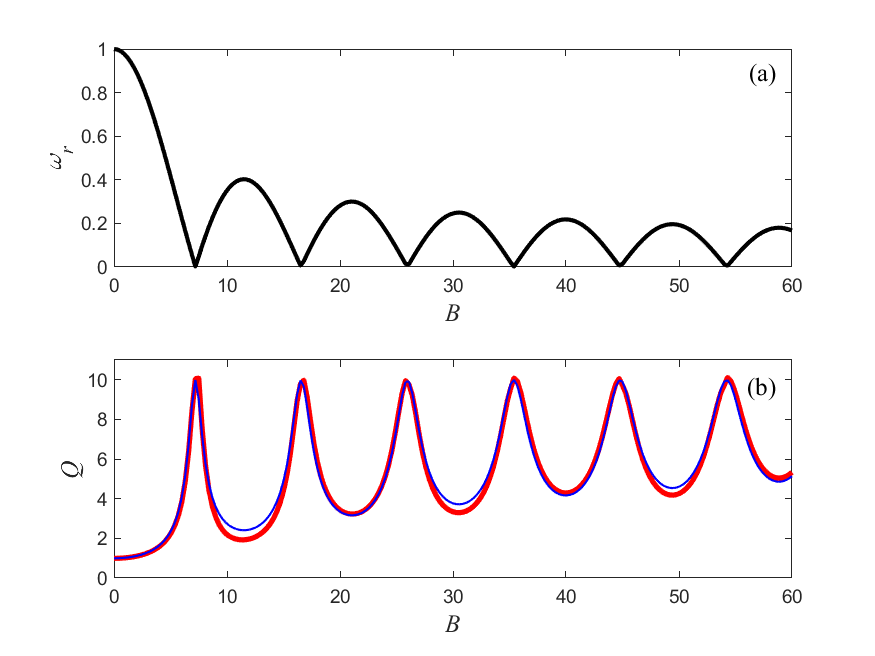}
\end{center}
\caption{The control parameter $\omega_r$ whose minimum value makes the resonance occur and the response amplitude $Q$ changing with the amplitude $B$ of the fast-varying auxiliary signal in the pendulum system of Eq.~(\ref{eq34}). (a) The minimum value of $\omega_r$ repeatedly appears as the signal amplitude $B$ increases. (b) The response amplitude $Q$ versus the signal amplitude $B$ presents multiple vibrational resonance. The thick solid line is plotted by using analytical results and the thin solid line is plotted by using numerical results. The parameters are $A=0.1$, $\omega=0.1$, $\Omega=3$.}
\label{pendulum}
\end{figure}

For the pendulum system with a periodical potential, taking  Eq.~(\ref{eq34}) as an example, by the method of direct separation of motions, eliminating the fast variable, the equivalent system is \cite{ref20}, \cite{ref293}
\begin{equation}
\label{eq168}
     \dot X + \frac{{d\bar V(X)}}{{dX}} = A \cos (\omega t).
\end{equation}
The effective potential function $\bar V(X) =  - {J_0}({F \mathord{\left/ {\vphantom {F \Omega }} \right. \kern-\nulldelimiterspace} \Omega })\cos X$, where ${\omega _r} = \left| {{J_0}({B \mathord{\left/ {\vphantom {F \Omega }} \right. \kern-\nulldelimiterspace} \Omega })} \right|$ and $J_0(\bullet)$ is the zero-order Bessel function of the first kind.
The stable equilibrium points of  Eq.~(\ref{eq168}) for $A=0$ are $ X^*=\pm 2n \pi , n= 1, 2, ...$. The response amplitude at the excitation frequency $\omega$ is
\begin{equation}
\label{eq169 }
     Q = \frac{1}{{\sqrt {\omega _r^2 + {\omega ^2}} }}.
\end{equation}
As depicted in Fig.~\ref{pendulum}, peaks emerge whenever $\omega_r$ reaches local minima. Owing to the characteristics of the zero-order Bessel function of the first kind, a series of resonance peaks is observed in the graph. This differs from the resonance mode observed in the bistable system.

\subsubsection{Vibrational resonance in coupled oscillators}

\begin{figure}[!h]
\begin{center}
\includegraphics[width=0.75\linewidth]{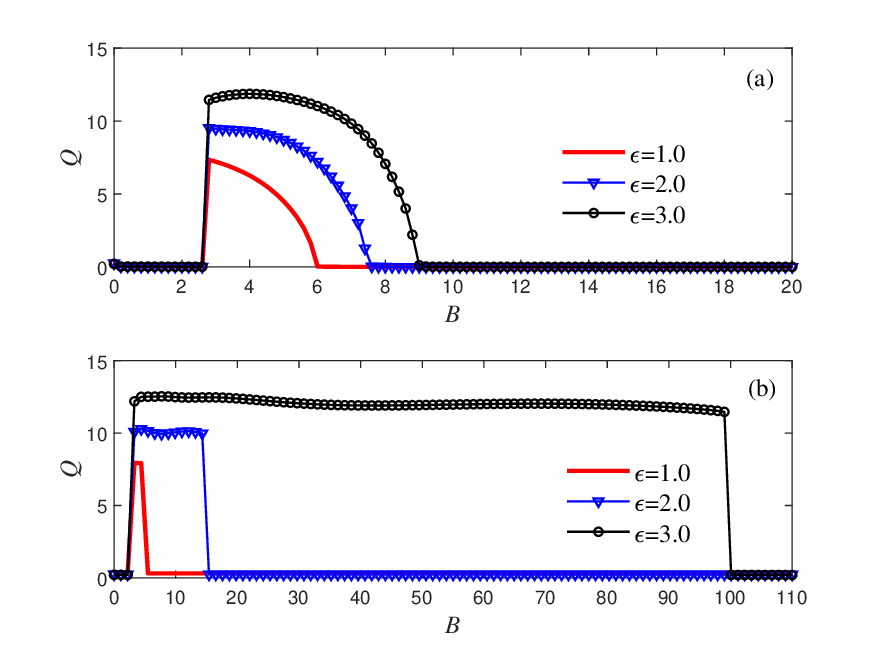}
\end{center}
\caption{Vibrational resonance of the last oscillator in the unidirectional coupled system of Eq.~(\ref{eq68}). (a) The high-frequency excitations are added to all oscillators. (b) The high-frequency excitation is added to the first oscillator only. The parameters are $a_1=1$, $b_1=1$, $A=0.2$, $\omega=0.1$, $\Omega=5.0$. The total number of the oscillators is $n=50$.}
\label{coupled}
\end{figure}

It is noteworthy that the coupling in Eq.~(\ref{eq68}) is unidirectional. Interestingly, when the coupling strength is significantly high, vibrational resonance occurs whether high-frequency excitations are applied to all oscillators or only to the first oscillator while the others remain free from excitation, as discussed in \cite{ref17}. In both scenarios, the response of the last oscillator at the frequency of the slowly varying excitation is notably robust, as depicted in Fig.\ref{coupled}. Apparently, there is a very wide resonance range in the subplot especially in Fig.~\ref{coupled}b. Especially, for the case $\epsilon=3$, the response amplitude $Q$ versus the amplitude $B$ almost presents a constant value in a large interval of $B$. In other words, a strong enough high-frequency signal makes the system achieve a saturated output of resonance in the corresponding interval. Here, we use the numerical simulation to obtain the results. According to the time series in \cite{ref17}, we find that the output of the $nth$ oscillator is approximated to a square waveform when $n$ is large enough. When using the method of direct separation of motions under these conditions, the deviation of slow motion from a harmonic waveform can result in a substantial error. Furthermore, when two excitations are exclusively applied to the first oscillator while leaving all others without excitation, the fast motion in the response of the $n$-th oscillator is nearly negligible, and the slow motion predominates. Therefore, in this case, we present only the numerical results.

\subsubsection{Vibrational resonance in the delayed system}

For the delayed system of Eq.~(\ref{eq115}) with a bistable potential, $\omega_0^2>0$, $\beta>0$ and $\gamma  <  - \omega _0^2$, by the method of direct separation of motions, it is easy to obtain the response amplitude at the frequency $\omega$ is
\begin{equation}
\label{eq170}
     Q = \frac{1} {{\sqrt {{{(\gamma \sin \omega \tau  - \omega )}^2} + {{(\gamma \cos \omega \tau  + {\omega _r})}^2}} }},
\end{equation}
where $\mu$ and $\omega_r$ have the same expressions with that in Eq.~(\ref{eq159}) and Eq.~(\ref{eq160}), but the equilibrium points are $X^* = 0$ and $X^*_\pm =\pm \sqrt{-(\mu +\gamma) /\beta}\,$. The amplitude of the fast motion is $C = \frac{B}{{\sqrt {{{(\gamma \sin \Omega \tau  - \Omega )}^2} + {{(\gamma \cos \Omega \tau  + \omega _0^2)}^2}} }}$.

If we choose the parameter $B$ to control the vibrational resonance, the resonance pattern depends on the feedback strength $\gamma$.

\begin{figure}[!h]
\begin{center}
\includegraphics[width=0.5\linewidth]{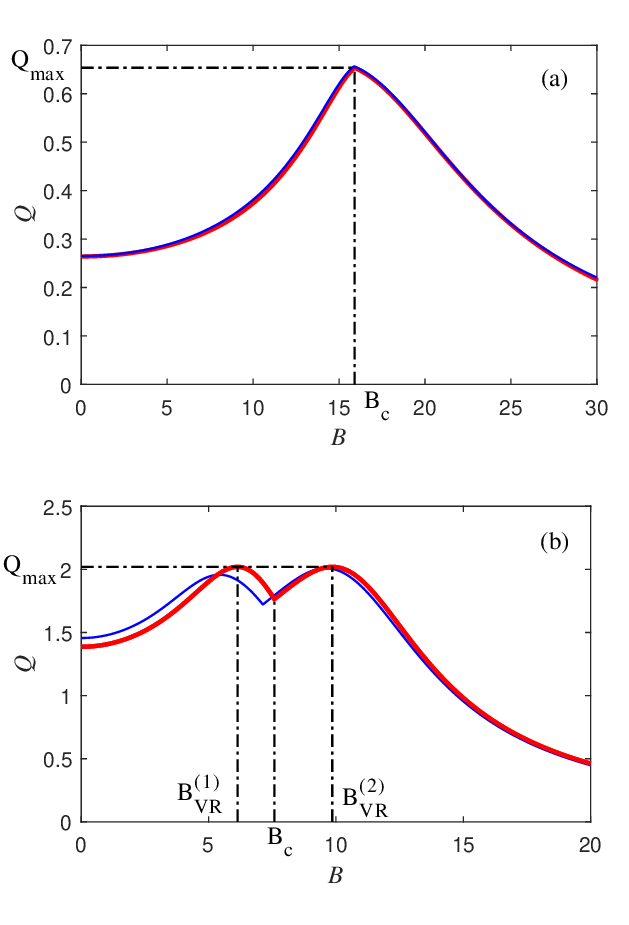}
\end{center}
\caption{Effect of the delayed strength $\gamma$ on vibrational resonance modal in the system of Eq.~(\ref{eq115}). (a) The single vibrational resonance appears when $\gamma=-0.6$. (b) The double vibrational resonance appears when $\gamma=0.6$. The thick solid line is plotted by using analytical results and the thin solid line is plotted by using numerical results. The parameters are $\omega_0^2=-1$, $\beta=1$, $\tau=1$, $A=0.1$, $\omega=1$, $\Omega=15$. }
\label{delay}
\end{figure}

\noindent (1) $\gamma  \leq 0$ \\
\indent The pitchfork bifurcation point is
\begin{equation}
\label{eq171}
   {B_c} = \sqrt { - \frac{{2 (\gamma  + \omega _0^2)}}{{3\beta }}\left[ {{{(\gamma \sin \Omega \tau  - \Omega )}^2} + {{(\gamma \cos \Omega \tau  + \omega _0^2)}^2}} \right]} .
\end{equation}
At $B=B_c$, the response amplitude is
\begin{equation}
\label{eq172}
     Q_c = \frac{1} {{\sqrt {{{(\gamma \sin \omega \tau  - \omega )}^2} + {{(\gamma \cos \omega \tau  - \gamma)}^2}} }},
\end{equation}
There is a single peak in the $Q-B$ curve, and the maximal value of the peak is $Q_{max}=Q_c$, as shown in Fig.~\ref{delay}(a).

\noindent (2) $0<\gamma<-\omega_0^2$ \\
\indent The pitchfork bifurcation point is still given in Eq.~(\ref{eq171}). Solving $\gamma \cos \omega \tau  + {\omega _r}=0$, when $0<B<B_c$, the resonance peak locates at
\begin{equation}
\label{eq173}
   B_{VR}^{(1)} = \sqrt {\frac{{(\gamma \cos \omega \tau  - 3\gamma  - 2\omega _0^2){\text{ }}}}
{{3\beta }}\left[ {{{(\gamma \sin \Omega \tau  - \Omega )}^2} + {{(\gamma \cos \Omega  + \omega _0^2)}^2}} \right]}.
\end{equation}
\indent When $ B > B_c$, the resonance peak occurs at
\begin{equation}
\label{eq174}
   B_{VR}^{(2)} = \sqrt { - \frac{{2(\gamma \cos \omega \tau  + \omega _0^2){\text{ }}}}
{{3\beta }}\left[ {{{(\gamma \sin \Omega \tau  - \Omega )}^2} + {{(\gamma \cos \Omega  + \omega _0^2)}^2}} \right]} .
\end{equation}
The corresponding peak value is ${Q_{\max }} = Q_{\max }^{(1)} = Q_{\max }^{(2)} = \frac{1}{{\left| {\gamma \sin \omega \tau  - \omega } \right|}}$. The double-resonance pattern is presented in Fig.~\ref{delay}(b). In addition, for the case $\tau=0$ or $\gamma=0$, i.e., when the time delay term is free, the system becomes the ordinary system, then we have $B_{VR}^{(1)}=B_{VR}^{(2)}$, and the double-peak disappear. Hence, the time delay is an important factor to induce the double-resonance pattern.

\begin{figure}[!h]
\begin{center}
\includegraphics[width=0.75\linewidth]{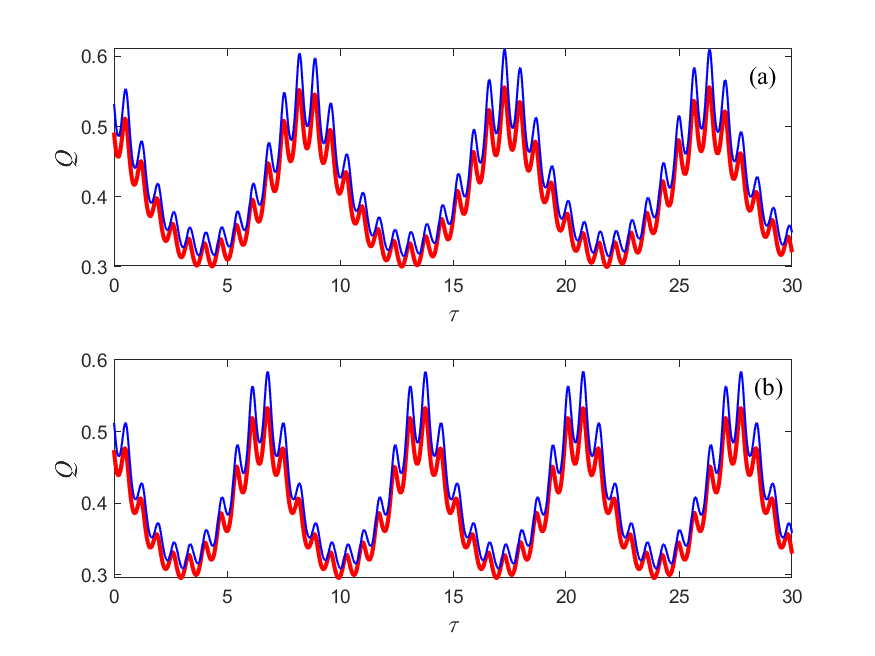}
\end{center}
\caption{Quasi-periodic or periodic vibrational resonance induced by the delay parameter $\tau$ in the system of Eq.~(\ref{eq115}). (a) The curve of $Q-\tau$ presents quasi-periodic vibrational resonance, $\omega=\sqrt(2)/2$. (b) The curve of $Q-\tau$ presents periodic vibrational resonance, $\omega=0.9$. The thick solid line is plotted by using analytical results and the thin solid line is plotted by using numerical results. The parameters are $\omega_0^2=-1$, $\beta=1$, $\gamma=-0.6$, $A=0.1$, $B=6$, $\Omega=9$. }
\label{delay-tau}
\end{figure}

When utilizing the delay parameter $\tau$ as the control parameter while keeping the remaining parameters fixed, resonance occurs at $\tau_{VR}$, which is the root of the equation $\frac{{dQ}}{{d\tau }} = 0$ and $\frac{{d^2Q}}{{d\tau^2 }} < 0$. Obtaining an accurate analytical solution for this equation can be challenging, as it may have numerous (or infinite) roots. If the ratio $\omega/\Omega$ is an irrational number, the delay $\tau$ will induce quasi-periodic vibrational resonance. Conversely, if the ratio $\omega/\Omega$ is a rational number, the delay $\tau$ will induce periodic vibrational resonance. Furthermore, the curve exhibits two periods, namely $2\pi/\omega$ and $2\pi/\Omega$, as depicted in Fig.~\ref{delay-tau}.

\subsection{The nonlinear vibrational resonance}

\begin{figure}[!h]
\begin{center}
\includegraphics[width=0.7\linewidth]{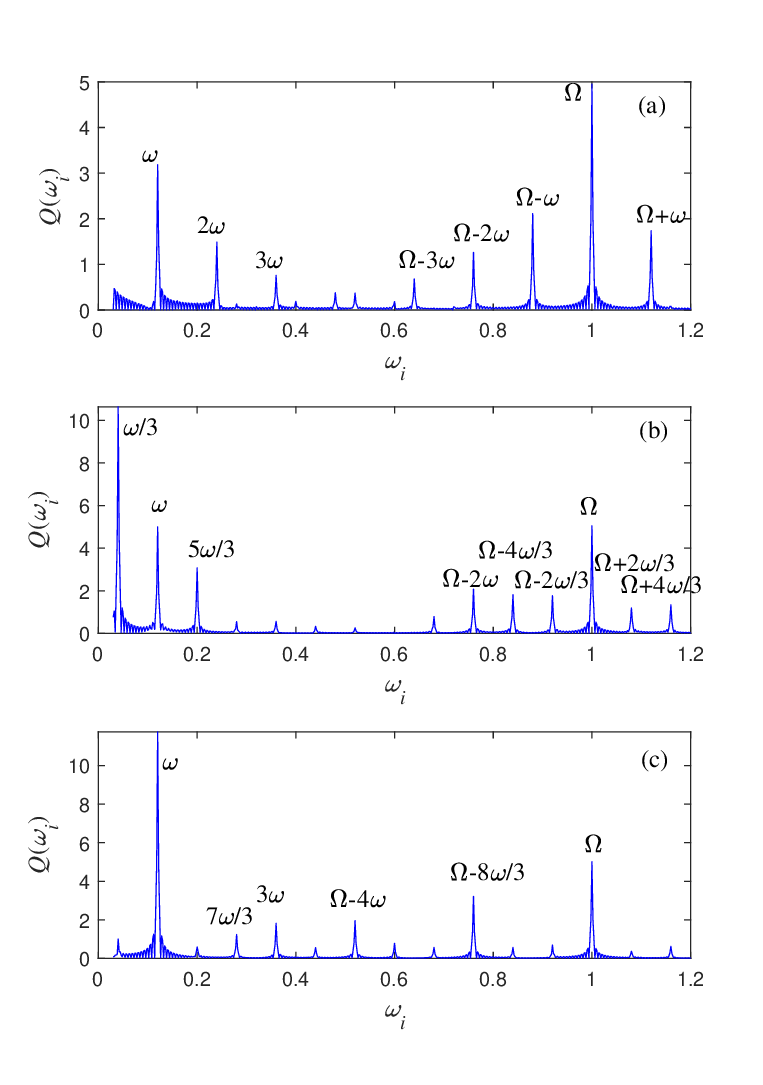}
\end{center}
\caption{Response amplitudes at different frequencies present nonlinear vibrational resonance in the system of Eq.~(\ref{eq16}) for different values of the amplitude $B$ of the fast-varying auxiliary signal. (a) $B=1.13$, (b) $B=1.14$, (c) $B=1.15$. The parameters are $\omega_0^2 =-1.5$, $\beta=1$, $A=0.1$, $\omega=0.12$, $\Omega=1$.}
\label{nlvr}
\end{figure}

As mentioned in Section~\ref{int}, different works on vibrational resonance at nonlinear frequencies have been carried out in different systems \cite{ref26}, \cite{ref94}-\cite{ref99}, \cite{ref254}. According to the linear response theory, the conventional vibrational resonance is obvious at the frequency of the characteristic signal. In addition, in Eq.~(\ref{eq16}), based on the nonlinear response theory, we know that there must be nonlinear frequencies in the response. Ghosh and Ray \cite{ref94} first investigated vibrational resonance at the frequency $2\omega$. Yang et al. \cite{ref98} expanded vibrational resonance to more nonlinear harmonic components, {\it i.e.}, subharmonic frequencies, superharmonic frequencies, or combined frequencies. Figure~\ref{nlvr} provides vibrational resonance at these frequencies. Especially in Fig.~\ref{nlvr}(b), the response amplitude at the subharmonic frequency component $\omega/3$ is much greater than that at $\omega$. In Fig.~\ref{nlvr}(a), the combined frequency $\Omega-\omega$ is also a large value. In fact, in Fig.~\ref{nlvr}, we know that the nonlinear vibrational resonance can occur at more nonlinear frequencies. The frequencies in the output can be controlled by the fast-varying excitation. In addition, the nonlinear vibrational resonance is found by Das and Ray at combined frequencies $\omega-\omega_0$ and $\omega+\omega_0$ in the underdamped Duffing oscillator Eq.~(\ref{eq17}) \cite{ref101}.

\subsection{Ultrasensitive vibrational resonance}

\begin{figure}[!h]
\begin{center}
\includegraphics[width=0.8\linewidth]{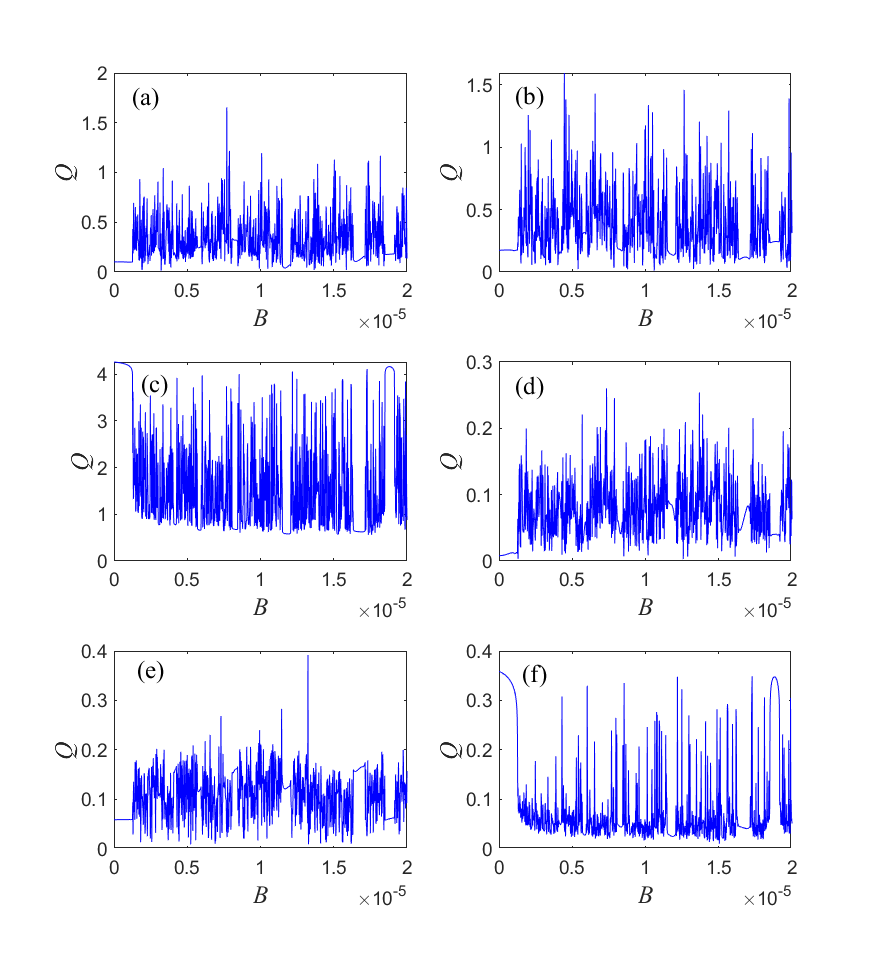}
\end{center}
\caption{The response amplitude $Q$ at some given frequencies versus the amplitude $B$ of the fast-varying auxiliary signal shows ultrasensitive vibrational resonance in a very small range of $B$ in the system of Eq.~(\ref{eq17}). The response amplitude $Q$ is calculated at the frequency $\omega/3$, $\omega/2$, $\omega$, $5\omega/3$, $2\omega$ and $3\omega$ from (a)-(f) successively. The parameters are $\delta = 0.15$, $\omega_0^2 = -1$, $\beta = 1$, $A=0.245$, $\omega = 0.9$ and $\Omega = 9.8$.}
\label{sensitive}
\end{figure}

Another peculiar resonance pattern emerges in the ultrasensitive response amplitude $Q$, induced by a phase space fractal structure \cite{ref205}. The system employed here is Eq.~(\ref{eq17}). When $B=0$, the response of the nonlinear system exhibits a basin of attraction, indicating a highly fractalized phase space. Upon introducing $B \ne 0$, the high-frequency perturbation leads to an ultrasensitive variation of $Q$. Within a very small range of $B$, this induces an ultrasensitive variation of $Q$, as depicted in Fig.~\ref{sensitive}. The response amplitude curve is characterized by multiple peaks in a fractal-like structure. Daza et al. \cite{ref205} referred to this phenomenon as \textit{ultrasensitive vibrational resonance}. Notably, ultrasensitive vibrational resonance occurs not only at the excitation frequency but also at various nonlinear frequencies. Moreover, the ultrasensitive response is highly dependent on initial conditions and calculation time. In Fig.~\ref{sensitive}, a total calculation time of $40T$ is used, with $T=2\pi/\omega$, and the initial simulation conditions are set to $x(0) = 0$ and $\dot x(0) = 0$.

\begin{figure}
\begin{center}
\includegraphics[width=0.7\linewidth]{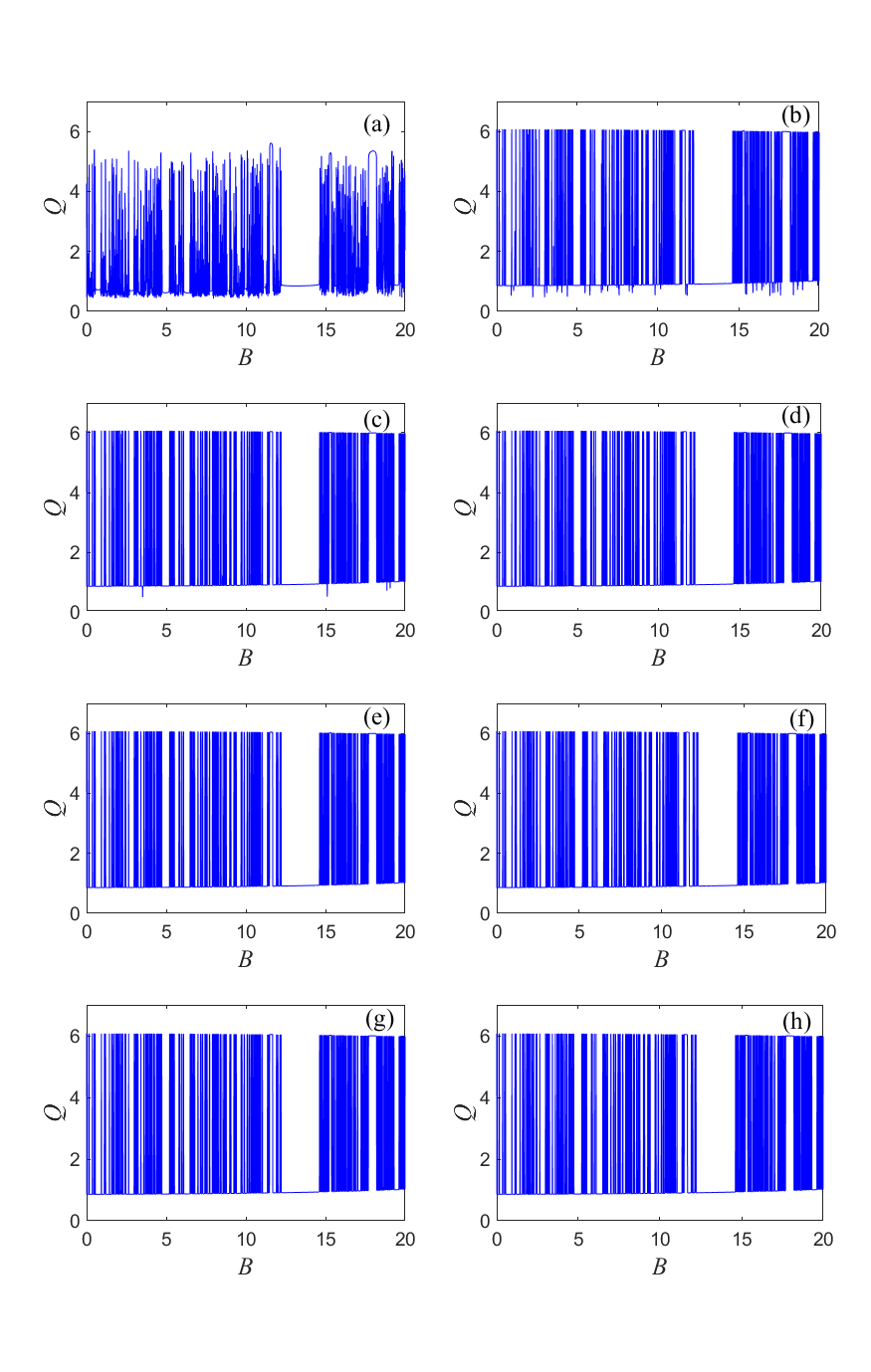}
\end{center}
\caption{The response amplitude $Q$ at the excitation frequency $\omega$ versus the amplitude $B$ of the fast-varying auxiliary signal shows ultrasensitive vibrational resonance in a relatively large range of $B$ in system of Eq.~(\ref{eq17}) in different time intervals. The total calculation time starts from the very beginning of $(40m+1)T$ to the end of $40jT$, $m=j-1$ and $j=1$-$8$ in turn from (a)-(h). The parameters are $\delta = 0.15$, $\omega_0^2 = -1$, $\beta = 1$, $A=0.245$, $\omega = 0.9$ and $\Omega = 9.8$. The initial conditions are $x(0) = 0$, $\dot x(0) = 0$.}
\label{sensitive-T}
\end{figure}

To further investigate whether ultrasensitive vibrational resonance is influenced by time, we present Fig.~\ref{sensitive-T}, which shows the patterns of ultrasensitive vibrational resonance at different time intervals. In these subplots, the ultrasensitive vibrational resonance always exists and the $Q-B$ curve becomes stable over time. Interestingly, while there are ultrasensitive vibrational resonance regions, there are also some non-ultrasensitive vibrational resonance regions in the figure.

\begin{figure}
\begin{center}
\includegraphics[width=0.8\linewidth]{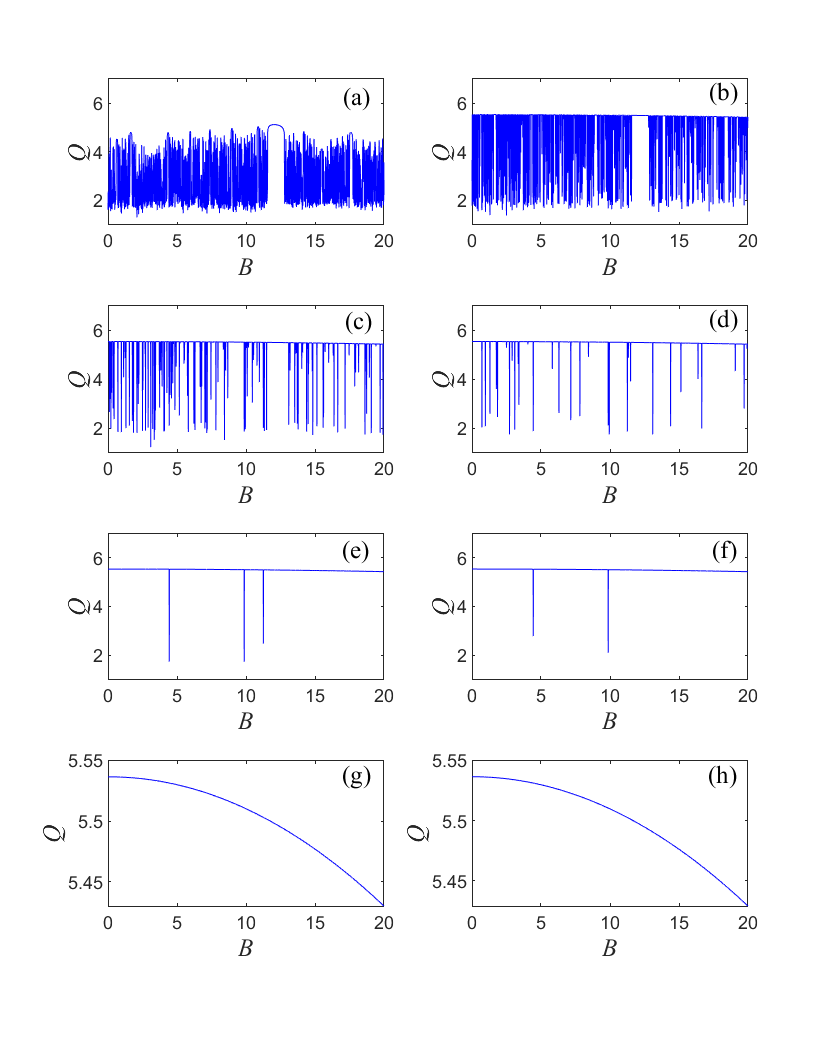}
\end{center}
\caption{The ultrasensitive vibrational resonance caused by the signal amplitude $B$ at the excitation frequency $\omega$ gradually disappears with the increase of time in the system of Eq.~(\ref{eq17}). The total calculation time starts from the very beginning of $(40m+1)T$ to the end of $40jT$, $m=j-1$ and $j=1$-$8$ in turn from (a)-(h). The parameters are $\delta = 0.15$, $\omega_0^2 = -0.6$, $\beta = 1$, $A=0.245$, $\omega = 0.9$ and $\Omega = 9.8$. The initial conditions are $x(0) = 0$, $\dot x(0) = 0$.}
\label{sensitive2-T}
\end{figure}

Although ultrasensitive vibrational resonance persists indefinitely in Fig.~\ref{sensitive-T}, it may transition to a deterministic response for other parameter configurations, as illustrated in Fig.~\ref{sensitive2-T}.

This transition occurs due to the system's response shifting from transient chaos to a periodic state over time. Such a phenomenon, where the system evolves from transient chaos to periodic behavior, has been extensively reported \cite{ref294}, \cite{ref295}. However, as illustrated in Fig.~\ref{sensitive-T}, the response remains chaotic for an extended period, leading to the prolonged presence of ultrasensitive vibrational resonance.

\begin{figure}
\begin{center}
\includegraphics[width=0.7\linewidth]{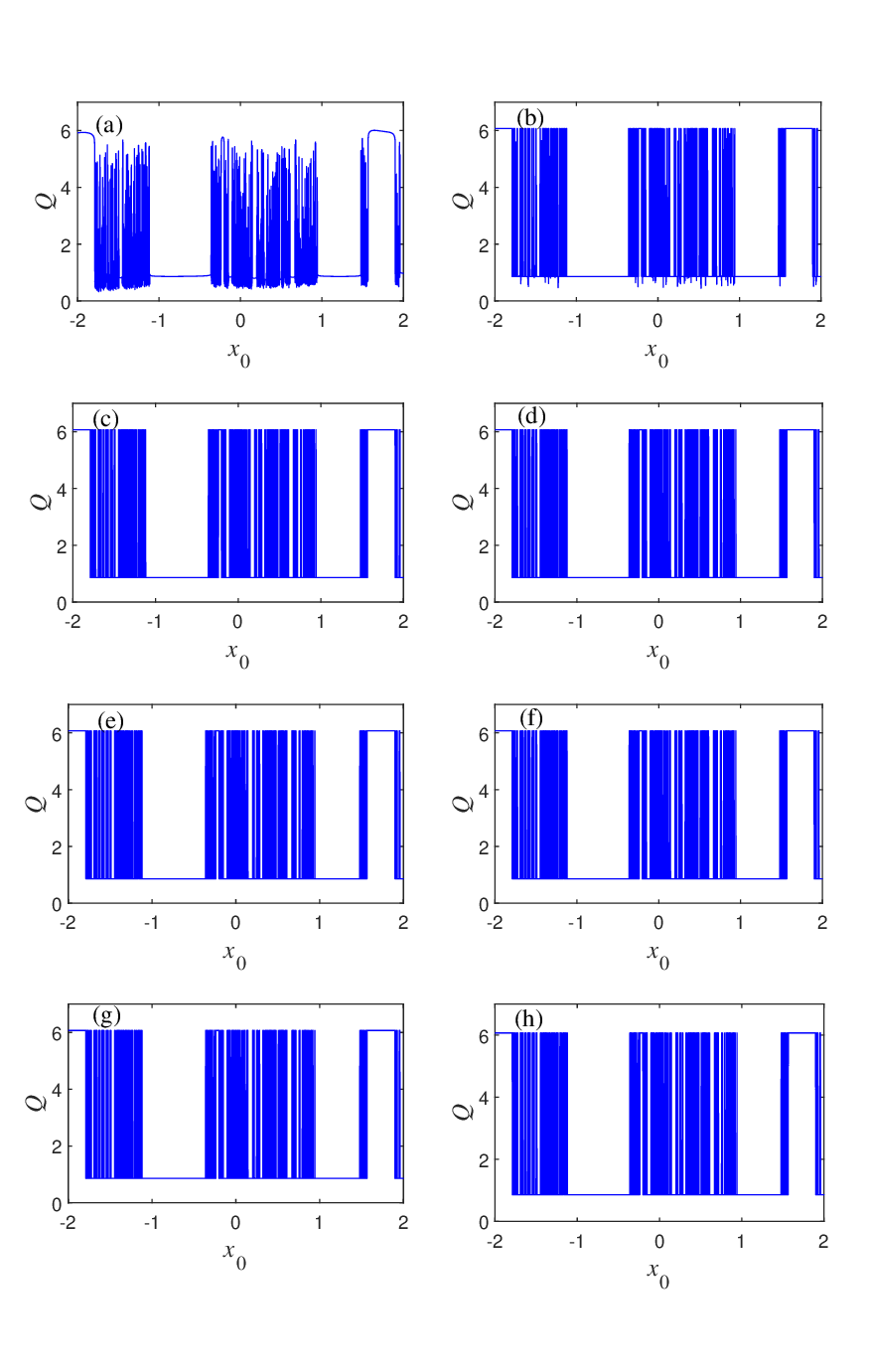}
\end{center}
\caption{The ultrasensitive vibrational resonance caused by the change of the initial condition $x_0$ at the excitation frequency $\omega$ in system of Eq.~(\ref{eq17}) in different time intervals. The total calculation time starts from the very beginning of $(40m+1)T$ to the end of $40jT$, $m=j-1$ and $j=1$-$8$ in turn from (a)-(h). The parameters are $\delta = 0.15$, $\omega_0^2 = -1$, $\beta = 1$, $A=0.245$, $\omega = 0.9$, $B=2$ and $\Omega = 9.8$.}
\label{sensitive-initlal}
\end{figure}

Ultrasensitive vibrational resonance is not solely induced by high-frequency disturbances; rather, it can also arise from variations in the initial simulation conditions, as confirmed in Fig.~\ref{sensitive-initlal}. The presence of distinct regions of ultrasensitive vibrational resonance and nonsensitive responses is evident in the figure. Interestingly, the pattern of ultrasensitive vibrational resonance response remains largely consistent across different time intervals.

\begin{figure}
\begin{center}
\includegraphics[width=0.8\linewidth]{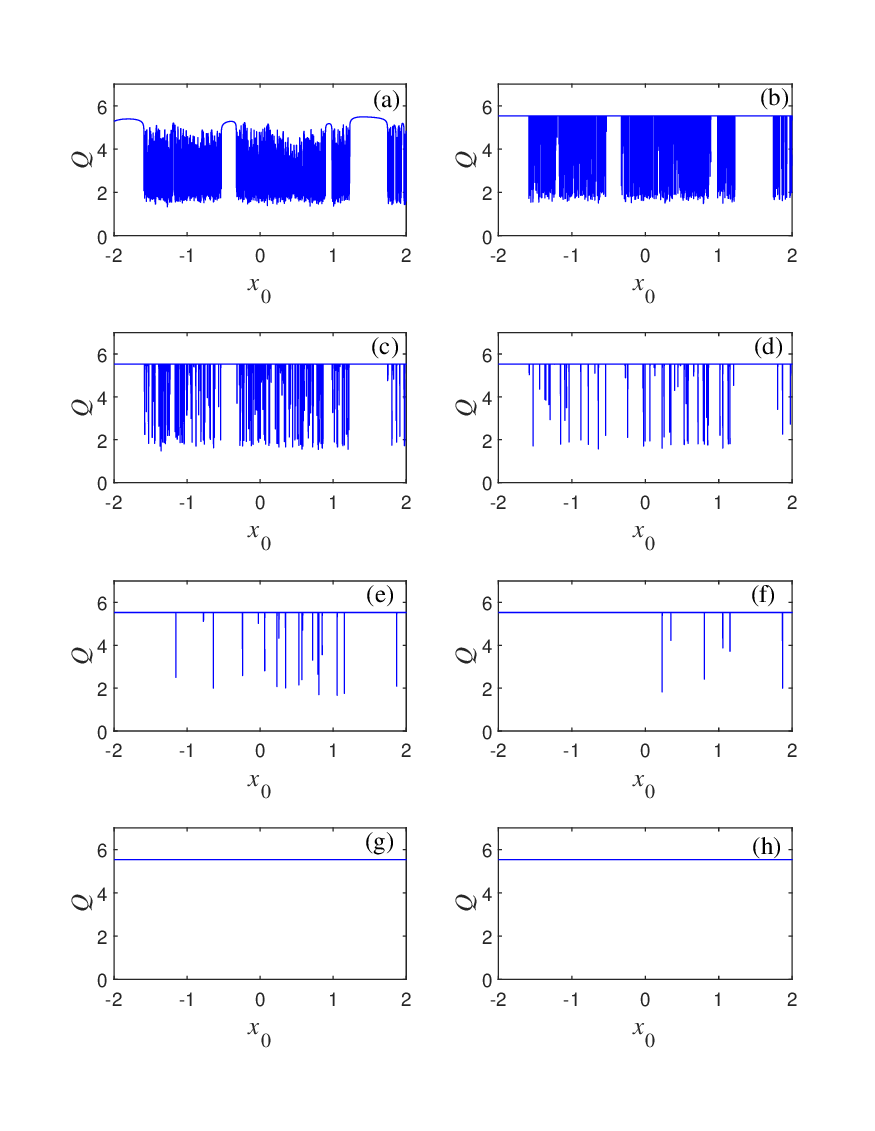}
\end{center}
\caption{The ultrasensitive vibrational resonance caused by the change of the initial condition $x_0$ at the excitation frequency $\omega$ gradually disappears with the increase of time in the system of Eq.~(\ref{eq17}). The total calculation time starts from the very beginning of $(40m+1)T$ to the end of $40jT$, $m=j$-$1$ and $j=1$-$8$ in turn from (a)-(h). The parameters are $\delta = 0.15$, $\omega_0^2 = -0.6$, $\beta = 1$, $A=0.245$, $\omega = 0.9$, $B=2$ and $\Omega = 9.8$.}
\label{sensitive2-initlal}
\end{figure}

Different parameter choices have been explored to examine ultrasensitive vibrational resonance induced by initial conditions, as depicted in Fig.~\ref{sensitive2-initlal}. Over time, the ultrasensitive vibrational resonance gradually diminishes until it transforms into a steady-state periodic response. The findings presented in Fig.~\ref{sensitive2-initlal} closely resemble those in Fig.~\ref{sensitive2-T}, suggesting a transition from transient chaos to periodic behavior within the system.

\begin{figure}[t]
\begin{center}
\includegraphics[width=0.7\linewidth]{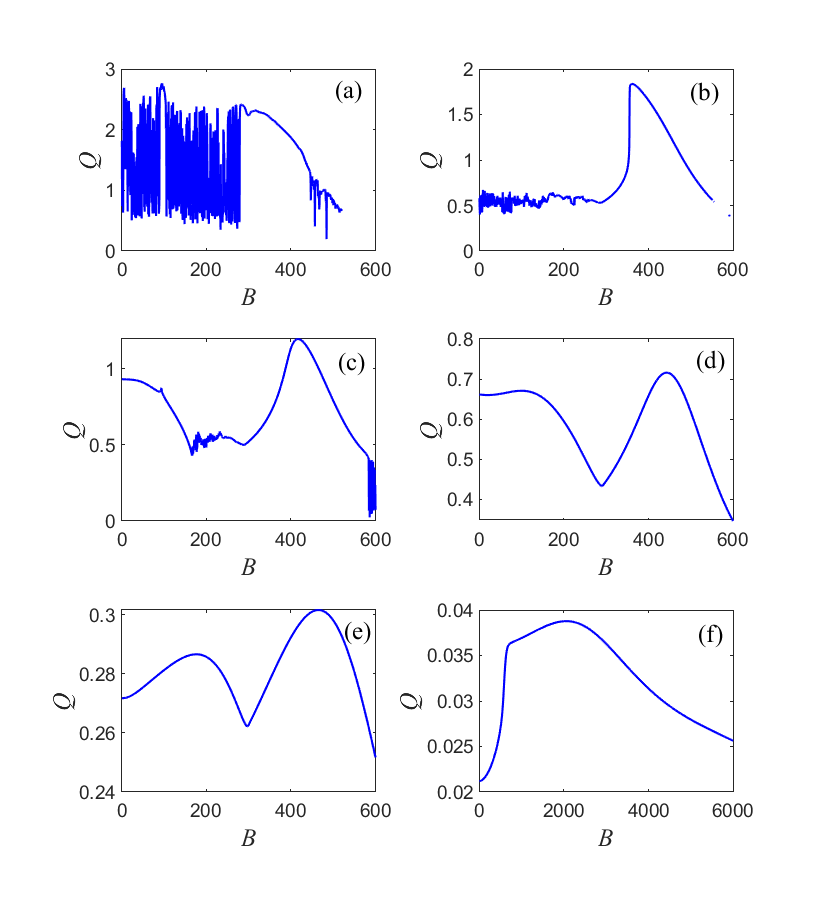}
\end{center}
\caption{The response amplitude of the first oscillator at the excitation frequency $\omega$ versus the signal amplitude $B$ presents ultrasensitive vibrational resonance and traditional vibrational resonance for different values of $\gamma_1$ and $\gamma_2$. In (a) - (f), $\gamma_1 =\gamma_2= 0.05$, $0.2$, $0.3$, $0.5$, $1.2$ and $10$ in turn. The parameters are $g_1=g_2=0.05$, $\beta_1=\beta_2=0.1$, $\omega_1=\omega_2=1.1$, $A=2$, $\omega=1.4$ and $\Omega=10$.}
\label{VRmul}
\end{figure}

Another instance of ultrasensitive vibrational resonance is observed in Eq.~(\ref{eq67}). The damping parameters $\gamma_1$ and $\gamma_2$ play a crucial role in shaping the vibrational resonance pattern. As depicted in Fig.~\ref{VRmul}(a), the $Q-B$ curve exhibits ultrasensitive vibrational resonance within certain ranges of $B$. Conversely, in Fig.~\ref{VRmul}(d), a trough-like feature is observed, which is referred to as \textit{vibrational antiresonance} by Sarkar and Ray \cite{ref16}. This phenomenon indicates that the strong fast-varying signal not only amplifies the weak low-frequency signal but also suppresses the response at the characteristic frequency to a significant extent. Figures~\ref{VRmul}(b) and (c) illustrate the transition between ultrasensitive vibrational resonance and vibrational antiresonance. In Fig.~\ref{VRmul}(f), an intriguing observation is made where the response amplitude $Q$ remains small despite the strong fast-varying signal. This occurrence is attributed to the damping effect, which effectively attenuates the vibration. It's worth noting that with time, the vibrational resonance pattern depicted in Fig.~\ref{VRmul} under different simulation parameters may transition from ultrasensitive vibrational resonance to the traditional vibrational resonance.

In fact, ultrasensitive vibrational resonance presents a highly complex phenomenon. The regions and patterns of ultrasensitive resonance, along with conventional resonance, are unpredictable. Moreover, variations in the system parameters and different forms of external excitation can lead to extremely complex responses. To gain deeper insights into these phenomena, it is essential to integrate various aspects of nonlinear dynamics, including bifurcation, chaos, fractals, and more. Investing further effort into studying ultrasensitive vibrational resonance holds promise for uncovering valuable insights and potential applications.

\subsection{The re-scaled vibrational resonance}
\label{rescvr}

The conventional approach to vibrational resonance assumes that the frequency of the slow-varying excitation is $\omega \ll 1$, or that the smallest width of the aperiodic binary signal is far larger than $1$. However, in many scientific and engineering applications, this assumption may not hold true. In such cases, achieving optimal vibrational resonance in the output becomes challenging because it is difficult to find a suitable correspondence between the characteristic signal and the system parameters. This challenge becomes even more pronounced when the characteristic signal takes on a complex frequency-modulated form. Identifying a matching parameter configuration that leads to vibrational resonance within the framework of conventional theory becomes nearly impossible. To address this issue, Liu et al. \cite{ref111}  proposed the normalized transformation method, while Yang et al. \cite{ref112}  introduced the generalized transformation method. These methods involve scale transformations that convert the fast-varying characteristic signal into a slow-varying signal. This transformation makes it easier for vibrational resonance to occur in the output. Through the scale transformations, the fast-varying characteristic signal can be transformed to a slow-varying signal, making it easy the vibrational resonance to occur in the output. The concept of re-scaled vibrational resonance draws inspiration from the re-scaled stochastic resonance method \cite{ref296}-\cite{ref299}, which has been explored in various studies. By employing these transformation methods, researchers can overcome the limitations of conventional vibrational resonance theory and enhance their ability to achieve resonance in real-world applications.

First, we still take the overdamped Duffing oscillator in Eq.~(\ref{eq16}) as an example. The difference is the frequency of the characteristic signal $\omega\gg 1$. If we do not use the re-scaled vibrational resonance theory, it is difficult to find that the system parameter matches the frequency $\omega$.

Introducing the generalized scale transformation
\begin{equation}
\label{eq175}
t_s = \kappa t, x(t) = z( t_s ),
\end{equation}
herein, $\kappa$ is the scale parameter and $t_s$ is the new time scale. Then, we obtain the equation in the new time scale
\begin{equation}
\label{eq176}
\frac{{dz(t_s)}}{{dt_s}} + \frac{\omega _0^2}{\kappa }z(t_s) + \frac{\beta }{\kappa }{z^3}(t_s) = \frac{A}{\kappa }\cos \left( {\frac{{\omega t_s}}{\kappa }} \right) + \frac{B}
{\kappa }\cos \left( {\Omega \frac{t_s }{\kappa }} \right).
\end{equation}
Let
\begin{equation}
\label{eq177}
{a} = \frac{{\omega _0^2}}{\kappa },\;{b} = \frac{\beta }{\kappa },\;{\omega _1} = \frac{\omega }{\kappa },\;{\Omega _1} = \frac{\Omega }{\kappa }
\end{equation}
and recover the signals to the original magnitude, then Eq.~(\ref{eq176}) turns to
\begin{equation}
\label{eq178}
\frac{{dz(t_s )}}{{dt_s }} + {a}z(t_s ) + {b}{z^3}(t_s ) = A\cos \left( {{\omega _1}t_s } \right) + B\cos \left( {{\Omega _1}t_s } \right).
\end{equation}
In Eq.~(\ref{eq178}), if we choose an appropriate scale parameter, we can make the frequency of the characteristic signal to be small enough to make the vibrational resonance to occur. Then, the analysis of vibrational resonance of Eq.~(\ref{eq178}) can be carried out by the method of direct separation of motions.

\begin{figure}[t]
\begin{center}
\includegraphics[width=0.8\linewidth]{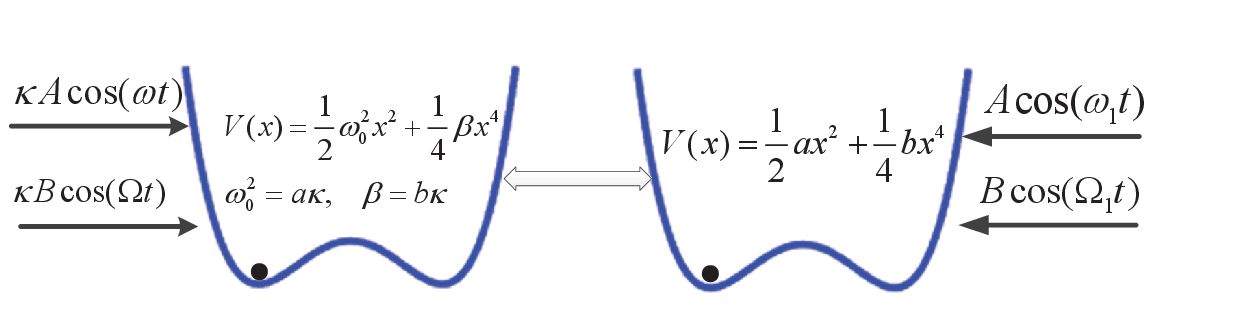}
\end{center}
\caption{The relationship between the re-scaled vibrational resonance and the conventional vibrational resonance.}
\label{rescale}
\end{figure}

Furthermore, for the numerical simulations, the original equation turns to the form
\begin{equation}
\label{eq179}
\frac{{dx}}
{{dt}} + \omega _0^2x + \beta {x^3} = \kappa A\cos (\omega t) + \kappa B\cos (\Omega t).
\end{equation}
Herein, $\omega_0^2$ and $\beta$ are large parameters with $\omega_0^2=a \kappa$, $\beta=b \kappa$. $a$ and $b$ are small parameters in the order of $1$. $\omega$ and $\Omega$ are large parameters. The relationship between the conventional vibrational resonance and the re-scaled vibrational resonance is shown in Fig.~\ref{rescale}. When $\kappa=1$, the re-scaled equation degenerates to the conventional vibrational resonance equation.

\begin{figure}[t]
\begin{center}
\includegraphics[width=0.6\linewidth]{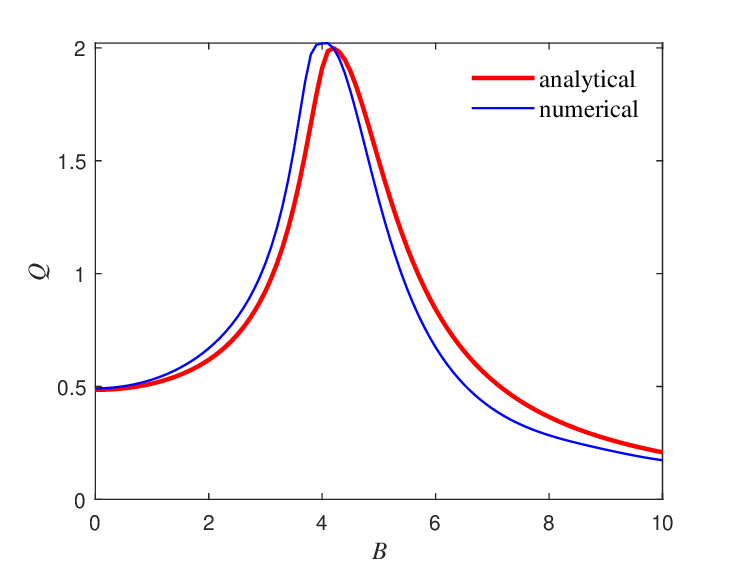}
\end{center}
\caption{The analytical and numerical results of the re-scaled vibrational resonance facilitated by a fast-varying characteristic signal and a faster-varying auxiliary signal. The parameters are $\omega=1500$, $\Omega=10\omega$, $A=0.1$, $\kappa=3000$, $\omega_0^2=-1$, $\beta=1$.}
\label{resVR}
\end{figure}

In Fig.~\ref{resVR}, we provide the analytical results and the corresponding numerical results of the re-scaled vibrational resonance. Because $\kappa=6000$, the re-scaled frequency is $\omega_1=0.5$. The response amplitude at $\omega=1500$ in Eq.~(\ref{eq179}) is the same as the response amplitude at $\omega_1=0.5$ in Eq.~(\ref{eq178}), i.e., Eq.~(\ref{eq16}) when $\omega=0.5$. As a result, the curves in Fig.~\ref{resVR} are the same as the ones in Fig.~\ref{numvsana} although their  $\omega$ values are in huge difference.

The re-scaled vibrational resonance theory can be used not only in the periodic characteristic signal case, but also in the aperiodic characteristic signal case, such as the aperiodic binary signal \cite{ref115}, \cite{ref116}, the linear frequency modulated signal \cite{ref118}, \cite{ref119}, among others. Especially, if the characteristic signal is a frequency-modulated signal, the piecewise idea should be introduced with the re-scaled treatment simultaneously. This is because the instantaneous frequency at a different time may vary greatly. Hence, fixed system parameters cannot always match the instantaneous frequency. In each segment, the instantaneous frequency varies within a small range. Fixed system parameters can match the instantaneous frequency corresponding to this segment. In other segment, we need to change the system parameters to different values to match the instantaneous frequency approximately. Certainly, the number of the segments will influence the vibrational resonance output. A new method needs to be developed to deal with complex frequency-modulated signals, such as the real-time scale transformation method \cite{ref300}-\cite{ref302}.

Another method similar to the re-scaled vibrational resonance is the twice-sampling method. The twice-sampling method was first used in studying stochastic resonance induced by a high-frequency characteristic signal in the bearing fault diagnosis field \cite{ref303}, \cite{ref304}. It was introduced in vibrational resonance analysis of the fast-varying characteristic signal case \cite{ref113}, \cite{ref115}, \cite{ref116}, \cite{ref155}. The results of the two methods are compared in detail in the above four references. These two different methods can achieve the same goal. For twice-sampling vibrational resonance, the following procedure is performed \cite{ref113}.

\renewcommand{\labelenumi}{\theenumi.}
\renewcommand{\theenumi}{\arabic{enumi}}
\begin{enumerate}
\item

Sampling the excitations by the first sampling frequency $f_\mathrm{s}$.
 \item
For the sampled signal in the first step, sampling the new time series by the twice-sampling frequency $f_\mathrm{s} / \gamma$ once more. Then, the frequency of the reconstructed signal will be $\gamma$ times of the original frequency. Here, $\gamma$ is the frequency reduced ratio.
\item
Input the reconstructed signal in the nonlinear system. Then, the time series of the output are obtained.
\item
Recovering the output to the first sampling frequency $f_\mathrm{s}$ according to the frequency reduced ratio $\gamma$.
\item
Calculating the index and tuning the control signal to make vibrational resonance occur.

\end{enumerate}

These two methods can handle not only the nonlinear system excited by two determined signals but also the system excited by a characteristic signal with different types of noise. Re-scaled vibrational resonance requires a system with large parameters to manage the mixed signal after amplifying it to a large amplitude. Conversely, the twice-sampling method necessitates converting the mixed fast-varying signal into a slow-varying one initially, followed by reverting the output to the original time scale through another sampling process. Both the re-scaled and twice-sampling methods have demonstrated good performance, albeit they may lack rigorous mathematical treatment.

\subsection{Role of noise on vibrational resonance}
In a nonlinear system excited by a bi-harmonic signal and noise simultaneously, an interplay between vibrational resonance and stochastic resonance typically occurs. In this subsection, we describe the role of noise in a bistable system, beginning with a pendulum system. Subsequently, we elucidate the impact of noise on logical vibrational resonance. We model the noise using Gaussian white noise denoted as $\xi (t)$, characterized by statistical properties $\left\langle {\xi (t)} \right\rangle  = 0$ and $\left\langle {\xi (t)\xi (0)} \right\rangle  = 2D\delta (t)$.

Baltan\'as et al.~\cite{ref125}, Lin and Huang~\cite{ref130} investigated stochastic resonance controlled by vibrational resonance in Eq.~(\ref{eq137}). Applying the method of direct separation of motions and assuming that the statistical properties of noise are invariant in the period $[0, 2\pi/\Omega]$, the equivalent equation for the slow variable $X$ is obtained as
\begin{equation}
\label{eq180}
         \dot X  + \mu X + \beta X^3  = A \cos (\omega t) + \xi (t) .
\end{equation}
The parameter $\mu$ shares the same expression as in Eq.~(\ref{eq159}). Consequently, both the Signal-to-Noise Ratio (SNR) and the spectrum amplification factor can be determined using stochastic resonance theory, regardless of whether the equivalent system described in Eq.~(\ref{eq180}) is bistable or monostable. The next step involves employing stochastic resonance theory to analyze various stochastic resonance measures of the output. Given the extensive literature on stochastic resonance \cite{ref2}, we refrain from delving further into the analysis of the influence of the fast-varying auxiliary signal on stochastic resonance. In fact, Eq.~(\ref{eq180}) indicates that the fundamental reason for stochastic resonance being controlled by vibrational resonance lies in the direct dependence of the equivalent system's parameter on the fast-varying auxiliary signal. Consequently, adjusting the fast-varying auxiliary signal is equivalent to tuning the system parameter in Eq.~(\ref{eq137}). Furthermore, controlling stochastic resonance through the fast-varying auxiliary signal offers additional advantages. For instance, if we were to control stochastic resonance by adjusting the system parameters using a hardware circuit, we would need to modify the values of the circuit elements. However, by leveraging vibrational resonance, we simply need to adjust the fast-varying auxiliary signal at the input, a task easily accomplished using a signal generator.

Baltanás et al.~\cite{ref125} investigated the phenomena of stochastic resonance and vibrational resonance in a forced bistable system. According to their findings, in the absence of the auxiliary signal, conventional stochastic resonance occurs. For low noise intensities, there is a noticeable vibrational resonance. When the noise is not overly strong, vibrational resonance can enhance the stochastic resonance effect. Specifically, an appropriate value of $B$ improves the maximal value of $Q$. However, in cases of strong noise, both vibrational resonance and stochastic resonance vanish. In such scenarios, the strongest resonance state can be achieved through the re-scaled method or the twice sampling method. Evidently, the auxiliary signal assists in adjusting the parameters of the equivalent system, thereby effectively controlling the strongest resonance through vibrational resonance. Chizhevsky and Giacomelli~\cite{ref128}, \cite{ref214} observed that the mechanism of vibrational resonance results in a higher Signal-to-Noise Ratio (SNR) compared to stochastic resonance.

\begin{figure}[t]
\begin{center}
\includegraphics[width=0.7\linewidth]{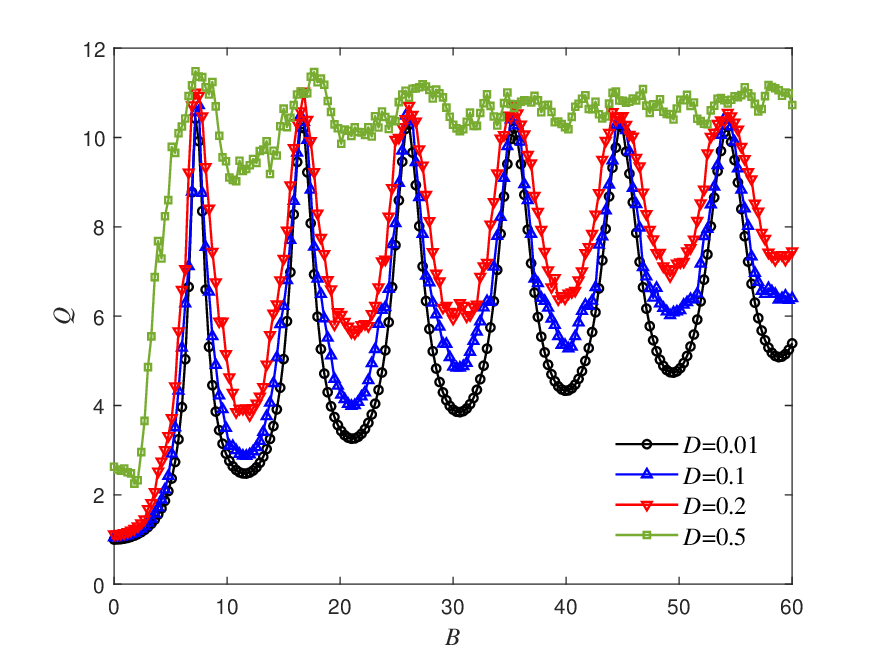}
\end{center}
\caption{The effect of noise on the multiple vibrational resonance in the pendulum system of Eq.~(\ref{eq141}). The parameters are $A=0.1$, $\omega=0.1$, $\Omega=3$.}
\label{VRpn}
\end{figure}

Considering the role of noise in the pendulum system described by Eq.~(\ref{eq141}), Fig.~\ref{VRpn} illustrates that the response amplitude $Q$ is augmented by noise, particularly evident at the valleys of the curve. At these points, stochastic resonance enhances vibrational resonance, underscoring the beneficial effect of noise. Conversely, at the peaks of the curves, the response amplitude undergoes minimal alteration, indicating the limited influence of vibrational resonance. Naturally, excessive noise weakens the response amplitude at the frequency $\omega$. Specifically, strong noise disrupts the coherence of the output, leading to a loss of order. This phenomenon is intuitively comprehensible, and thus we omit providing an example for this case here.

\begin{figure}
\begin{center}
\includegraphics[width=0.65\linewidth]{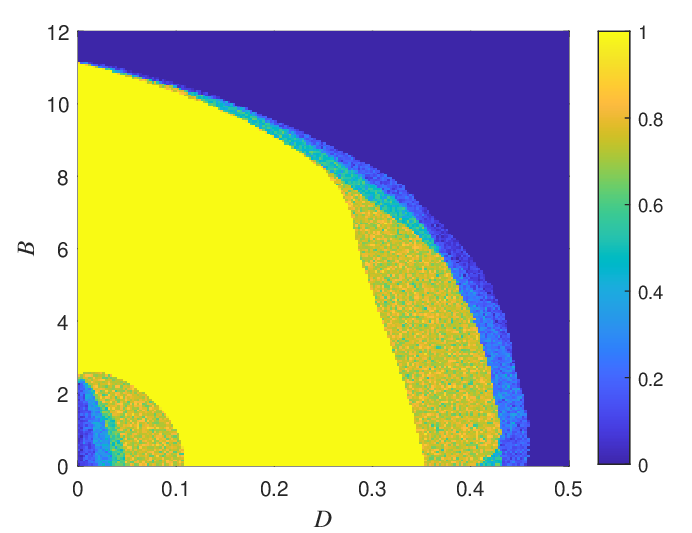}
\end{center}
\caption{The region of the logical vibrational resonance ($\bf OR$ operation) in the two-dimensional plane of the success probability with the noise intensity $D$ and the signal amplitude $B$ in the system of Eq.~(\ref{eq140}). The color coding represents the success probability. The parameters are $\omega_0^2 =-2$, $\beta = 4$, $r = 0.5$, $A=0.1$, $\omega = 1$, $\Omega=20$.}
\label{VRlogn}
\end{figure}

The role of noise on logical vibrational resonance, as described by Eq.~(\ref{eq140}), is depicted in Fig.~\ref{VRlogn}. Notably, both logical stochastic resonance and logical vibrational resonance are evident simultaneously in the figure. With the presence of a small noise, logical resonance occurs at smaller values of $B$. Even when logical stochastic resonance is observed, logical vibrational resonance persists across a wide range of $B$. However, strong noise disrupts the phenomenon of logical vibrational resonance. Through logical vibrational resonance, effective control over logical stochastic resonance can be achieved. In addition, if a chaotic signal replaces the fast-varying signal, the output will manifest a logical chaotic signal \cite{ref305}-\cite{ref307}.

\section{Applications of vibrational resonance}
\label{Appl}
In this section, we show applications of vibrational resonance to image perception enhancement and bearing fault diagnosis.
\subsection{Vibrational resonance applied in image perception}
In a recent study, Morfu et al.~\cite{ref158}, \cite{ref159} explored the application of vibrational resonance in image perception. They devised a resonant threshold detector based on stochastic resonance and vibrational resonance principles to process images. In this approach, a specific image serves as the characteristic signal input to the system. The choice of either noise or a high-frequency sinusoidal signal as the auxiliary signal determines the nature of the detector: when noise is selected, it operates as a stochastic resonance-based detector, whereas employing a high-frequency sinusoidal signal transforms it into a vibrational resonance-based detector. Given that images inherently contain noise, this process essentially represents a vibrational resonance-enhanced stochastic resonance phenomenon. The detector's operation mode can be toggled between stochastic resonance and vibrational resonance through switch control. The effectiveness of the detector is evaluated using the cross-correlation coefficient between the output image and the noise-free source image. Maximal values of the cross-correlation coefficient indicate optimal noise reduction, particularly in the most significant regions of the image, potentially leading to improved perception of image contours.

For a noisy image, the grey levels of the pixels of coordinates $P_{i,j}$ are given by
\begin{equation}
\label{eq181}
{P_{i,j}} = {I_{i,j}} + \sigma {\eta _{i,j}},\quad i = 1 \cdots M,\quad j = 1 \cdots N,
\end{equation}
where $I_{i,j}$ is the pure image without noise, $\eta _{i,j}$ is noise with strength $\sigma$, $M$ and $N$ denote the image size. Image noise may have different types for different backgrounds. Here, for simplicity, we choose the standard Gaussian white noise. For a given image, we use the vibrational resonance detector as below
\begin{equation}
 \label{eq182}
{X_{i,j}} = {P_{i,j}} + B\cos \left( {\frac{{2\pi ij\Omega  \times 0.5(M + N)}}
{{M \times N}} + {\varphi _{i,j}}} \right),\quad i = 1 \cdots M,\quad j = 1 \cdots N,
\end{equation}
where $X_{i,j}$ is the output image processed by the vibrational resonance detector, $B$ and $\Omega$ are the amplitude and frequency of the high-frequency perturbation, $\varphi _{i,j}$ is the noise added to the detector, and $\varphi _{i,j} = \sqrt{2D} \xi(t)$ with $\xi(t)$ is a standard Gaussian white noise.

Once we have the value of $X_{i,j}$, we compare it to a threshold value, $T_h$. If $X_{i,j}$ exceeds this threshold ($X_{i,j} > T_h$), we set it to $X_{i,j}=1$, representing the white level. Otherwise, if $X_{i,j}$ is less than or equal to the threshold ($X_{i,j} \le T_h$), we set it to $X_{i,j}=0$, indicating the black level. Evidently, the final output, $X$, is strongly influenced by the chosen threshold value.

\begin{figure}[!h]
\begin{center}
\includegraphics[width=0.7\linewidth]{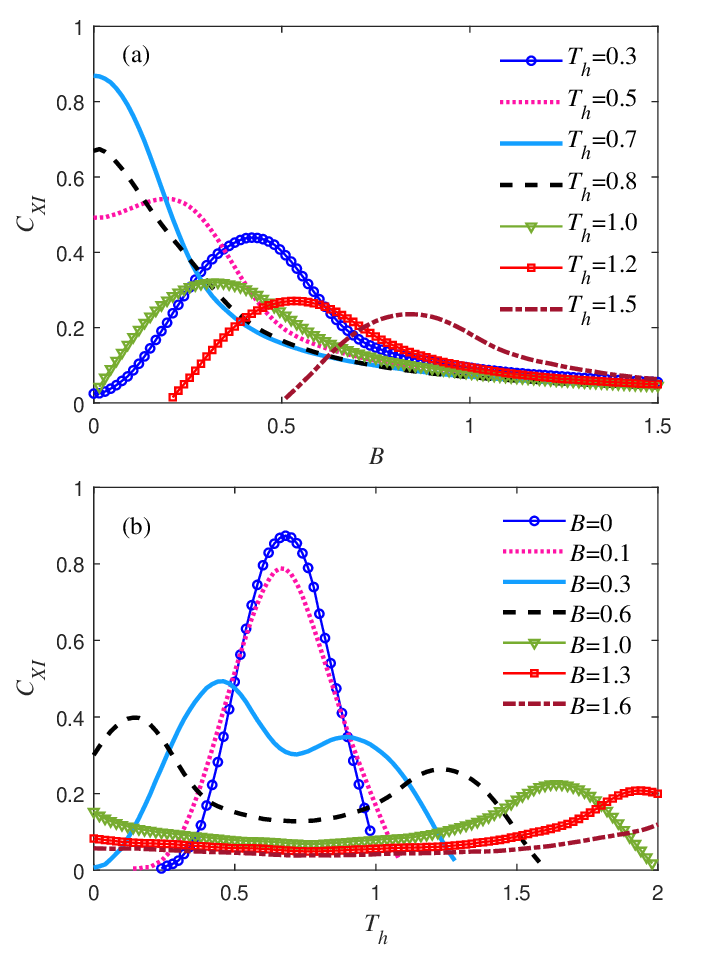}
\end{center}
\caption{The cross-correlation coefficient $C_{XI}$ between the output image and the pure original image. (a) $C_{XI}$ versus the signal amplitude $B$ under different values of threshold value $T_h$. (b) $C_{XI}$ versus the threshold value $T_h$ under different values of the signal amplitude $B$. The parameters are $\sigma=0.1$, $D=10$.}
\label{VRCorr}
\end{figure}

\begin{figure}[!h]
\begin{center}
\includegraphics[width=0.9\linewidth]{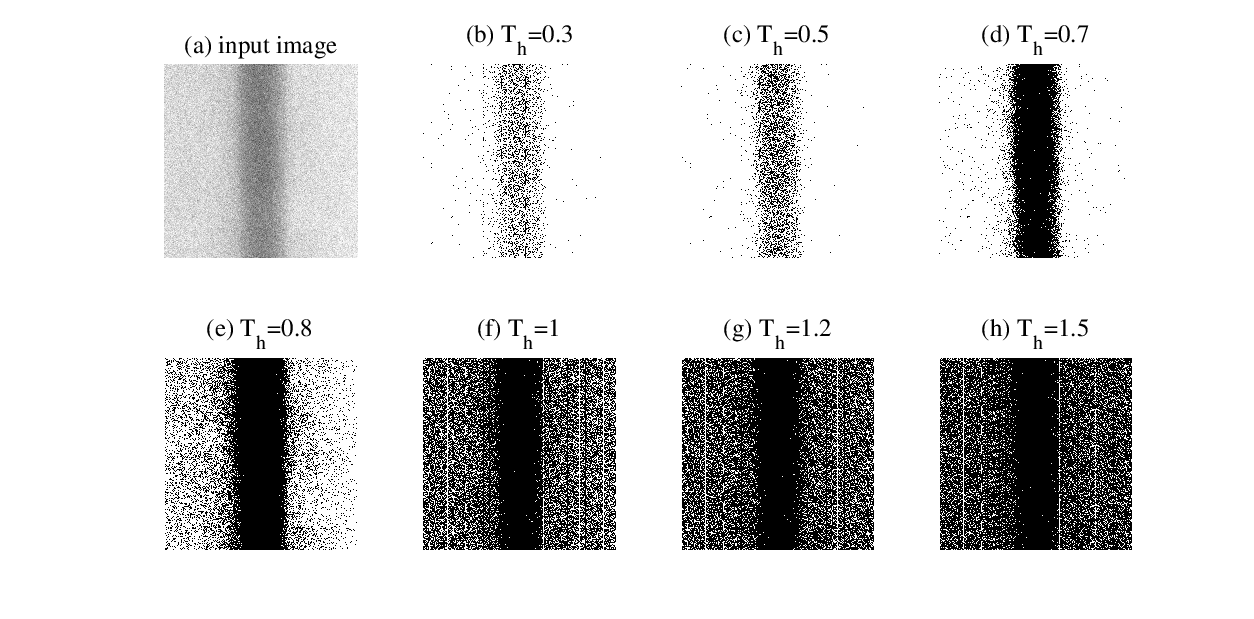}
\end{center}
\caption{The image contour perception based on the vibrational resonance detector of the noisy image under different threshold values. (a) The input noisy image, (b)-(h) The extraction effect of the weld seam image contour. The parameters are $\sigma=0.1$, $D=10$, and $B=0.43$, $0.2$, $0$, $0$, $0.32$, $0.54$ and $0.84$ from (b) to (h) successively.}
\label{VRimage}
\end{figure}

As an index to measure the aperiodic vibrational resonance, the cross-correlation coefficient $C_{XI}$ between the final output $X$ and the original pure image source without noise can be calculated to observe the resonance phenomenon. In Fig.~\ref{VRCorr}(a), for different values of $T_h$, the cross-correlation coefficient $C_{XI}$ versus the amplitude of the fast-varying perturbation $B$ is shown. The image for calculation in Fig.~\ref{VRCorr} is a weld flaw detection image, which is given in Fig.~\ref{VRimage}(a). Again in Fig.~\ref{VRCorr}, for $T_h=0.7$ and $T_h=0.8$, $C_{XI}$ decreases monotonously with the increase of $B$. The cross-correlation coefficient $C_{XI}$ is large due to the dither phenomenon here \cite{ref308}, \cite{ref309}. Especially for $T_h=1.2$ and $T_h=1.5$, it is found that the values of the cross-correlation coefficient disappear. It is because $X_{i,j}=0$ is zero for a large value of $T_h$. If the denominator of the cross-correlation coefficient formula in Eq.~(\ref{eq151}) becomes zero, it renders the coefficient undefined, thereby making its interpretation meaningless. Incidentally, to get a better perceived image, in addition to the cross-correlation coefficient, we also suggest using some other specific metrics of the image. To further investigate the effect of the threshold value on the cross-correlation coefficient, we provide Fig.~\ref{VRCorr}(b) which illustrates the dependence of the cross-correlation coefficient on the threshold value directly. Generally, the cross-correlation coefficient versus the threshold value presents a nonlinear correlation. For different values of $B$, there is always an optimal threshold to make the cross-correlation coefficient to achieve the maximum. There is also ``single-resonance" or ``double-resonance" in Fig.~\ref{VRCorr}(b). The curves in Fig.~\ref{VRCorr} highlight the importance of the threshold selection. In fact, from Fig.~\ref{VRCorr}, we find that the dither phenomenon cannot be ignored. It may bring a better perception effect. The selection of a threshold value can make the image darker or whiter, which is meaningful in image contour detection.

Corresponding to the maximal value of each curve in Fig.~\ref{VRCorr}(a), we present the images processed by the vibrational resonance detector in Fig.~\ref{VRimage}(b) - (h) respectively. The effective perception of image contours is facilitated by vibrational resonance or dithering techniques. By analyzing the curves in Fig.~\ref{VRCorr}, we can identify the optimal amplitude of $B$ for different values of $T_h$. Utilizing the peaks of the cross-correlation coefficient curve, we obtain the optimal output image corresponding to the specific threshold value. Notably, the weld contour becomes more distinct in some images of Fig.~\ref{VRimage}(b)-(h). Vibrational resonance plays a crucial role in enhancing the perception of feature contours in noisy images, particularly for certain values of $T_h$. It is essential to recognize that $T_h$ dictates the black-and-white degree of the image, consequently impacting the efficacy of vibrational resonance processing in contour perception. Therefore, selecting appropriate thresholds is vital for refining image features in specific applications. Clearly, the choice of threshold significantly influences the contour of the weld seam area in Fig.~\ref{VRimage}.

To illustrate the advantages of using vibrational resonance to process noisy images, we employ two commonly used image quality evaluation indicators: the mean absolute error and the peak signal-to-noise ratio. The mean absolute error (MAE) is defined as follows
\begin{equation}
 \label{eq183}
        MAE = \frac{{\sum\limits_{i = 1}^M {\sum\limits_{j = 1}^N {\left| {I(i,j) - X(i,j)} \right|} } }}{{M \times N}}.
\end{equation}
The peak signal-to-noise ratio (PSNR) is calculated by
\begin{equation}
 \label{eq184}
        PSNR = 10{\log _{10}}[\frac{{{{255}^2} \times M \times N}}{{\sum\limits_{i = 1}^M {{{\sum\limits_{j = 1}^N {\left[ {I(i,j) - X(i,j)} \right]} }^2}} }}].
\end{equation}
A smaller value of the mean absolute error indicates a smaller deviation from the original image, implying better image quality. Conversely, a higher value of the peak signal-to-noise ratio signifies better image quality.\\
\indent The results of the mean absolute error and the peak signal-to-noise ratio before and after processing by vibrational resonance are presented in Table 2. The original noisy signal is depicted inimage in Fig.~\ref{VRimage}(a), while  Figs.~\ref{VRimage}(b) - (h) show images processed by vibrational resonance.

Analysis of Table~\ref{t2} reveals that vibrational resonance processing effectively improves both the mean absolute error and the peak signal-to-noise ratio, enhancing image quality and perceptibility. Selecting appropriate thresholds based on the specific characteristics of the image can significantly improve image perception, particularly regarding contour features.

\begin{table}
\begin{center}
\caption{Results of the mean absolute error and the peak signal-to-noise ratio for noisy images in Fig.~\ref{VRimage}(a) and images processed by vibrational resonance in Fig~\ref{VRimage} (b) - (h).}
 \vskip 3pt
 \label{t2}
\begin{tabular}{ c c c c c c c c c}
 \hline    \noalign{\smallskip}
   &  (a) & (b) & (c) & (d) & (e) & (f) & (g) & (h) \\   \noalign{\smallskip}
 \hline    \noalign{\smallskip}
 MAE &  198.464 & 26.352 & 26.296 & 26.133 & 26.234 & 26.717 & 26.736 & 26.736 \\  \noalign{\smallskip}
 PSNR &  2.062 & 17.848 & 17.878 & 17.948 & 17.937 & 17.856 & 17.852 & 17.852  \\  \noalign{\smallskip}
 \hline
\end{tabular}
\end{center}
\end{table}

The analysis presented above highlights that the primary role of vibrational resonance in image processing extends beyond conventional noise reduction. Instead, it introduces a novel approach for extracting image contour features, a crucial aspect of image processing. While noise reduction remains an essential component, vibrational resonance offers a distinctive mechanism for enhancing image perception and facilitating feature extraction.
\subsection{Vibrational resonance applied in fault diagnosis}

As a successful application example, the re-scaled vibrational resonance method in the bearing fault diagnosis field has been used in several studies \cite{ref152}-\cite{ref157}.

\begin{figure}[!h]
\begin{center}
\includegraphics[width=0.7\linewidth]{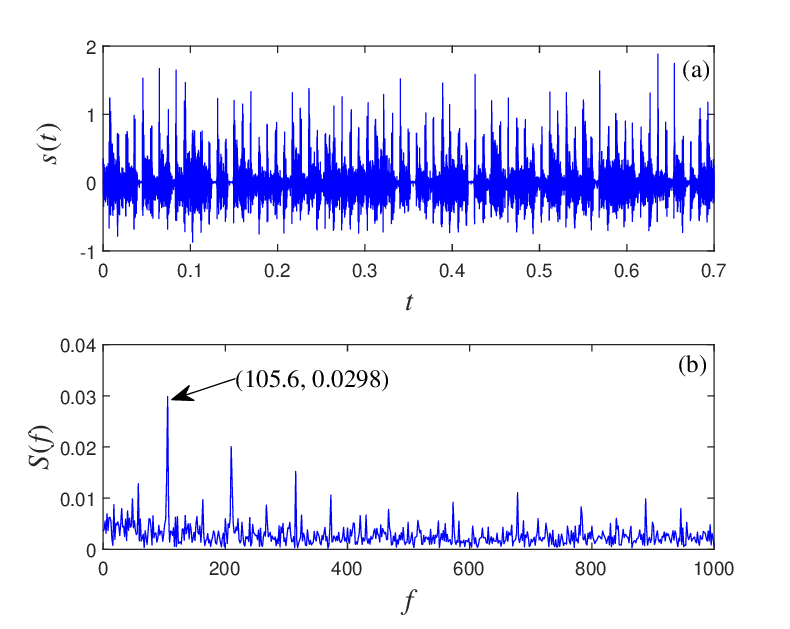}
\end{center}
\caption{Vibration signal of a bearing with a scratch fault in the outer ring. (a) The time domain waveform. (b) The amplitude spectrum versus the frequency of the signal.}
\label{fault1}
\end{figure}

For a vibration signal obtained from an experimental system of a bearing test rig, Fig.~\ref{fault1} presents both the time series in the time domain and the corresponding amplitude spectrum. The characteristic frequency, observed at $105.6\,\mathrm{Hz}$, exhibits a weak amplitude. However, alongside the characteristic frequency, numerous other interference frequency components are evident in the spectrum, rendering the signal considerably complicated, what may pose challenges in realizing vibrational resonance.

\begin{figure}[!h]
\begin{center}
\includegraphics[width=0.7\linewidth]{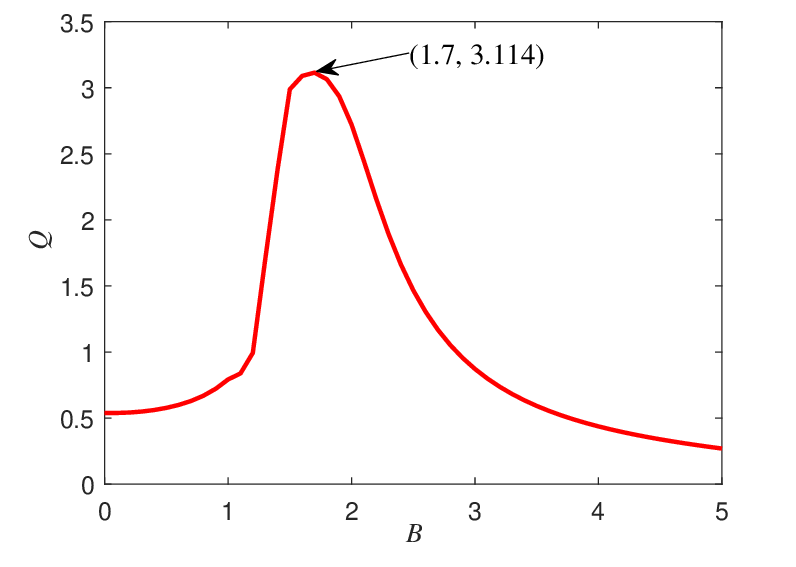}
\end{center}
\caption{The response amplitude at the characteristic frequency versus the amplitude $B$ of the fast-varying auxiliary signal presents the re-scaled vibrational resonance phenomenon.}
\label{fault2}
\end{figure}

Using a classic overdamped bistable model to process the vibration signal, the excitation in the system contains the vibration signal of the bearing and a harmonic auxiliary signal with frequency $2500 Hz$. The re-scaled vibrational resonance is achieved by adjusting the amplitude of high-frequency auxiliary signal, as is shown in Fig.~\ref{fault2}. Although the input signal has complex frequency components as shown in Fig.~\ref{fault1}, the characteristic frequency is amplified to a great extent when the re-scaled vibrational resonance occurs in Fig.~\ref{fault2}. In addition, the parameters herein used are $a=0.02$, $b=0.0001$, $\kappa=100000$, and the meanings of these parameters are given in Eq.~(\ref{eq177}).

\begin{figure}[!h]
\begin{center}
\includegraphics[width=0.7\linewidth]{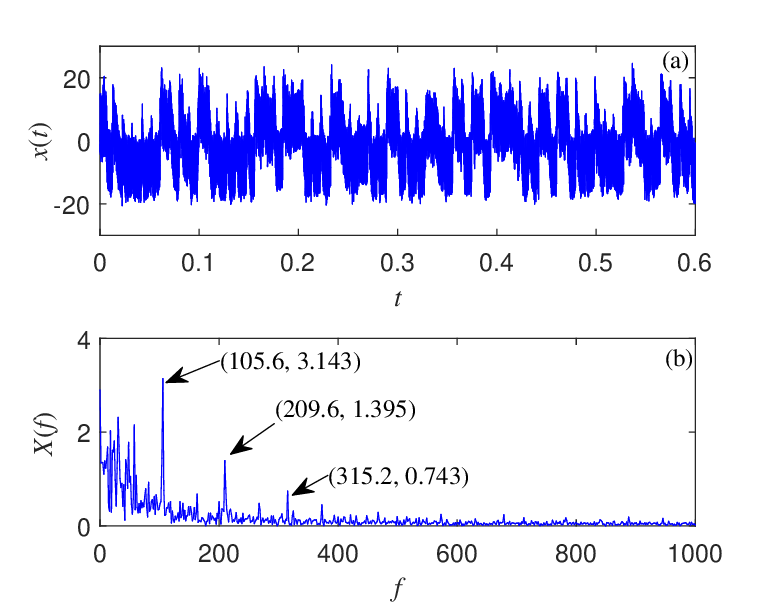}
\end{center}
\caption{The energy improves greatly at the characteristic frequency of the output corresponding to the optimal re-scaled vibrational resonance. (a) The time domain waveform. (b) The amplitude spectrum versus the frequency of the output.}
\label{fault3}
\end{figure}

Figure ~\ref{fault3} gives the time series and the amplitude spectrum of the system output corresponding to the peak in Fig.~\ref{fault2}. In Figs.~\ref{fault1}-\ref{fault3}, the amplitude is calculated by Eq.~(\ref{eq148}) without dividing by the amplitude $A$. From Fig.~\ref{fault3}, the amplification of the characteristic frequency is verified once again. More importantly, compared with Fig.~\ref{fault1}, we find that the interference frequency components are suppressed. The advantage of employing the vibrational resonance method is its capability to amplify a complex signal containing multiple interference frequency components. In contrast, when using a signal amplifier, all signal components are amplified simultaneously.

In this subsection, the vibrational resonance is applied successfully in the bearing fault diagnosis under the constant speed condition. In addition, this method can also be used in bearing fault diagnosis under the time-varying speed condition, such as the vibration signal in Fig.~\ref{figsignal}(l). For this special case, there are two ways to realize the vibrational resonance. The one is using the aperiodic vibrational resonance directly to deal with the engineering signal which has a frequency-modulated characteristic. The other one is using some technology such as order analysis to transmit from the non-stationary signal (corresponding to the time-varying speed condition) to a stationary signal (corresponding to the constant speed condition) \cite{ref310}, \cite{ref311}. Then, the re-scaled vibrational resonance is used to enhance or extract the characteristic frequency of the bearing fault.

\section{Conclusions and outlooks}
\label{conc}

In this review, we provide a broad overview of the vibrational resonance phenomenon, theory, applications and related topics. Vibrational resonance was proposed twenty years ago and focused on the response of a nonlinear system driven by a slow-varying characteristic signal and fast-varying auxiliary signal. With the development of vibrational resonance, it has been extended from the conventional vibrational resonance to the aperiodic vibrational resonance, the nonlinear vibrational resonance, the ultrasensitive vibrational resonance, the re-scaled vibrational resonance, the twice-sampling vibrational resonance, the logical vibrational resonance, etc. Furthermore, the vibrational resonance theory and related topics is still developing rapidly.

At present, there are a large number of nonlinear models for studying vibrational resonance. In this review paper, we have summarized the nonlinear models into five categories. The models described by ordinary differential equations, maps, fractional differential equations, delayed systems, and stochastic systems. Each category, contains nonlinear models associated to numerous disciplines.

In the field of signal processing, the purpose of the vibrational resonance research is to process the characteristic signal. The form of the characteristic signal is one of the most important factors in the study of vibrational resonance. We have given some typical types of the characteristic signal. Among them, vibrational resonance when the characteristic signal in a simple harmonic signal or an aperiodic binary signal have been widely studied in the past two decades. Some complex signals, such as a frequency-modulated signal, bearing vibration signal are also introduced. While certain typical characteristic signals have been used in research, they are insufficient to address the requirements of complex signal processing in engineering. Complex signals often comprise numerous intricate frequency components that can evolve over time. Moreover, the rules governing these changes may be uncertain. When a nonlinear system is subjected to such complex signals, it exhibits considerably more intricate and diverse dynamical phenomena. Achieving vibrational resonance under the influence of such complex signals is a challenging yet significant and pressing issue to address. This necessitates the application of advanced signal processing techniques. Achieving this objective is demanding using only conventional methods for vibrational resonance research. In certain situations, the presence of strong noise makes it challenging to determine the weak characteristic signal in advance. In this situation, the stochastic resonance and vibrational resonance may be combined to realize the optimal resonance in the future. Corresponding to different characteristic signals, the choice of auxiliary signals determines the quality of the system output. If we blindly choose a harmonic signal or a periodic signal as an auxiliary signal, we may not get the strongest resonance output. These are also problems worth of study. With regards to the characteristic signal, we hope to pay much more attention to the complex signal especially for the signals acquired from real systems.

For studying vibrational resonance by using analytical tools, the method of direct separation of motions is mainly used. Needless to say, further new methods are necessary to develop  to identify the vibrational resonance occurring at the subharmonic, superharmonic, or combined frequencies.

In the past studies, we have only paid attention to the forward problem. We know the nonlinear system and the characteristic signal and the auxiliary signal at first, then we investigate the occurrence of vibrational resonance. However, in applied science, there are a lot of inverse problems. Specifically, if we know the nonlinear system and the response, or we know the characteristic signal and the output, how to solve the unknown factor? The inverse problems have been studied in dynamics for decades. Whereas, the vibrational resonance theory is different from the general vibration formulation. The vibrational resonance focuses on the response of a nonlinear system with a very weak signal. It may solve the inverse problem corresponding to some special issues, such as the very small fault feature recognition and dynamic evolution of characteristic signals in machinery fault diagnosis applications, among others.

To conclude, the vibrational resonance investigations in the last decades have been mainly focused on analyzing nonlinear systems with simple characteristic signals in order to study the resonance response. In the future, much attention needs to be paid to find new nonlinear models, complex characteristic signals, and the corresponding inverse problems. Especially, key aspects of the research should be the study of vibrational resonance in experimental settings along with physical and engineering applications.\\

\vspace{0.5em}
{\Large \bf \noindent Declaration of competing interest}
\vspace{0.5em}

The authors declare that they have no known competing financial interests or personal relationships that could have appeared to influence the work reported in this paper.\\

\vspace{0.5em}
{\Large \bf \noindent Author contributions}
\vspace{0.5em}

{\bf Jianhua Yang:} Conceptualization, Writing - original draft, review \& editing, Funding acquisition. {\bf S. Rajasekar:} Conceptualization, Writing - review \& editing. {\bf Miguel A. F. Sanjua\'n:} Conceptualization, Writing - review \& editing, Funding acquisition.\\

\vspace{0.5em}
{\Large \bf \noindent Acknowledgements}
\vspace{0.5em}

The project was supported by the National Natural Science Foundation of China (Grant Nos. 12072362 and 12311530053), the Priority Academic Program Development of Jiangsu Higher Education Institutions, the Spanish State Research Agency (AEI) and the European Regional Development Fund (ERDF, EU) under Project No. PID2019- 105554GB-I00 (MCIN/AEI/10.13039/501100011033). We thank Prof. Chenggui Yao in Jiaxing University and Dr. Jinjie Zhu in Nanjing University of Aeronautics and Astronautics for their useful discussions. We also thank the graduate students Shengping Huang and Tao Gong in CUMT for some numerical calculations.\\

\vspace{0.5em}
{\Large \bf \noindent Appendix}
\vspace{0.5em}

\setcounter{equation}{0}
\renewcommand\theequation{A\arabic{equation}}

The fractional-order derivative of a function $f(t)$ usually has a fractional order $\alpha \in {R^ + }$. For the Riemann-Liouville definition of the fractional-order derivative, it is given as
\begin{equation}
\label{eqA1}
{}_{RL}{D^\alpha }f(t) \buildrel \Delta \over = \frac{{{d^m}}}{{d{t^m}}}\left[ {\frac{1}{{\Gamma (m - \alpha )}}\int_0^t {\frac{{f(\tau )}}{{{{(t - \tau )}^{\alpha  - m + 1}}}}d\tau } } \right].
\end{equation}
\indent For the Caputo definition of the fractional-order derivative, it is defined as
\begin{equation}
\label{eqA2}
{}_C{D^\alpha }f(t) \buildrel \Delta \over = \frac{1}{{\Gamma (m - \alpha )}}\int_0^t {\frac{{{f^{(m)}}(\tau )}}{{{{(t - \tau )}^{\alpha  - m + 1}}}}d\tau }.
\end{equation}
In Eq.~(\ref{eqA1}) and Eq.~(\ref{eqA2}), $\Gamma ( \bullet )$ is the Gamma function, and $m - 1 < \alpha  < m$, $m \in N$.\\
\indent The Gr\"unwald-Letnikov definition of the fractional-order derivative facilitates discretization of numerical calculations and is described in the form
\begin{equation}
\label{eqA3}
{\left. {{D^\alpha }f(t)} \right|_{t = kh}} = \mathop {\lim }\limits_{h \to 0} \frac{1}{{{h^\alpha }}}\sum\limits_{j = 0}^k {{{( - 1)}^j}\left( \begin{array}{l}
 \alpha  \\
 j \\
 \end{array} \right)} f(kh - jh),
\end{equation}
where ${\left( \begin{array}{l} \alpha  \\ j \\ \end{array} \right)}$ is the binomianl coeffcients.\\
\indent More details for the three fractional-order derivative difinations mentioned above are given in \cite{ref261}.\\


\end{document}